\documentclass{aa}  

\usepackage{graphicx}
\usepackage{gensymb}
\usepackage{verbatim}
\usepackage{siunitx}
\usepackage{amsmath}
\usepackage{caption, subcaption}
\usepackage[normalem]{ulem} 
\usepackage{txfonts}
\usepackage[colorlinks,allcolors=blue]{hyperref}
\usepackage{placeins}
\newcommand{\add}[1]{#1}
\newcommand{\addnew}[1]{#1}
\newcommand{\remove}[1]{\textcolor{red}{\sout{}}}

\begin{document}

   \title{The SEDIGISM survey: Molecular cloud morphology}
   \subtitle{II. Integrated source properties}
   \author{K. R. Neralwar\inst{1}\thanks{Member of the International Max Planck Research School (IMPRS) for Astronomy and Astrophysics at the Universities of Bonn and Cologne.},
          D. Colombo\inst{1},
          A. Duarte-Cabral\inst{2},
          J. S. Urquhart\inst{3},
          M. Mattern\inst{4},
          F. Wyrowski\inst{1},
          K. M. Menten\inst{1},
          P. Barnes\inst{5,6},
          \'A. S\'anchez-Monge\inst{7},
          A. J. Rigby\inst{2},
          P. Mazumdar\inst{1},
          D. Eden\inst{8},
          T. Csengeri\inst{9},
          C.L. Dobbs\inst{10},
          V. S. Veena\inst{7},
          S. Neupane\inst{1},
          T. Henning\inst{11},
          F. Schuller\inst{1,12},
          S. Leurini\inst{13},
          M. Wienen\inst{1},
          A. Y. Yang\inst{1},
          S.\,E. Ragan\inst{2},
          S. Medina\inst{1},
          Q. Nguyen-Luong\inst{14}
          }%

   \institute{Max-Planck-Institut f\"ur Radioastronomie, Auf dem H\"ugel 69, 53121 Bonn, Germany \\ email: \texttt{kneralwar@mpifr-bonn.mpg.de}
   \and
   School of Physics \& Astronomy, Cardiff University, Queen’s building, The parade, Cardiff CF24 3AA, UK
   \and
   Centre for Astrophysics and Planetary Science, University of Kent, Canterbury, CT2\,7NH, UK
   \and
   Laboratoire d’Astrophysique (AIM), CEA, CNRS, Universit{\'e} Paris-Saclay, Universit{\'e} Paris Diderot, Sorbonne Paris Cit{\'e}, 91191 Gif-sur-Yvette, France
   \and
   Space Science Institute, 4765 Walnut St Suite B, Boulder, CO 80301, USA
   \and
   School of Science and Technology, University of New England, Armidale NSW 2351, Australia
   \and
   I.\ Physikalisches Institut, Universität zu K\"oln, Z\"ulpicher Strasse 77, 50937 Cologne, Germany
   \and
   Astrophysics Research Institute, Liverpool John Moores University, IC2, Liverpool Science Park, 146 Brownlow Hill, Liverpool, L3 5RF, UK
   \and
   Laboratoire d’astrophysique de Bordeaux, CNRS, Univ. Bordeaux, B18N, allée Geoffroy Saint-Hilaire, F-33615 Pessac, France
   \and
   School of Physics and Astronomy, University of Exeter, Stocker Road, Exeter EX4 4QL, UK
   \and
   Max-Planck-Institut f\"ur Astronomie, K\"onigstuhl 17, D-69117 Heidelberg, Germany
   \and
   Leibniz-Institut für Astrophysik Potsdam (AIP), An der Sternwarte 16, D-14482 Potsdam, Germany
   \and
   INAF – Osservatorio Astronomico di Cagliari, Via della Scienza 5, I-09047 Selargius (CA), Italy
   \and
   The American University of Paris, 2bis, Passage Landrieu 75007 Paris, France
}

  \date{Received XXX; accepted XXX}

  \abstract 
  {\add{The Structure, Excitation, and Dynamics of the Inner Galactic InterStellar Medium (SEDIGISM) survey has produced high (spatial and spectral) resolution $^{13}$CO (2--1) maps of the Milky Way.
  It has allowed us to investigate the molecular interstellar medium in the inner Galaxy at an unprecedented level of detail and characterise it into molecular clouds. 
  In a previous paper, we have classified the SEDIGISM clouds into four morphologies. However, how the properties of the clouds vary for these four morphologies is not well understood.
  Here, we use the morphological classification of SEDIGISM clouds to find connections between the cloud morphologies, their integrated properties, and their location on scaling relation diagrams. We observe that ring-like clouds show the most peculiar properties, having, on average, higher masses, sizes, aspect ratios and velocity dispersions compared to other morphologies. We speculate that this is related to the physical mechanisms that regulate their formation and evolution, for example, turbulence from stellar feedback can often results in the creation of bubble-like structures. We also see a trend of morphology with virial parameter whereby ring-like, elongated, clumpy and concentrated clouds have virial parameters in a decreasing order. Our findings provide a foundation for a better understanding of the molecular cloud behaviour based on their measurable properties.}}

   \titlerunning{SEDIGISM: molecular cloud morphology II}
   \authorrunning{K. R. Neralwar, D. Colombo et al.}
   \maketitle
%

\section{Introduction}

Molecular clouds (MCs) are some of the densest and coldest regions in the interstellar medium (ISM) and the exclusive sites of star formation in galaxies \citep{cloud_review_krumholz, intro_refer_1, miville_2017}.
They are turbulent structures with masses of order 10$^2$--10$^7 \, \mathrm{M_\odot}$ and most of this mass is concentrated in high-mass giant molecular cloud complexes (> 10$^5 \, \mathrm{M_\odot}$ for Milky Way) \citep{roman_duval_2010}. Most MCs are generally magnetically super-critical, though not by a large margin \citep{crutcher_2010, crutcher_2012, cloud_review_krumholz}. Moreover, magnetic fields vary from cloud to cloud and some clouds may be magnetically supported \citep{chapman_2011, barnes_2015}.
MCs span a range of sizes between $\sim$ 1 and $\sim$ 200 pc and present a hierarchical structure, with dense clumps and denser (proto-stellar or pre-stellar) cores \citep{blitz_1986, dendrograms, ballesteros_2020}.
The surface densities of molecular clouds have a wide range; i.e. $\sim$ 1 to more than 1000 $\mathrm{M_\odot \; pc^{-2}}$ \citep{barnes_2018}. A typical value for the surface densities of clouds is $\sim$ 100 $\mathrm{M_\odot \; pc^{-2}}$ \citep{roman_duval_2010, rebolledo_2012} and it may change with the galactic environmental (e.g. M51; \citealp{colombo_2013, sun_2018}).

The \add{correlations} between velocity dispersion and other cloud properties \add{such as} surface density and the virial parameter have led to various interpretations explaining their structure. The historically widely used \citet{larson_realtion} paper concludes that clouds are self-gravitating objects in virial equilibrium and the gravity acting on clouds is counterbalanced by internal forces giving an average virial parameter around unity \citep{solomon_1987, fukui_2008, roman_duval_2010}. \add{Another work, } \citet{ballesteros_paredes_2011} argues that clouds are in free-fall collapse\add{, but the uncertainties/scatter in the size-linewidth relation allow for virial equilibrium to be a consistent mechanism.} \citet{field_2011} present clouds as marginally self-gravitating pressure-confined objects.  
A more recent work, \citet{vazquez_2019}, presents the global hierarchical collapse (GHC) scenario as a description of the processes involved throughout the life cycle of molecular clouds, starting from their formation in the diffuse medium to their destruction by the massive stars. It proposes a `moderated' free-fall to maintain the cloud structure, that takes into account the interplay at different scales and includes processes, e.g. stellar feedback, that affect the overall cloud evolution.

\add{The filamentary nature of molecular clouds is ubiquitous in the Galaxy and }
most of the star-forming clumps are situated in large scale filamentary structures \citep{globular_filament_ref_nessie_Stru, globular_filaments_2, polychroni_2013, hacar_2013, ATLASGAL_filaments, olmi_2016, kainulainen_2017, bresnahan_2018, mass_vel_scaling_relation, ballesteros_2020, arzoumanian_2022}. \add{The evolution of filaments is influenced by various mechanisms, e.g., ISM turbulence, shocked flows, supernovae feedback, galactic shear} \citep{chen_2020, colombo_2021}. These filaments \add{can be} further classified into various types \add{ -- e.g. bones, giant molecular filament (GMF) --} based on their properties (e.g. aspect ratio; \citealp{filaments_catherine_zucker}). \add{The variations in filament properties might be a result of the Galactic environment and the scales at which they evolve. For example, }
recent studies attribute the formation of GMFs to galactic shear and large-scale gas motions, considering these filaments as a subset of giant molecular clouds \citep{GMF_formation_1, filaments_simulation_ana_16}.

The formation and evolution of molecular clouds is still not completely understood \citep{ballesteros_2020}. Similarly, there is no unique global property of molecular clouds that can define its ability to form high-mass stars. 
Modern approaches towards studying molecular clouds include an analysis of their three-dimensional structure \citep{zucker_2021} and their morphological classification \citep{yuan_2021} into filamentary and non-filamentary structures.
Relating the cloud morphology to its properties and Galactic environment may provide us a better understanding of molecular clouds and star formation. It could also help us understand the physical processes behind the structure of clouds \citep{arzoumanian_2022}.

\add{Stellar feedback plays an important role in regulating the star formation in a galaxy and shaping the neighbouring molecular clouds. Stars can impart mechanical energy in the form of turbulence, stellar winds and shocks, and radiative energy, which heats up the gas, leading to the creation and destruction of molecular clouds and their substructures. Feedback drives the mass, energy, momentum and metal enrichment into the surrounding ISM, inducing new star formation events} \citep{ballesteros_2020, schneider_2020, beuther_2022}. \add{On the other hand, it can also shread the natal molecular gas, limiting further star formation} \citep{geen_2016, james_paper}\add{. Feedback often results in the formation of interstellar bubbles }\citep{deharveng_2010, jayasingghe_2019}\add{, which are some of the most morphologically complex structures in the Galaxy.} For instance, the bubble RCW 49 \citep{schneider_2020, rodgers_1960} shows an expanding shell that is decoupled from the ambient ISM \citep{tiwari_2021}.
Similarly, a fragmented ring of molecular gas is detected around the ionised region surrounding the bubble RCW 120 \citep{zavagno_2007, rodgers_1960}, the presence of which is attributed to the collect and collapse mode \citep{elmegreen_1977, zavagno_2006, jianjun_2020, luisi_2021}. \add{These bubbles/shells are prime examples of the ring-like clouds discussed in this work.}

In this paper, we try to find a connection between the morphology of molecular clouds and their properties. The paper is organised as follows: Sec. \ref{sec: data} provides a brief overview of the survey data and the cloud catalogue used for the analysis. \add{Sec. \ref{sec: morphological cloud classification} gives a brief description of the methods used for classification of clouds into different morphologies, and the resulting morphological classes and cloud samples. We have discussed these in detail in \citet[hereafter Paper I]{neralwar_2022}.
In sec. \ref{sec: results} we analyse the integrated properties of molecular clouds for the four morphological classes. We also check if the results from the different cloud samples agree with each other for the different morphologies. 
In Sec. \ref{sec: scaling relation} we study the Larson's and Heyer's scaling relations, and in Sec. \ref{sec: discussions} we discuss the results and the physical processes leading to the observed morphologies. We summarise our main findings in Sec. \ref{sec: summary}. 
}

\section{Data}\label{sec: data}


SEDIGISM (Structure, Excitation, and Dynamics of the Inner Galactic InterStellar Medium) is a southern hemisphere inner Galaxy survey that probes the \add{moderately} dense ($\approx 10^{3} \; \mathrm{cm}^{-3}$) interstellar medium. 
It covers $-60 \degree \leq l \leq +18 \degree$ and  $|b| \leq 0.5 \degree$ in several molecular lines, mainly the $J = 2$--$1$ transitions of $^{13}$CO and C$^{18}$O. The observations for the survey were carried out using the 12m Atacama Pathfinder EXperiment (APEX, \citealt{gusten_2006}) telescope.
A detailed description of the observations, data reduction and data quality checks is provided in \citet{schuller_2017} and \citet{SEDIGISM_1}.

\citet[][hereafter called DC21]{ana_paper} used the first data release (DR1) \citep{SEDIGISM_1} of the SEDIGISM survey to obtain a catalogue of 10663 clouds using the $\sc{scimes}$\footnote{\url{https://github.com/Astroua/SCIMES}} algorithm \citep[detailed description in][]{SCIMES, scimes_2}. The data has a FWHM beam size of \ang{;;28} with a 1 $\sigma$ sensitivity of 0.8--1.0 K at $0.25 \; \mathrm{km \, s^{-1}}$ spectral resolution. In Paper I, we classify these clouds based on their morphology \add{(see Sec. \ref{sec: morphological cloud classification})} and thus add more information to the original catalogue\footnote{\url{https://sedigism.mpifr-bonn.mpg.de/cgi-bin-seg/SEDIGISM_DATABASE.cgi}}.

This cloud catalogue (hereafter called SEDIGISM cloud catalogue) presents the directly measured properties and the derived properties for the clouds identified with $\sc{scimes}$. Each cloud is assigned an ID from $\sc{scimes}$ and a name using its Galactic coordinates. The directly measured properties include the clouds' position, velocity, velocity dispersion, size and the intensity associated with each cloud. The catalogue also lists near, far and adopted kinematic distances to clouds, their reliability and uncertainties, and the presence of star formation tracers. These properties are used to derive other properties like the aspect ratio, mass, surface density and the virial parameter. The catalogue also provides the values of the properties obtained after beam deconvolution, which are used in the current work.


\section{Morphological cloud classification}\label{sec: morphological cloud classification}

We study the integrated cloud properties for different morphologies and this demands that the clouds \add{be} highly resolved and have reliable distance measurements. Thus, we use the clouds belonging to the science sample from DC21 to study their properties. The clouds in the science sample have reliable distance measurements, are well resolved (cloud area > 3$\Omega_{beam}$) and do not lie on the latitude edges of the survey (edge = 0; in SEDIGISM catalogue).

\begin{figure*}
    \centering
    \begin{subfigure}[b]{0.45\textwidth}
    \includegraphics[width = \textwidth, keepaspectratio]{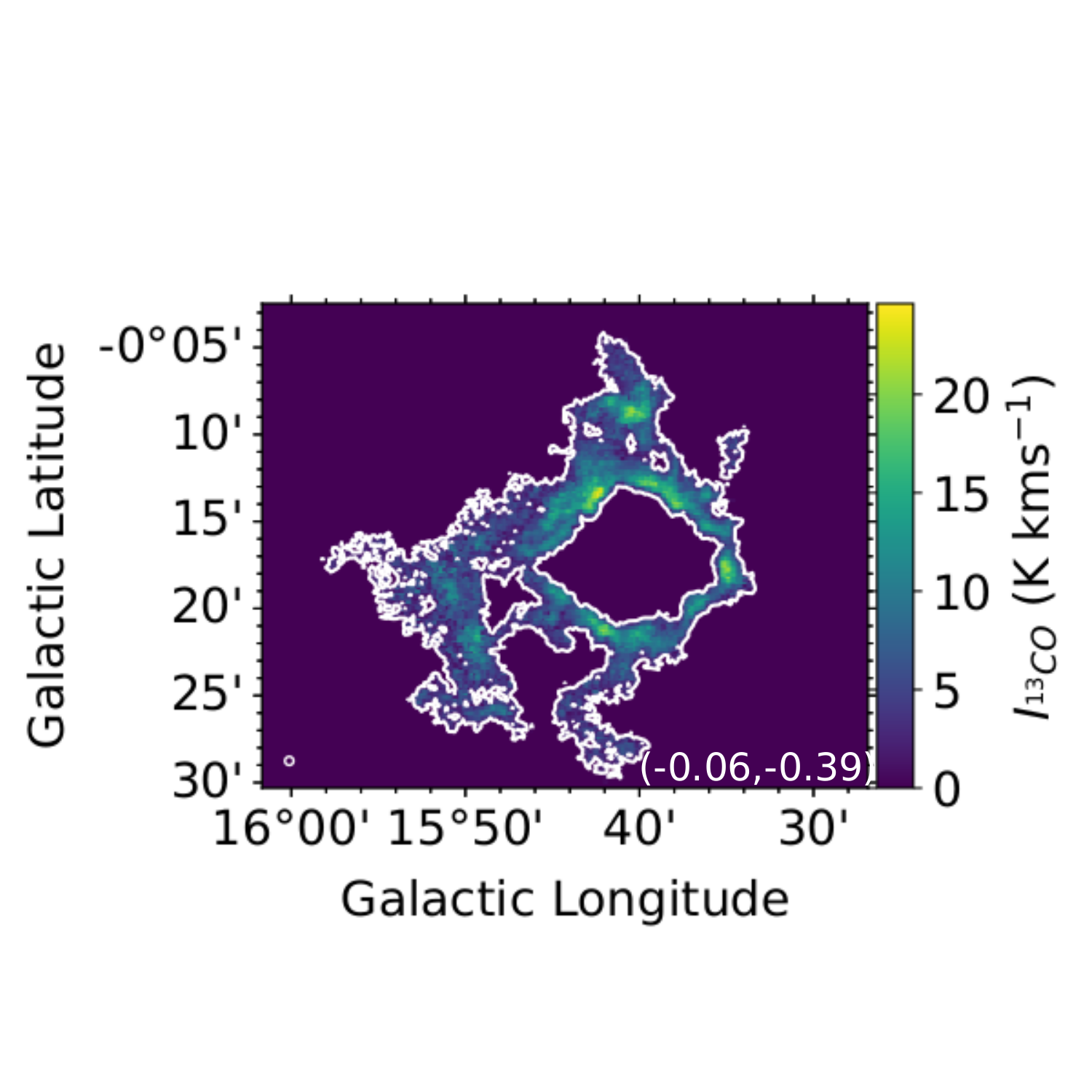}
    \caption{Ring-like cloud ($J$-bubble; cloud id: 10326)}
    \label{fig: ring-like cloud image}
    \end{subfigure}
    \begin{subfigure}[b]{0.45\textwidth}
    \includegraphics[width = \textwidth, keepaspectratio]{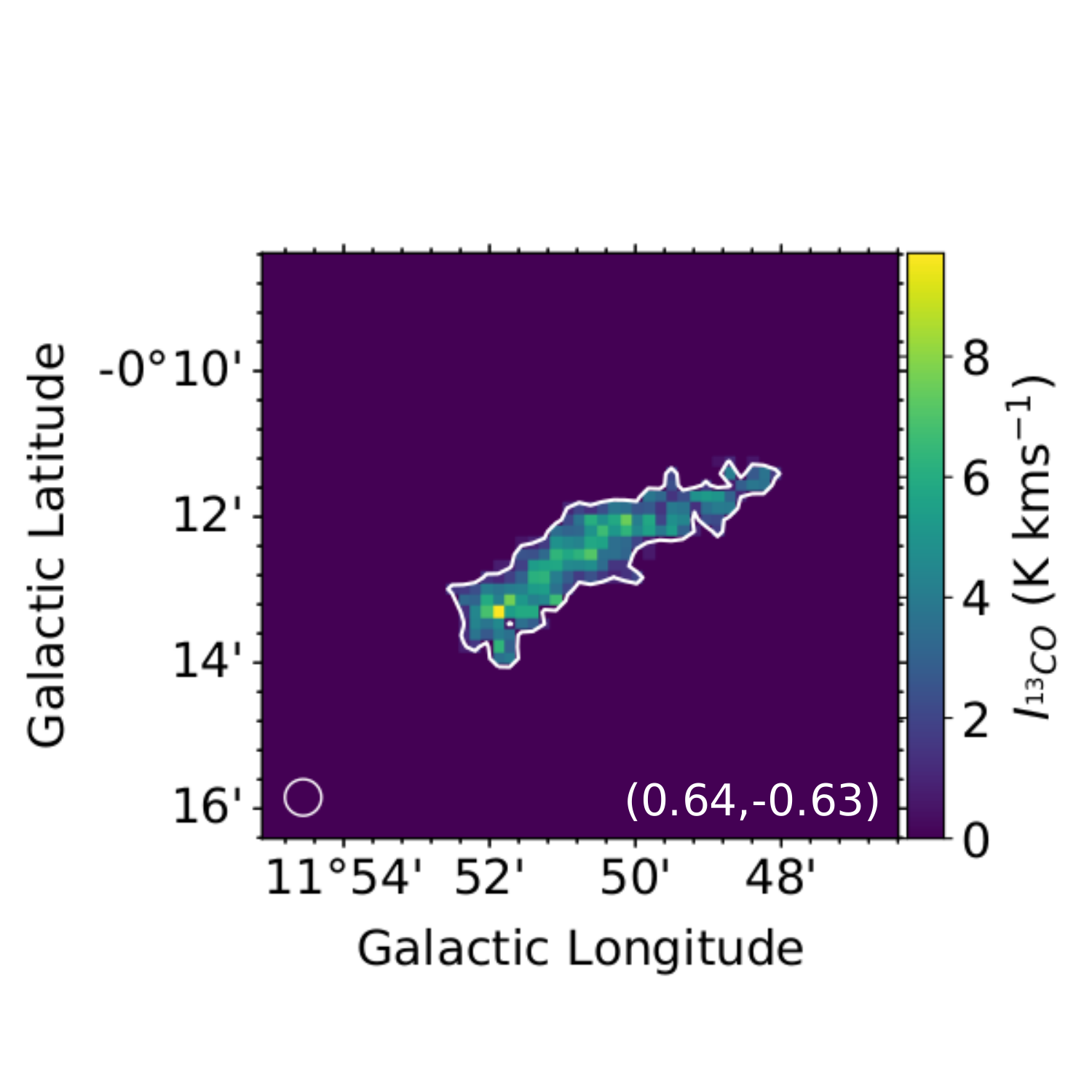}
    \caption{Elongated cloud ($J$-filament; cloud id 9600)}
    \label{fig: elongated cloud image}
    \end{subfigure}
    \begin{subfigure}[b]{0.45\textwidth}
    \includegraphics[width = \textwidth, keepaspectratio]{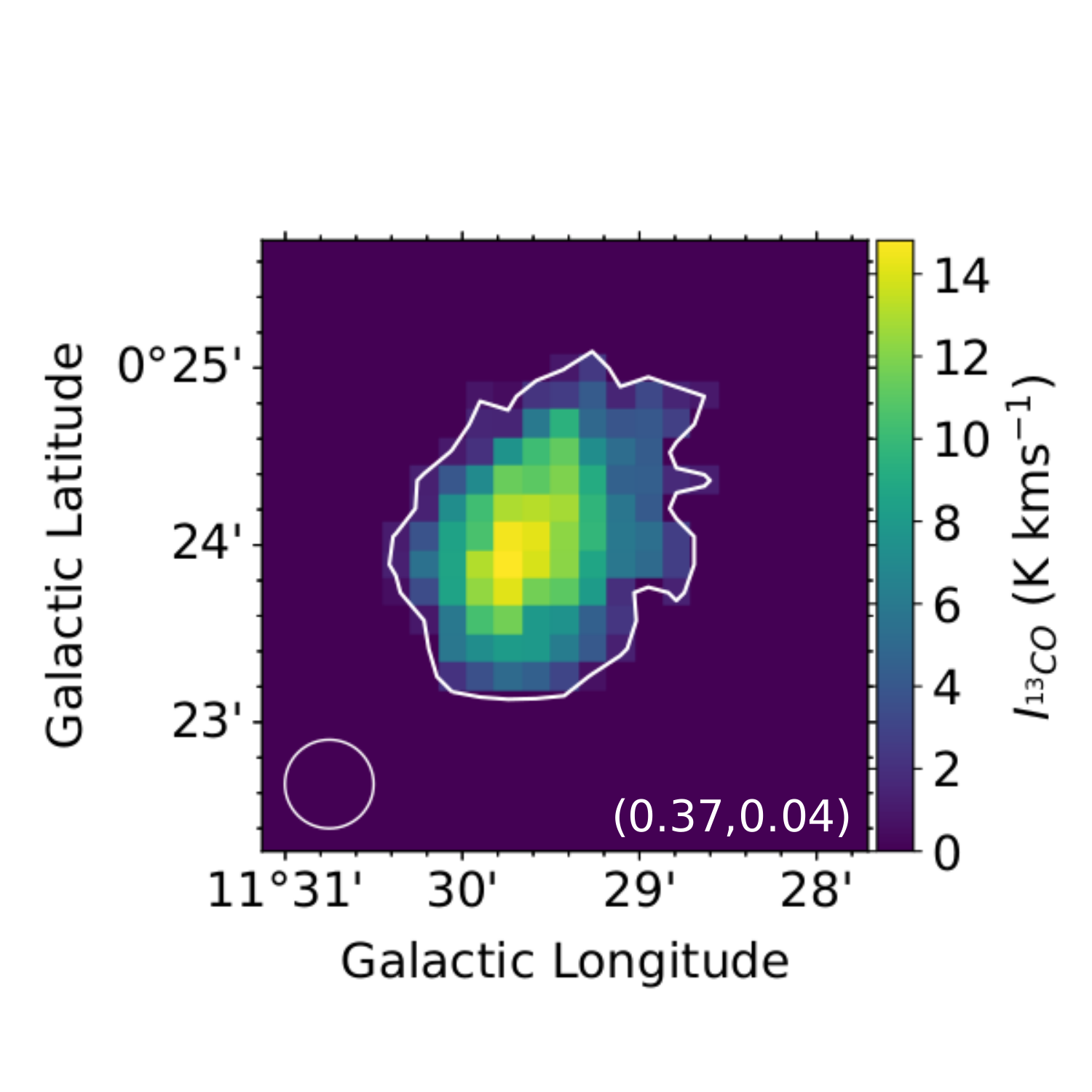}
    \caption{Concentrated cloud ($J$-core; cloud id 9440)}
    \label{fig: concentrated cloud image}
    \end{subfigure}
    \begin{subfigure}[b]{0.45\textwidth}
    \includegraphics[width = \textwidth, keepaspectratio]{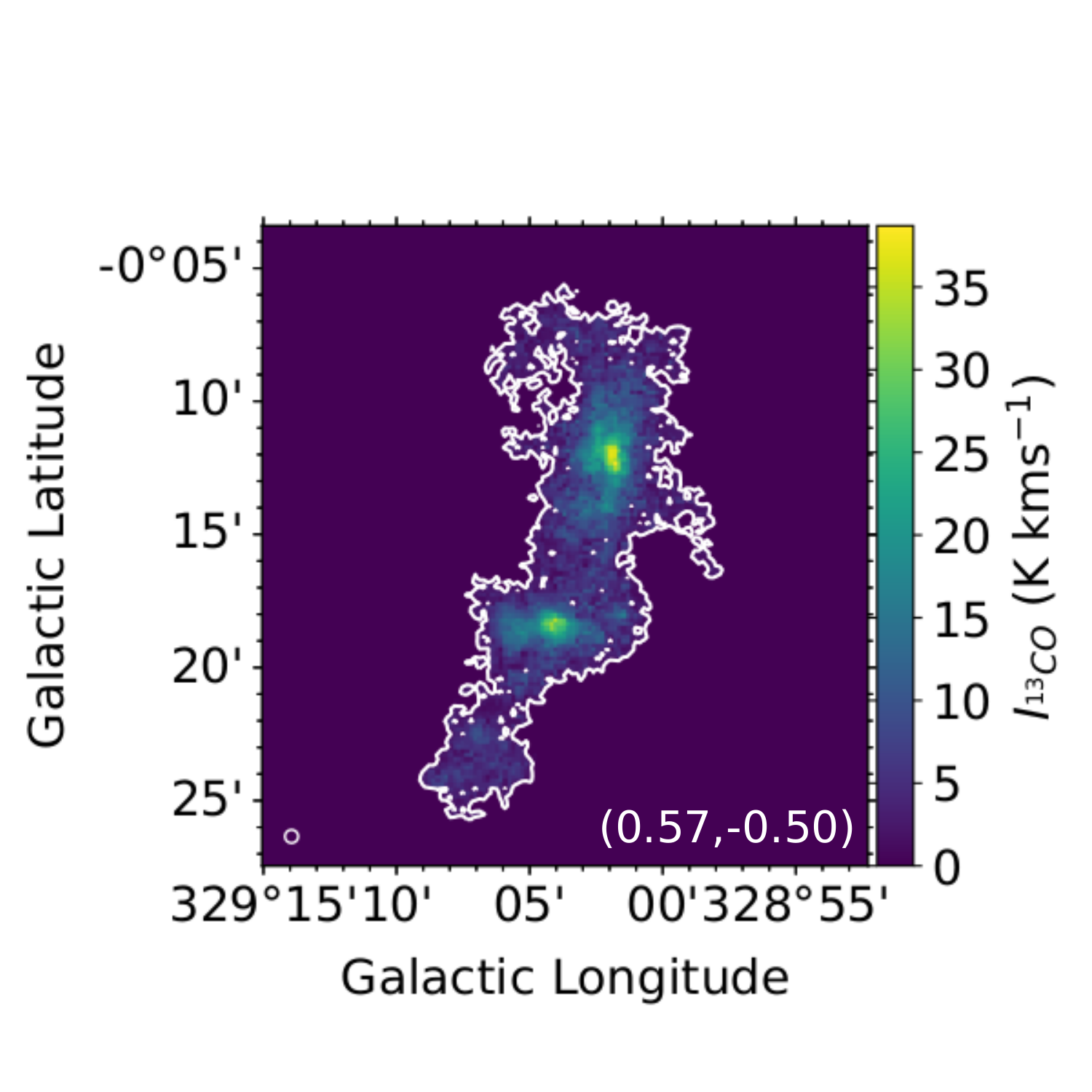}
    \caption{Clumpy cloud ($J$-filament; cloud id 2802)}
    \label{fig: clumpy cloud image}
    \end{subfigure}

    \caption{\add{Examples of a different cloud morphologies. 
    The images are integrated intensity (moment 0) maps of the $^{13}$CO (2 -- 1) transition. The numbers in the right bottom corner of an image represent the $J_1$ and the $J_2$ moments ($J$ plots algorithm) respectively. The colour bars represent the $^{13}$CO integrated intensity in K $\mathrm{km \; s}^{-1}$. The white contours represent the cloud edge. The white circles at the bottom left of the figures represent the telescope beam size.}}
    \label{fig: cloud image combined}
\end{figure*}

Paper I presents the classification of molecular clouds from the SEDIGISM cloud catalogue based on their morphology. It uses two methods for this: $J$ plots \citep{jaffa_2018} and by-eye (visual) classification. $J$ plots is an automated algorithm that uses the principal moment of inertia of a structure to determine its morphology. It distinguishes the structures based on their degree of elongation and degree of central concentration thus classifying them into three types: cores, filaments and bubbles (rings). In Paper I, we also conducted a visual analysis of the integrated intensity maps of the clouds which resulted in four morphological classes:
(i) ring-like clouds \add{(e.g., Fig. \ref{fig: ring-like cloud image})},
(ii) elongated clouds \add{(e.g., Fig. \ref{fig: elongated cloud image})},
(iii) concentrated clouds \add{(e.g., Fig. \ref{fig: concentrated cloud image})},
(iv) clumpy clouds \add{(e.g., Fig. \ref{fig: clumpy cloud image})}.

There are 298 clouds that could not be classified into any of the above classes.

The two classification methods were used to identify two samples from the SEDIGISM cloud catalogue. These are the visually classified (VC) and morphologically reliable (MR) samples. Both these samples are obtained from the SEDIGISM full sample -- a sample of 10663 molecular clouds (detailed description in DC21). The VC sample is obtained by categorising the clouds into the four morphologies using by-eye classification (i.e. SEDIGISM full sample excluding unclassified clouds) whereas the MR sample contains clouds that have the same morphology under both $J$ plot and by-eye classifications. \add{It means that MR sample includes ring-like, elongated and concentrated clouds that are identified by $J$ plots as bubbles, filaments and cores, respectively. The clumpy clouds that are recognised as either filaments or cores by the $J$ plots algorithm also belong to the MR sample.} A detailed description of the classifications and the resulting samples is provided in Sec. 3 and 4.1 of Paper I. 

We restrict our MR and VC samples in this paper (Table \ref{tab: quantitative vc and mr samples}) only to the clouds from the science sample (DC21), and use this sub-sample of clouds for further analysis.
These samples are used to understand the connection between cloud properties, their morphologies and the effects due to Galactic environment. It is worth noting that the quality of classification for the MR sample is better than for the VC sample, as it is classified using the aid of both visual and automated classifications. However, this leads to the rejection of a large amount of data and results in a small sample size for some morphologies (e.g. ring-like clouds). Hence, we use both samples\footnote{We remind the readers that the VC and MR samples used in this paper contain only the science sample clouds and therefore differ from the VC and MR samples presented in Paper I.} for our analysis (Table \ref{tab: quantitative vc and mr samples}).

 \begin{table}[]
 \caption{Quantitative description of the clouds.}
\label{tab: quantitative vc and mr samples}
\centering
\begin{tabular}{lllll}
\hline
Morphological class & MR sample         & VC sample \\ \hline
Ring-like cloud     & 156               & 961   \\
Elongated cloud     & 3607              & 3676  \\
Concentrated cloud  & 528               & 1094  \\
Clumpy cloud        & 729               & 733   \\
Total               & 5020              & 6464  \\ \hline
\end{tabular}
\tablefoot{The table presents the clouds in visually classified (VC) and morphologically reliable (MR) samples for different morphologies. We only present the clouds contained in the science sample. For the full SEDIGISM sample, refer to Paper I.}
\end{table}

\section{Results}\label{sec: results}


In this section, we analyse the connection between cloud morphologies and their other properties i.e. cloud mass, surface density, radius, velocity dispersion, length, aspect ratio and virial parameter. 
\add{The properties used throughout this paper are the deconvolved properties from DC21.} The cloud radius (effective radius and deconvolved radius) is calculated (by DC21) from the exact area of the cloud, assuming a spherical geometry.
Therefore, the concept of radius may be inefficient in correctly describing the size of clouds with complex structures, which have a non-spherical geometry.
An alternative measure of the major and the minor axes of a cloud could be the length and width obtained using the geometric medial axis. The longest running central spine through a 2D projected cloud mask is considered to determine the cloud length, whereas the width is twice the average distance from the spine to the cloud edge. The cloud length ($length_\mathrm{MA}$) and the aspect ratio ($AR_\mathrm{MA}$) used in this work were obtained using the medial axis technique (DC21).
The medial axis length is free of the assumption that clouds have a particular geometry and hence provides a different estimate of the cloud size (especially for elongated structures), as compared to the cloud radius.
Our implementation of the medial axis analysis does not consider the intensity distribution of the cloud and therefore we do not trace the density-weighted spine. 

The average integrated intensity of a cloud is obtained from the observed flux and it is converted into the cloud mass $M$ using the $^{13}$CO (2-1) to H$_2$ conversion factor; $X_\mathrm{CO} = 1^{+1}_{-0.5} \times 10^{21} \mathrm{cm}^{-2} (\mathrm{K \; km \; s}^{-1})^{-1}$ \citep{schuller_2017}.
This mass is further used to obtain the surface density and the virial parameter for the clouds (DC21). The virial parameter is calculated assuming that clouds are spheres  \citep{crops}, but this \add{is not consistent} with the different cloud morphologies that we see. Moreover, the virial parameter is a crude property calculated using the balance between the kinetic and gravitational energies in the clouds, and might not be \add{an} accurate representation of the gravitational state of all the clouds. Nevertheless, we discuss it for the different morphologies as it is a widely used parameter in the star formation literature (e.g. DC21).

\begin{figure}
    \centering
    \includegraphics[width = 0.5\textwidth, keepaspectratio]{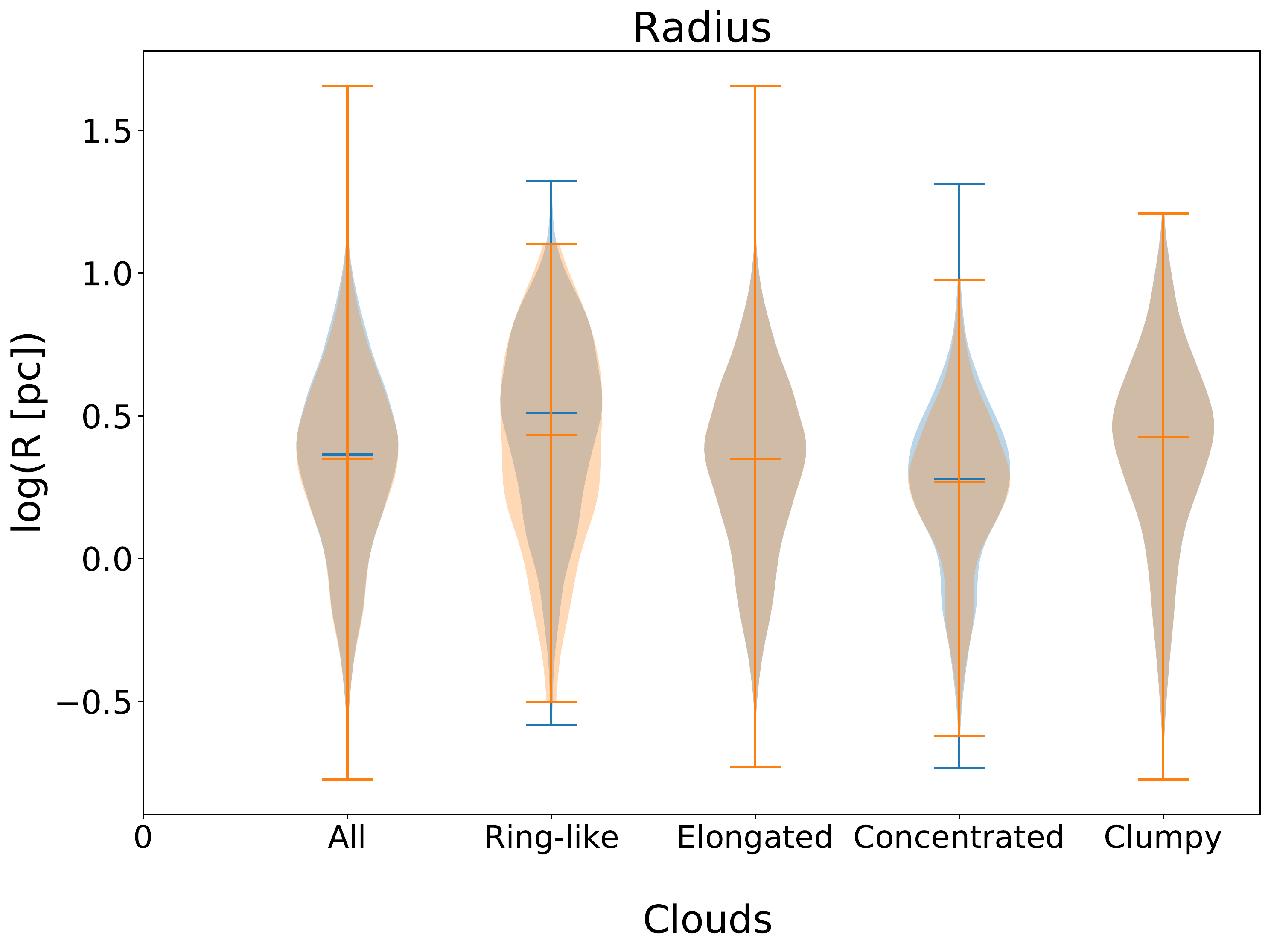}
    
    \vspace{0.2 cm}
    
    \includegraphics[width = 0.5\textwidth, keepaspectratio]{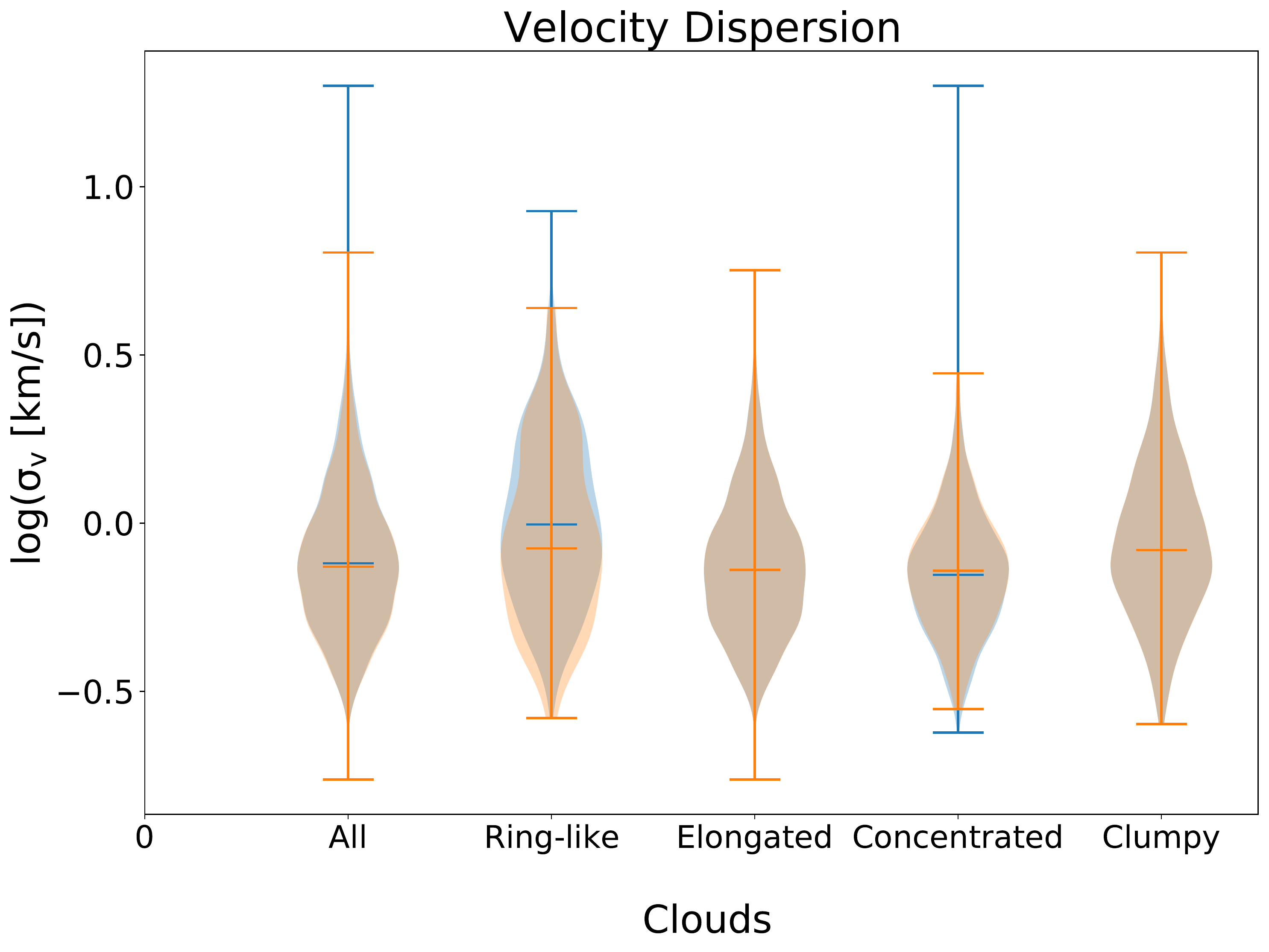}
    \caption{Distribution of radius (top) and velocity dispersion (bottom) for the VC sample (blue) and the MR sample (orange). The violin plots present the density of the data at different values, which is smoothed through kernel density estimator. The horizontal lines at the ends and middle of the plots represent the extreme and the median values of the distributions respectively.}
    \label{fig: prop rad vel dis}
\end{figure}

\begin{figure}
    \centering
    \includegraphics[height = 0.4\textwidth, width = 0.5\textwidth]{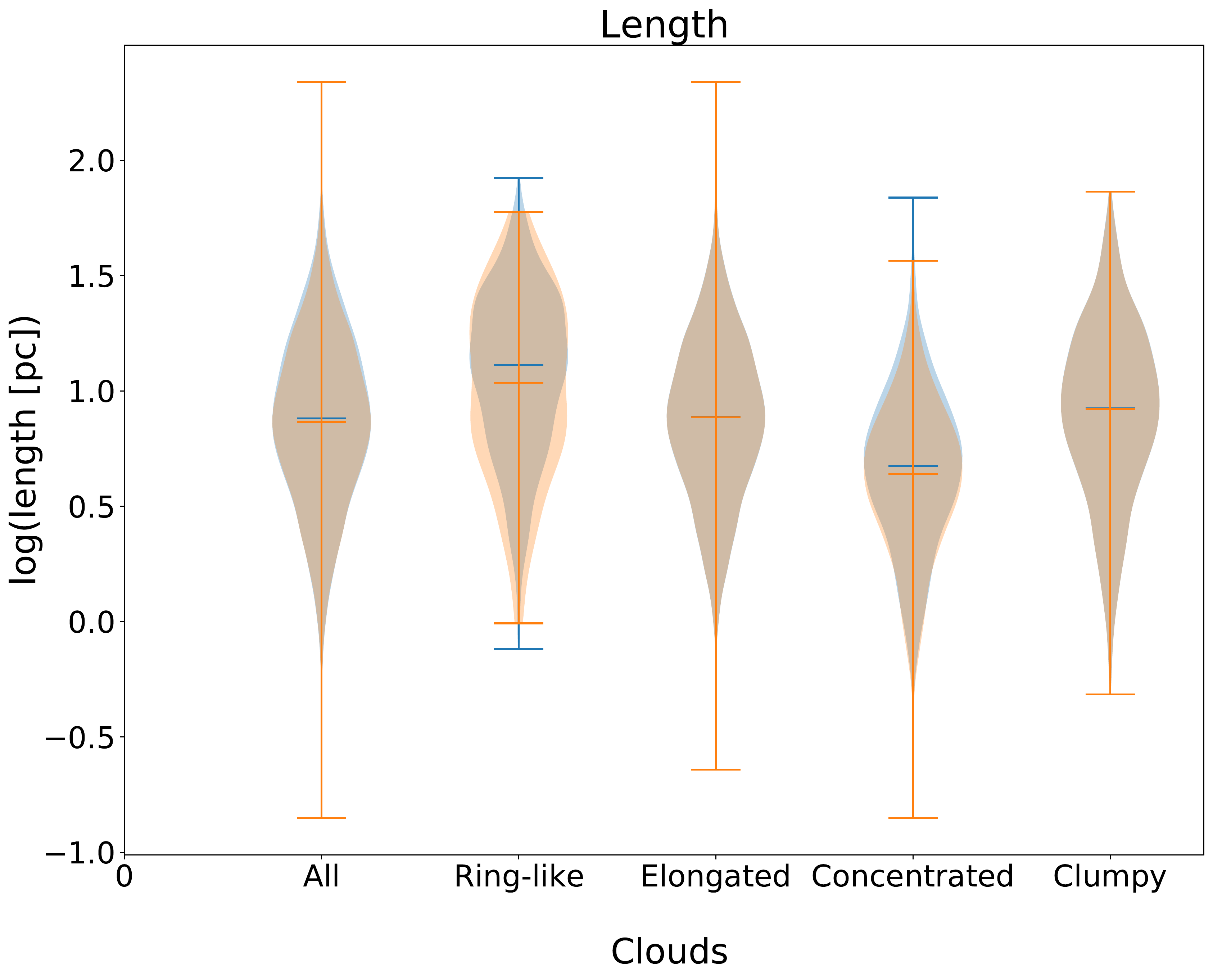}
    
    \vspace{0.2 cm}
    
    \includegraphics[height = 0.4\textwidth, width = 0.5\textwidth]{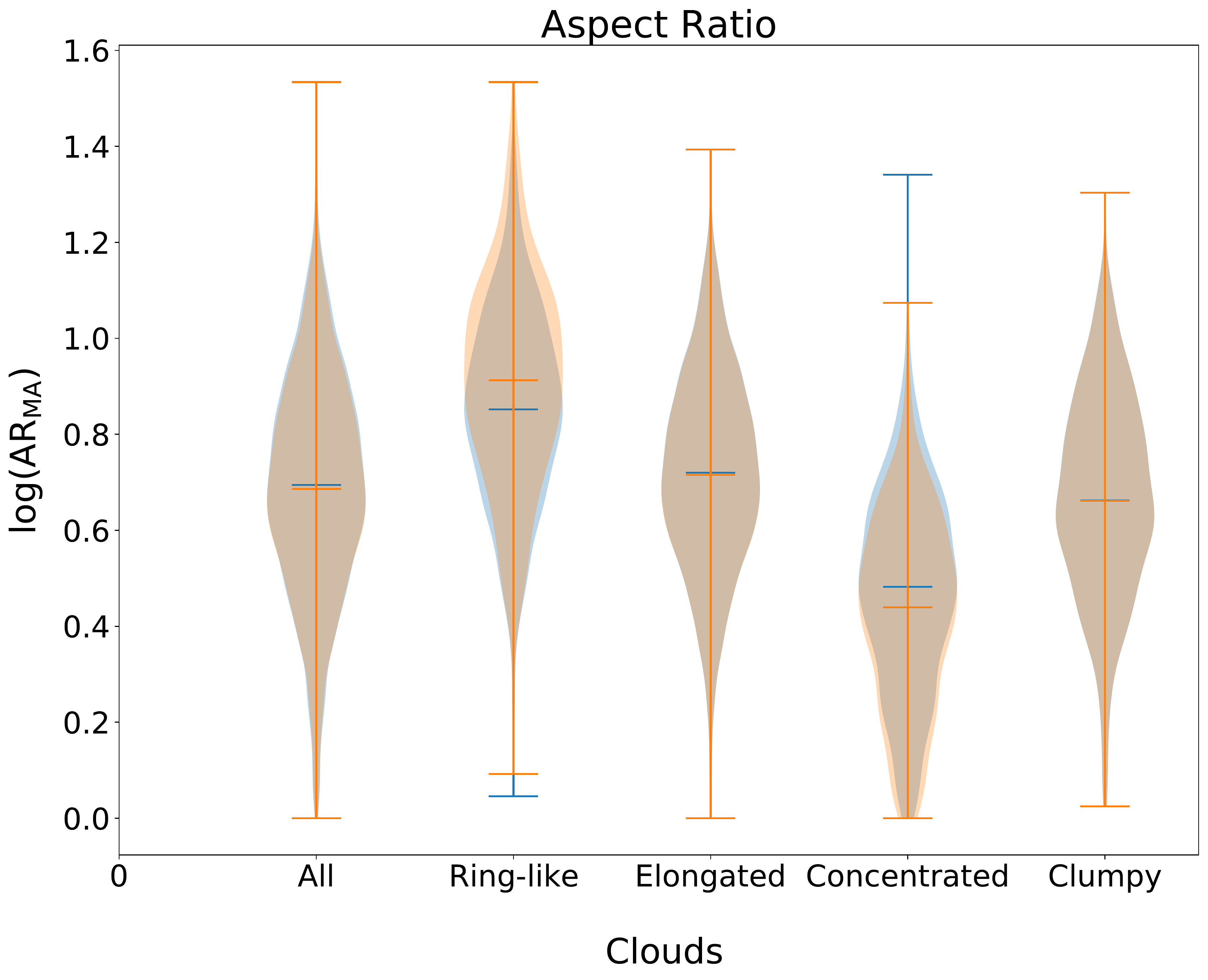}
    \caption{Distribution of length (top) and aspect ratio (bottom). The symbols and conventions follow Fig. \ref{fig: prop rad vel dis}.}
    \label{fig: prop len asp}
\end{figure}

\begin{figure}
    \centering
    \includegraphics[height = 0.4\textwidth, width = 0.5\textwidth]{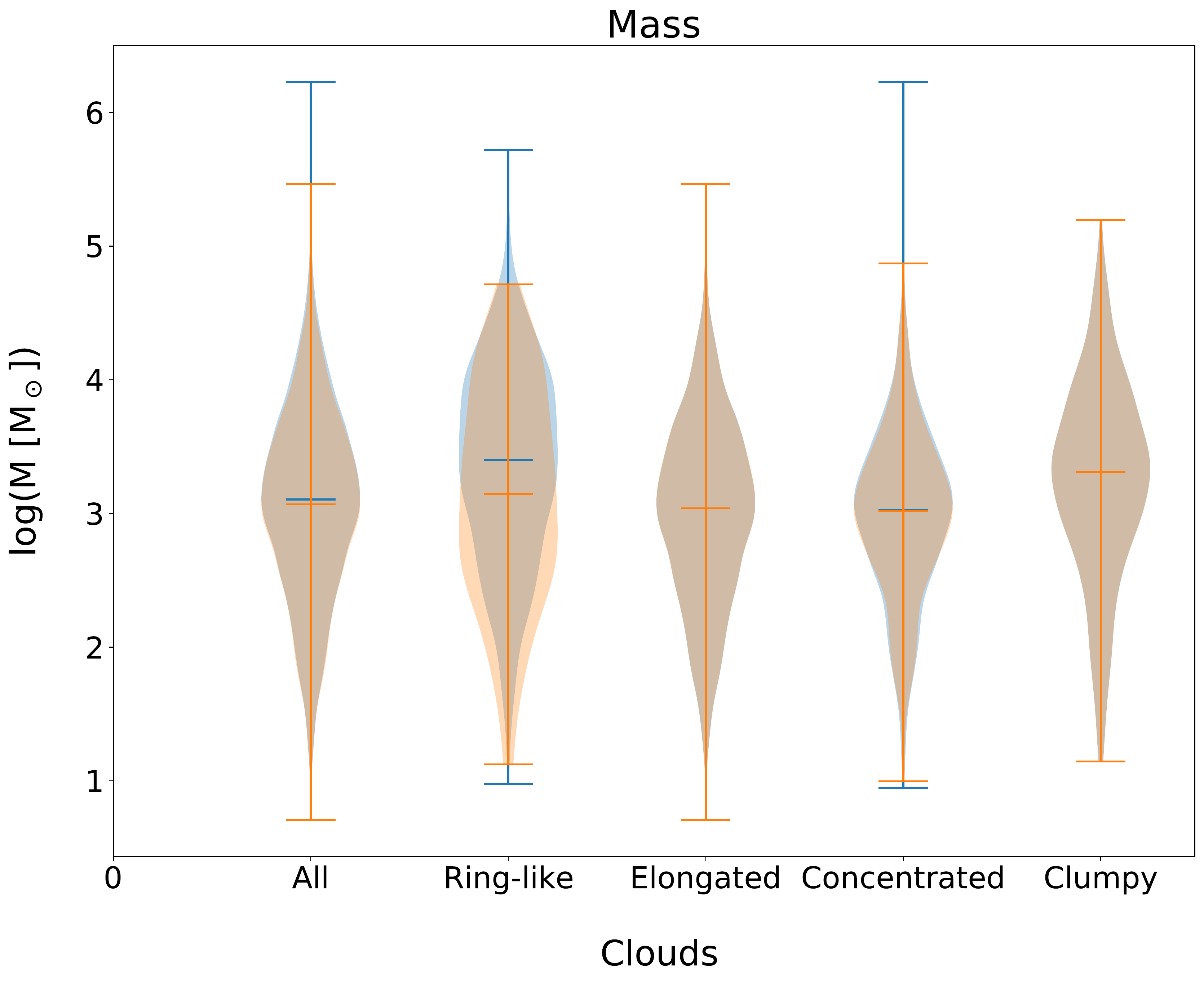}
    
    \vspace{0.2 cm}

    \includegraphics[height = 0.4\textwidth, width = 0.5\textwidth]{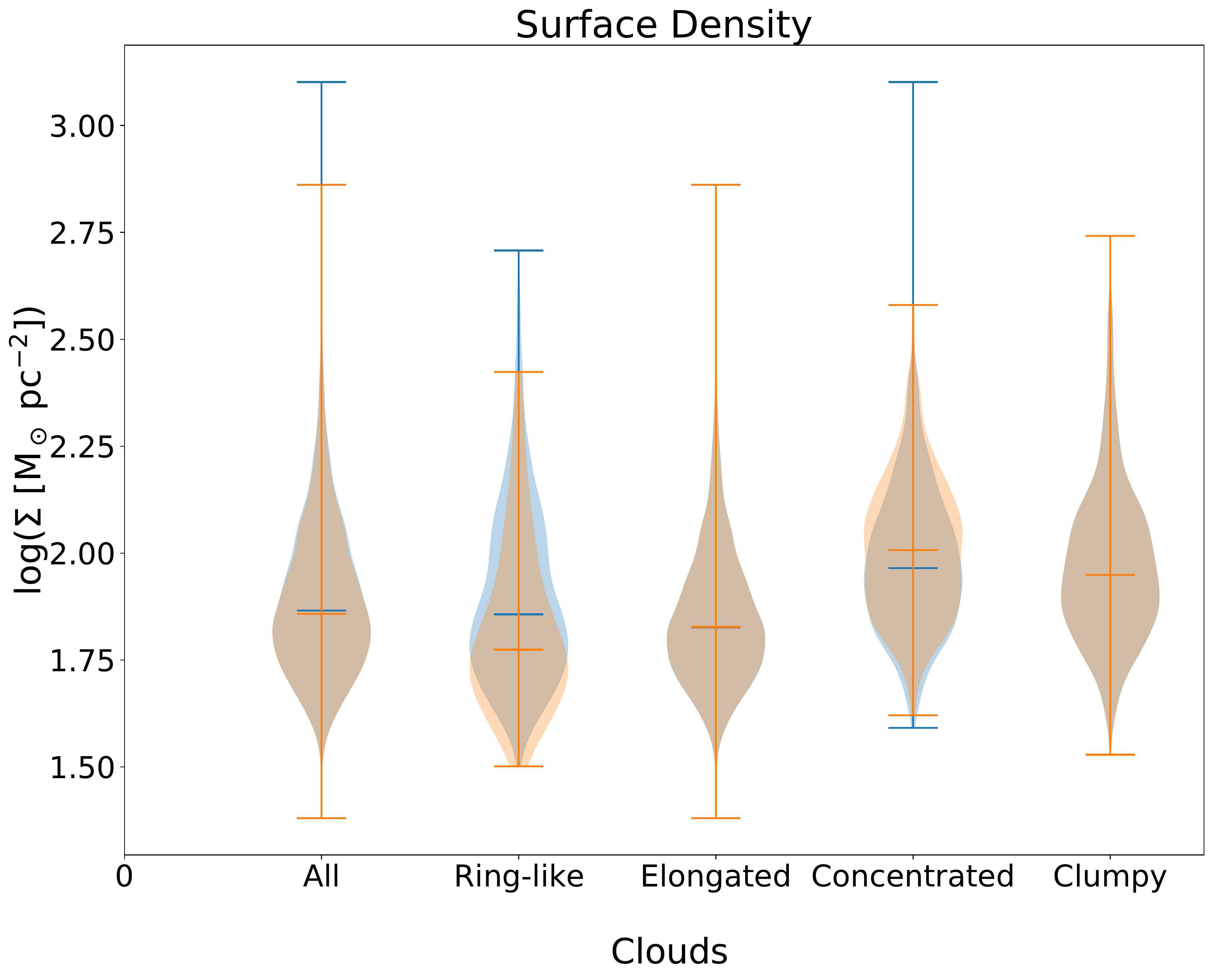}
    \caption{Distribution of mass (top) and surface density (bottom). The symbols and conventions follow Fig. \ref{fig: prop rad vel dis}.}
    \label{fig: prop mass surf dens}
\end{figure}

\begin{figure}
    \centering
    \includegraphics[width = .5\textwidth,height = 0.4\textwidth]{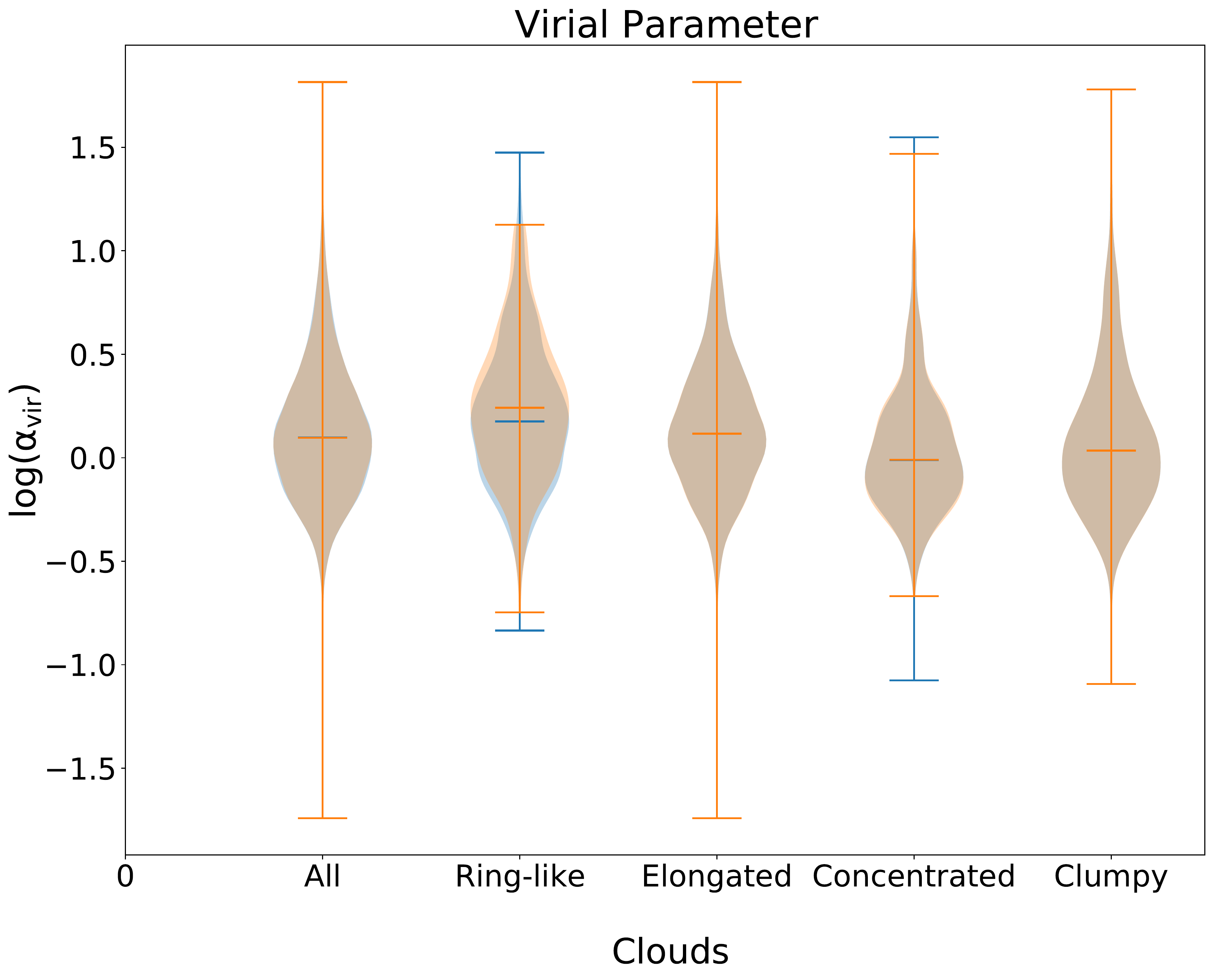}
    \caption{Distribution of virial parameter for the VC \add{and MR samples}. The symbols and conventions follow Fig. \ref{fig: prop rad vel dis}.}
    \label{fig: prop vir}
\end{figure}

The violin plots in Fig. \ref{fig: prop rad vel dis}--\ref{fig: prop vir} present distributions of the properties for different cloud morphologies using the VC and MR samples. The main statistics (i.e. the median, interquartile range, skewness and kurtosis) of cloud properties for different morphologies for both samples are reported in Table \ref{Table: Properties}. The similar values (especially for median) between the two samples in most cases \add{indicate} that they follow each other closely. The two samples show the most similar distributions for elongated and clumpy clouds (p-value = 1, Table \ref{tab: KS test VC vs MR}).
The p-values presented throughout this work are obtained \add{using} the \add{two sample Kolmogorov–Smirnov (}KS) test. A detailed description of the KS test and p-values is presented in appendix \ref{app: ks test}.
\addnew{We compare the p-values between the VC and MR samples using the distributions of cloud properties in Table \ref{tab: KS test VC vs MR}. We find a p-value > 0.01 in $\sim$ 85\% of the cases, which suggests that the VC and MR samples usually show a similar distribution.}

The differences between the distributions \add{of the cloud properties for the various morphological classes} can be identified visually from the violin plots \add{(Fig. \ref{fig: prop rad vel dis} -- \ref{fig: prop vir})} and confirmed using the \add{p-values from the KS} test \add{(Table \ref{tab: KS test diff morphologies})}. 
The distributions for the different morphologies differ from each other for each cloud property (typically p-value < 0.01, Table \ref{tab: KS test diff morphologies}), e.g., the mass distribution of concentrated and ring-like clouds differ from each other.

Ring-like clouds show the most \addnew{different properties} as compared to other morphologies. These clouds show higher average values and different distributions of radius, velocity dispersion, aspect ratio and length as compared to other morphologies (Fig. \ref{fig: prop rad vel dis} and \ref{fig: prop len asp}\add{; see `Median' values in Table \ref{Table: Properties}}). They also have higher average cloud mass than elongated and clumpy clouds, but the average surface density is comparable to elongated clouds and lower than other morphologies (Fig. \ref{fig: prop mass surf dens}). \add{Ring-like clouds also show slightly higher average values of virial parameter compared to other morphologies (Fig. \ref{fig: prop vir}). However, we again caution the readers that virial parameter might not correctly represent a cloud's gravitational state for all the morphologies.}

Concentrated clouds present a morphology quite different to the ring-like clouds. These clouds have a high density at the centre leading to a near-circular structure and therefore exhibit smaller (average) lengths and aspect ratios \add{(Fig. \ref{fig: prop len asp})}. They also have lower average mass than other morphologies but their small size results in higher average surface density than other structures \add{(Fig. \ref{fig: prop mass surf dens})}.

Elongated clouds and clumpy clouds have comparable radius, velocity dispersion, length and aspect ratio\footnote{There are 25 elongated and 12 clumpy clouds with aspect ratio < 1.5 in the VC sample. These clouds are low resolution structures which might have been classified incorrectly or have unreliable medial axes measurements. We do not exclude these structures as they constitute only a very small percentage of the total sample and do not affect our analysis.} distributions \add{(Fig. \ref{fig: prop rad vel dis} and \ref{fig: prop len asp})}; with elongated clouds having more tailed distributions towards both lower and higher values (see kurtosis in Table \ref{Table: Properties}). The lower kurtosis of clumpy clouds may be a result of their smaller sample size. The length distributions \add{(Fig. \ref{fig: prop len asp})} of these two morphologies are similar to that of all clouds, which might be due to most of the clouds having an elongated structure. Clumpy clouds have more mass but similar lengths to elongated clouds, resulting in higher surface densities (Fig. \ref{fig: prop mass surf dens}).

\begin{table*}[]
\caption{Statistics of the integrated properties for various cloud morphologies.}
\label{Table: Properties}
\centering
\begin{tabular}{l|rrrr|rrrr}
\hline
Subset                    & \multicolumn{4}{c|}{Visually classified sample}                                                                     & \multicolumn{4}{c}{Morphologically reliable sample}                                                               \\
                                                  & \multicolumn{1}{l}{Median} & \multicolumn{1}{l}{IQR} & \multicolumn{1}{l}{Skewness} & \multicolumn{1}{l|}{Kurtosis} & \multicolumn{1}{l}{Median} & \multicolumn{1}{l}{IQR} & \multicolumn{1}{l}{Skewness} & \multicolumn{1}{l}{Kurtosis} \\ \hline
M [$\times 10^3 \; \mathrm{M_\odot}$]             & \multicolumn{1}{l}{}       & \multicolumn{1}{l}{}    & \multicolumn{1}{l}{}         & \multicolumn{1}{l|}{}         & \multicolumn{1}{l}{}       & \multicolumn{1}{l}{}    & \multicolumn{1}{l}{}         & \multicolumn{1}{l}{}         \\
All                                               & 1.27                       & 3.25                    & 52.3                         & 3451.1                        & 1.17                       & 2.83                    & 13.2                         & 294.6                        \\
Ring-like                                         & 2.51                       & 7.96                    & 14.1                         & 283.3                         & 1.40                       & 5.89                    & 2.8                          & 8.3                          \\
Elongated                                         & 1.09                       & 2.67                    & 16.6                         & 422.3                         & 1.09                       & 2.65                    & 16.9                         & 437.0                        \\
Concentrated                                      & 1.06                       & 2.01                    & 32.1                         & 1046.9                        & 1.05                       & 1.93                    & 7.2                          & 71.4                         \\
Clumped                                           & 2.04                       & 4.71                    & 5.4                          & 39.8                          & 2.04                       & 4.71                    & 5.5                          & 41.2                         \\ \hline
$\Sigma$ [$\mathrm{M_\odot} \; \mathrm{pc}^{-2}$] & \multicolumn{1}{l}{}       & \multicolumn{1}{l}{}    & \multicolumn{1}{l}{}         & \multicolumn{1}{l|}{}         & \multicolumn{1}{l}{}       & \multicolumn{1}{l}{}    & \multicolumn{1}{l}{}         & \multicolumn{1}{l}{}         \\
All                                               & 73.42                      & 42.40                   & 5.0                          & 67.1                          & 72.15                      & 40.29                   & 3.7                          & 28.1                         \\
Ring-like                                         & 71.94                      & 52.35                   & 2.7                          & 10.7                          & 59.49                      & 34.30                   & 2.2                          & 4.9                          \\
Elongated                                         & 67.04                      & 32.09                   & 5.1                          & 52.6                          & 67.22                      & 32.03                   & 5.1                          & 53.6                         \\
Concentrated                                      & 92.29                      & 49.68                   & 8.2                          & 144.7                         & 101.73                     & 55.33                   & 1.8                          & 4.9                          \\
Clumped                                           & 88.99                      & 47.33                   & 2.8                          & 11.6                          & 88.99                      & 47.33                   & 2.8                          & 11.5                         \\ \hline
R [pc]                                            & \multicolumn{1}{l}{}       & \multicolumn{1}{l}{}    & \multicolumn{1}{l}{}         & \multicolumn{1}{l|}{}         & \multicolumn{1}{l}{}       & \multicolumn{1}{l}{}    & \multicolumn{1}{l}{}         & \multicolumn{1}{l}{}         \\
All                                               & 2.32                       & 2.26                    & 3.1                          & 30.7                          & 2.23                       & 2.15                    & 3.7                          & 43.6                         \\
Ring-like                                         & 3.24                       & 3.44                    & 1.4                          & 3.3                           & 2.71                       & 3.25                    & 1.1                          & 0.8                          \\
Elongated                                         & 2.24                       & 2.17                    & 4.3                          & 56.2                          & 2.23                       & 2.17                    & 4.3                          & 57.6                         \\
Concentrated                                      & 1.90                       & 1.51                    & 3.2                          & 24.3                          & 1.86                       & 1.40                    & 1.7                          & 4.7                          \\
Clumped                                           & 2.67                       & 2.37                    & 1.7                          & 4.1                           & 2.67                       & 2.34                    & 1.7                          & 4.1                          \\ \hline
$\sigma_\varv \mathrm{[km\; s^{-1}]}$             & \multicolumn{1}{l}{}       & \multicolumn{1}{l}{}    & \multicolumn{1}{l}{}         & \multicolumn{1}{l|}{}         & \multicolumn{1}{l}{}       & \multicolumn{1}{l}{}    & \multicolumn{1}{l}{}         & \multicolumn{1}{l}{}         \\
All                                               & 0.76                       & 0.53                    & 6.9                          & 158.5                         & 0.74                       & 0.49                    & 2.6                          & 12.0                         \\
Ring-like                                         & 0.99                       & 0.89                    & 2.2                          & 10.5                          & 0.84                       & 0.86                    & 1.9                          & 4.3                          \\
Elongated                                         & 0.73                       & 0.48                    & 2.6                          & 11.7                          & 0.73                       & 0.48                    & 2.5                          & 11.3                         \\
Concentrated                                      & 0.70                       & 0.39                    & 19.4                         & 522.1                         & 0.72                       & 0.38                    & 1.9                          & 5.5                          \\
Clumped                                           & 0.83                       & 0.60                    & 2.5                          & 11.5                          & 0.83                       & 0.60                    & 2.6                          & 11.6                         \\ \hline
$AR_\mathrm{MA}$                                  & \multicolumn{1}{l}{}       & \multicolumn{1}{l}{}    & \multicolumn{1}{l}{}         & \multicolumn{1}{l|}{}         & \multicolumn{1}{l}{}       & \multicolumn{1}{l}{}    & \multicolumn{1}{l}{}         & \multicolumn{1}{l}{}         \\
All                                               & 4.95                       & 3.66                    & 1.6                          & 4.6                           & 4.85                       & 3.46                    & 1.7                          & 5.3                          \\
Ring-like                                         & 7.11                       & 4.72                    & 1.5                          & 3.9                           & 8.18                       & 5.71                    & 1.6                          & 4.4                          \\
Elongated                                         & 5.25                       & 3.45                    & 1.5                          & 3.5                           & 5.19                       & 3.41                    & 1.5                          & 3.6                          \\
Concentrated                                      & 3.03                       & 2.04                    & 2.9                          & 18.7                          & 2.75                       & 1.89                    & 1.5                          & 4.6                          \\
Clumped                                           & 4.60                       & 3.25                    & 1.3                          & 2.7                           & 4.59                       & 3.21                    & 1.3                          & 2.8                          \\ \hline
$\alpha_\mathrm{vir}$                             & \multicolumn{1}{l}{}       & \multicolumn{1}{l}{}    & \multicolumn{1}{l}{}         & \multicolumn{1}{l|}{}         & \multicolumn{1}{l}{}       & \multicolumn{1}{l}{}    & \multicolumn{1}{l}{}         & \multicolumn{1}{l}{}         \\
All                                               & 1.25                       & 1.31                    & 8.4                          & 125.7                         & 1.25                       & 1.29                    & 9.1                          & 144.1                        \\
Ring-like                                         & 1.50                       & 1.68                    & 4.3                          & 27.1                          & 1.74                       & 1.80                    & 2.4                          & 6.1                          \\
Elongated                                         & 1.31                       & 1.34                    & 8.9                          & 140.4                         & 1.31                       & 1.33                    & 8.9                          & 139.5                        \\
Concentrated                                      & 0.98                       & 0.90                    & 8.5                          & 105.7                         & 0.98                       & 0.93                    & 7.8                          & 94.1                         \\
Clumped                                           & 1.08                       & 1.18                    & 9.6                          & 135.7                         & 1.08                       & 1.18                    & 9.6                          & 135.0                        \\ \hline
Length [pc]                                       & \multicolumn{1}{l}{}       & \multicolumn{1}{l}{}    & \multicolumn{1}{l}{}         & \multicolumn{1}{l|}{}         & \multicolumn{1}{l}{}       & \multicolumn{1}{l}{}    & \multicolumn{1}{l}{}         & \multicolumn{1}{l}{}         \\
All                                               & 7.60                       & 9.43                    & 3.6                          & 37.2                          & 7.32                       & 8.77                    & 4.4                          & 53.9                         \\
Ring-like                                         & 12.96                      & 15.21                   & 1.5                          & 2.9                           & 10.84                      & 15.20                   & 1.3                          & 1.5                          \\
Elongated                                         & 7.71                       & 8.87                    & 5.0                          & 66.2                          & 7.69                       & 8.82                    & 5.0                          & 67.5                         \\
Concentrated                                      & 4.73                       & 4.87                    & 3.8                          & 25.1                          & 4.38                       & 4.13                    & 2.4                          & 9.9                          \\
Clumped                                           & 8.39                       & 10.00                   & 2.1                          & 6.2                           & 8.34                       & 9.95                    & 2.1                          & 6.2                          \\ \hline
\end{tabular}
\tablefoot{ The properties are mass ($M$), surface density ($\Sigma$) , radius ($R$), velocity dispersion ($\sigma_\varv$), aspect ratio ($AR_\mathrm{MA}$), virial parameter ($\alpha_\mathrm{vir}$) and length ($length_\mathrm{MA}$). The median is an estimate of the average of a distribution. The Interquartile Range (IQR) is the difference between the 75$^\mathrm{th}$ percentile and the 25$^\mathrm{th}$ percentile of the distribution and is a measure of the variability in the data. Skewness is a measure of symmetry of a distribution and the positive and negative values signify the data being skewed left and right respectively. Kurtosis describes the outliers or extremities in the distribution with a high value representing a highly-tailed distribution.}
\end{table*}

\begin{figure}[!h]
    \centering
    \includegraphics[width = .4\textwidth, keepaspectratio]{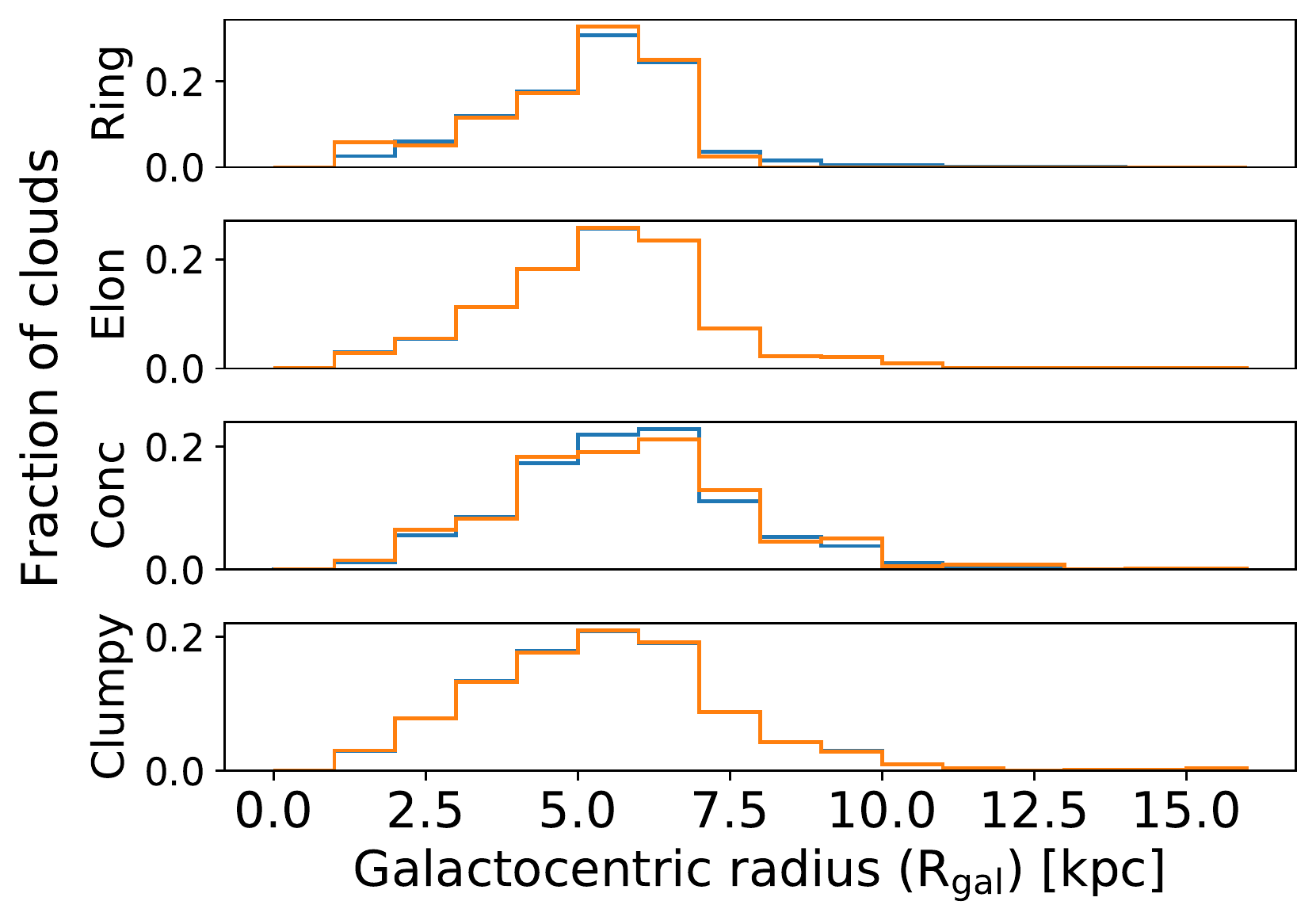}
    \includegraphics[width = .4\textwidth, keepaspectratio]{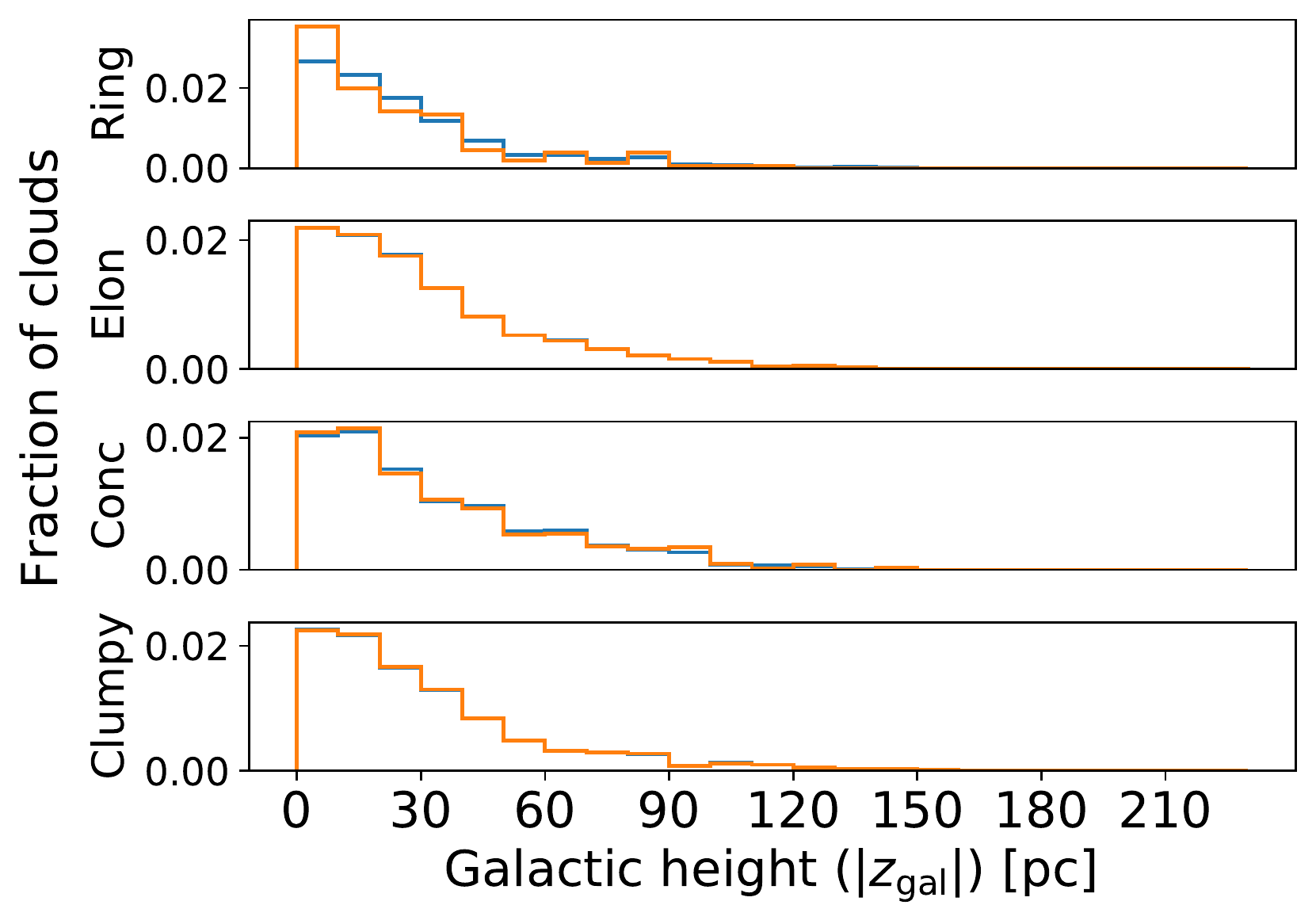}
    \includegraphics[width = .4\textwidth, keepaspectratio]{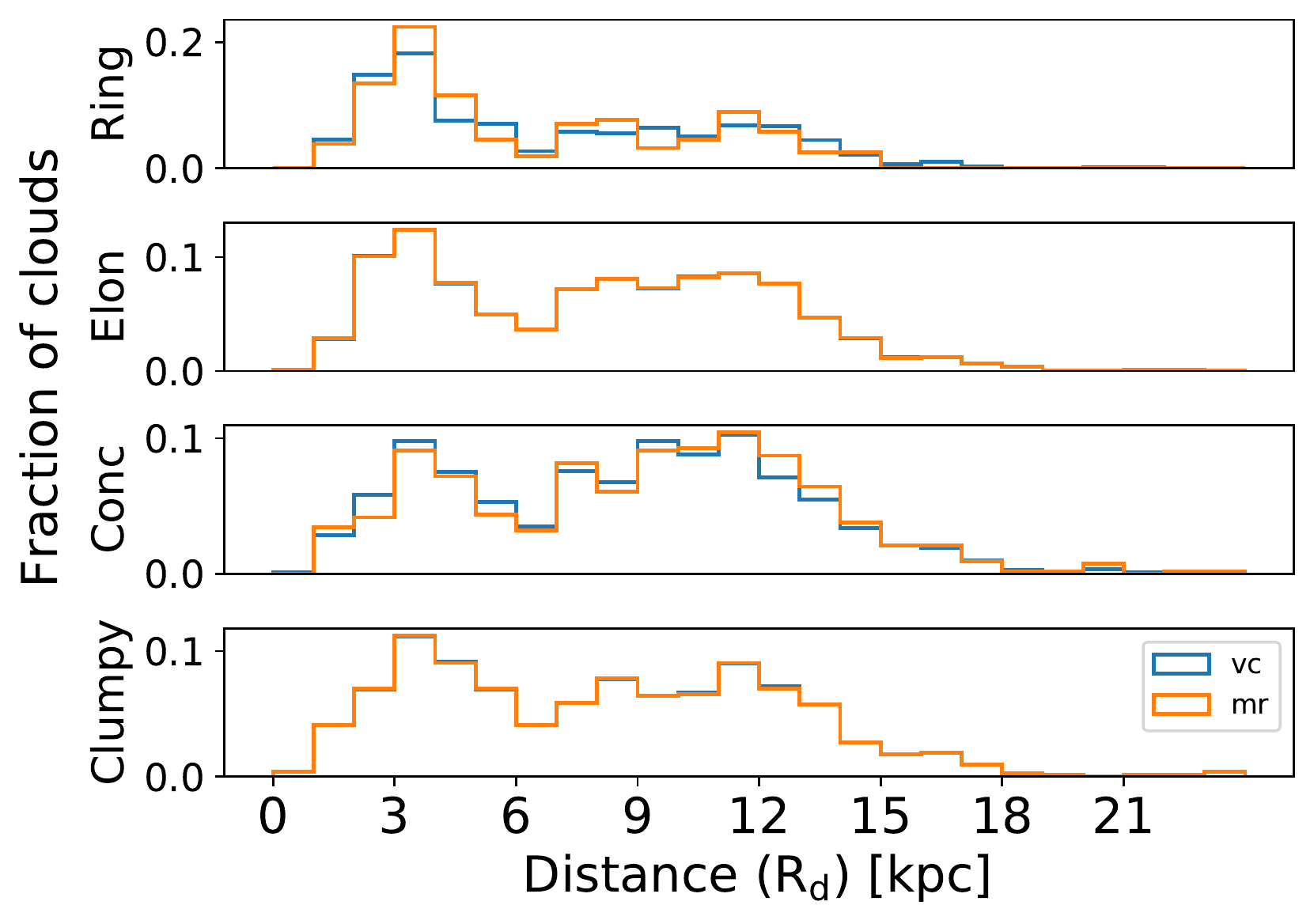}
    \caption{Histogram (normalised) for distribution of clouds with respect to Galactocentric radius [top] Galactic height ($z_\mathrm{gal}$) [middle], and heliocentric distance [bottom]. \add{The histograms represent the ratio of clouds in each bin to the total number of clouds in all the bins. }The bin width is 1 kpc for the heliocentric distance and Galactocentric radius plots and 1 pc for the $z_\mathrm{gal}$ histogram. The blue colour represents the VC sample and the orange colour represents the morphologically reliable (MR) sample. The cloud morphologies from top to bottom represent ring-like clouds (Ring), elongated clouds (Elon), concentrated clouds (Conc) and clumpy clouds (Clumpy).}
    \label{fig: dist rad z histogram}
\end{figure}

\add{We also check if the distribution of the clouds in the Galaxy is influenced by environmental factors, namely Galactocentric radius/distance ($R_\mathrm{gal}$) and Galactic height ($z_\mathrm{gal}$ (see Fig. \ref{fig: dist rad z histogram}). }
\add{We do not find any specific environmental trend that separates a particular cloud morphology from others. }
\add{All cloud morphologies show a gaussian like distribution in the $R_\mathrm{gal}$ histograms (Fig. \ref{fig: dist rad z histogram}, top).} The small sample size at smaller $R_\mathrm{gal}$ (Fig. \ref{fig: dist rad z histogram}; top) is due to rejection of clouds near the Galactic centre (the science sample in DC21 excludes clouds with unreliable distances) whereas the small sample size at higher $R_\mathrm{gal}$ (which for SEDIGISM corresponds to large heliocentric distances) is mostly due to the sensitivity and resolution limitations of the telescope. 
\add{Unsurprisingly, a large number of molecular clouds are situated at low Galactic heights (Fig. \ref{fig: dist rad z histogram}, middle), which may be a consequence of the high density of molecular gas in the Galactic plane. Moreover, all the morphologies seem to be distributed similarly across the Galactic height.} 

\add{In order to identify any potential observational biases, we have also investigated how the different morphologies are distributed as a function of heliocentric distance ($R_\mathrm{d}$). The $R_\mathrm{d}$ histograms show that ring-like clouds have a higher concentration closer to us (Fig. \ref{fig: dist rad z histogram}; bottom)}. \addnew{It could be due to a projection effect leading to distortion in the observed structure of the clouds at large distances. Also, the most distant ring-like clouds could be unresolved and thus get classified into other morphologies.} This is possibly reflected in the distribution of concentrated clouds, which shows slightly higher number densities at larger $R_\mathrm{d}$. Hence, this is \add{most likely} a consequence of observational bias and \add{not a physical phenomenon.}
The distribution of other morphologies is similar along the heliocentric distance, with a dip in the number of clouds at $R_\mathrm{d} \approx 6 ~ \mathrm{kpc}$. The clouds in this region lie very close to the tangent velocity and are assigned a higher distance value (i.e. the tangent distance) to avoid the distance ambiguity (see Sec. 4.1 in DC21).

We observe that clouds of all morphologies are abundant at all $R_\mathrm{d}$ (Fig. \ref{fig: dist rad z histogram}; bottom), however the slight trends in the distributions indicate the presence of observational biases. \add{We discuss the observational limitations potentially affecting our analysis in appendix \ref{app: telescope limitations}.} \add{The influence of biases is most visible in the directly measured properties, i.e., mass, radius and length. For example, smaller and less massive clouds can be detected in our vicinity but the technical limitation of telescopes minimise their detection at large distances.} \add{We also discuss the influence of the environmental factors, i.e. R$_\mathrm{gal}$ and z$_\mathrm{gal}$ in appendix \ref{app 1}. We find that smaller clouds are usually seen at low Galactic heights. However, clouds at all Galactic heights show similar aspect ratios. A general observation is that the clouds properties follow the global trends (as seen in Fig. \ref{fig: prop rad vel dis} -- \ref{fig: prop vir}) irrespective of the Galactic environment and the observational biases.}

\section{Scaling Relations}\label{sec: scaling relation}

Larson's size-linewidth relation \citep{larson_realtion} or Larson's first law is one of the most \add{studied} scaling relations, which has provided observational constraints on the dynamics of molecular clouds for the past 40 years. The relation $\sigma_\varv$ = 1.10 $L^{0.38}$ was obtained by analysing a large number of molecular clouds from sub-parsec size to few hundred parsecs. It \add{has a similar form} to Kolmogorov's law for turbulence and implies that larger clouds have higher  \add{velocity dispersion}. It was further updated for molecular clouds using analysis from \citet{solomon_1987} as discussed in \citet{scimes_2}. 

\begin{figure*}
    \centering
    \includegraphics[width = .47 \textwidth, keepaspectratio]{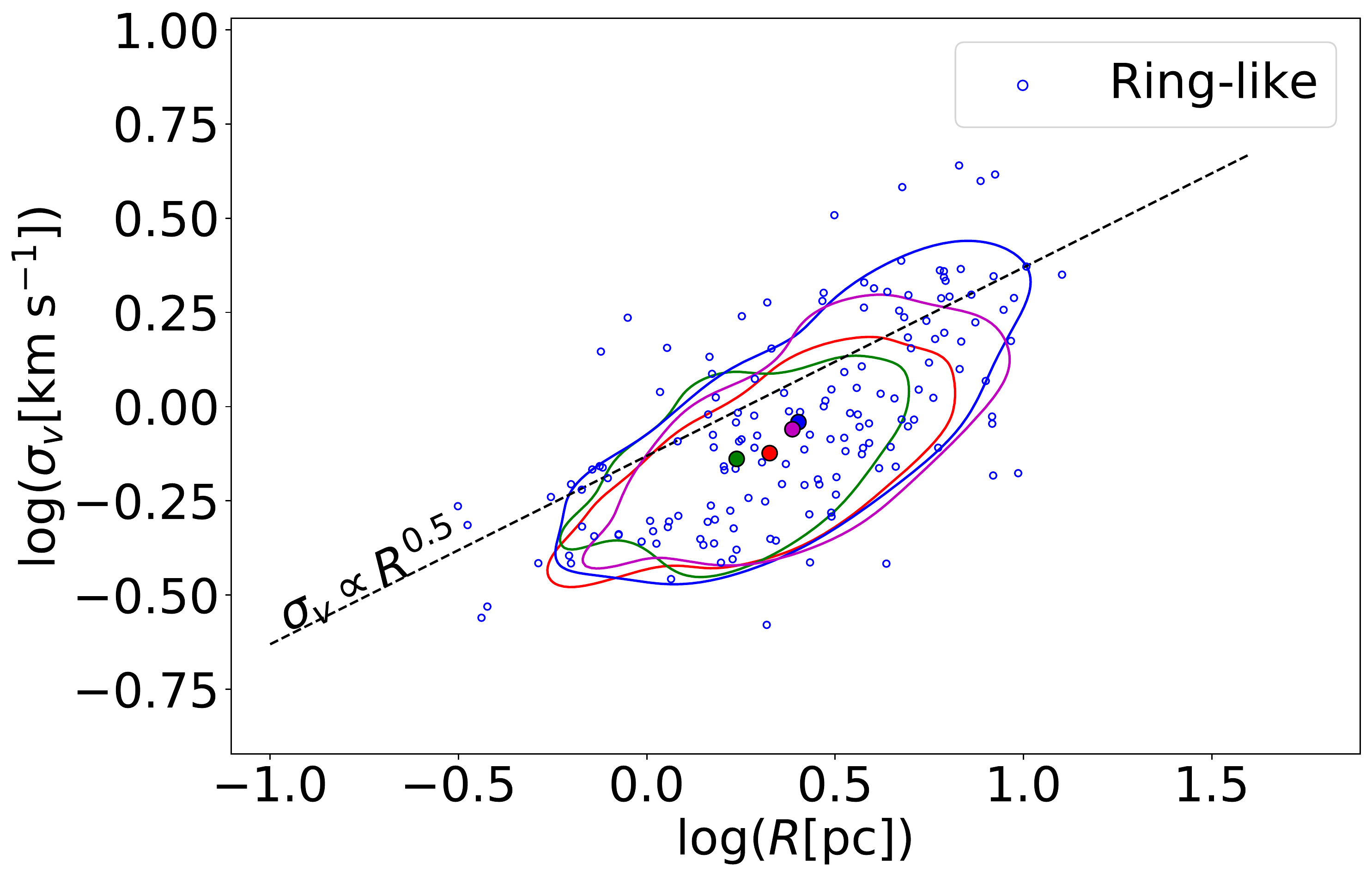}
    \includegraphics[width = .47 \textwidth, keepaspectratio]{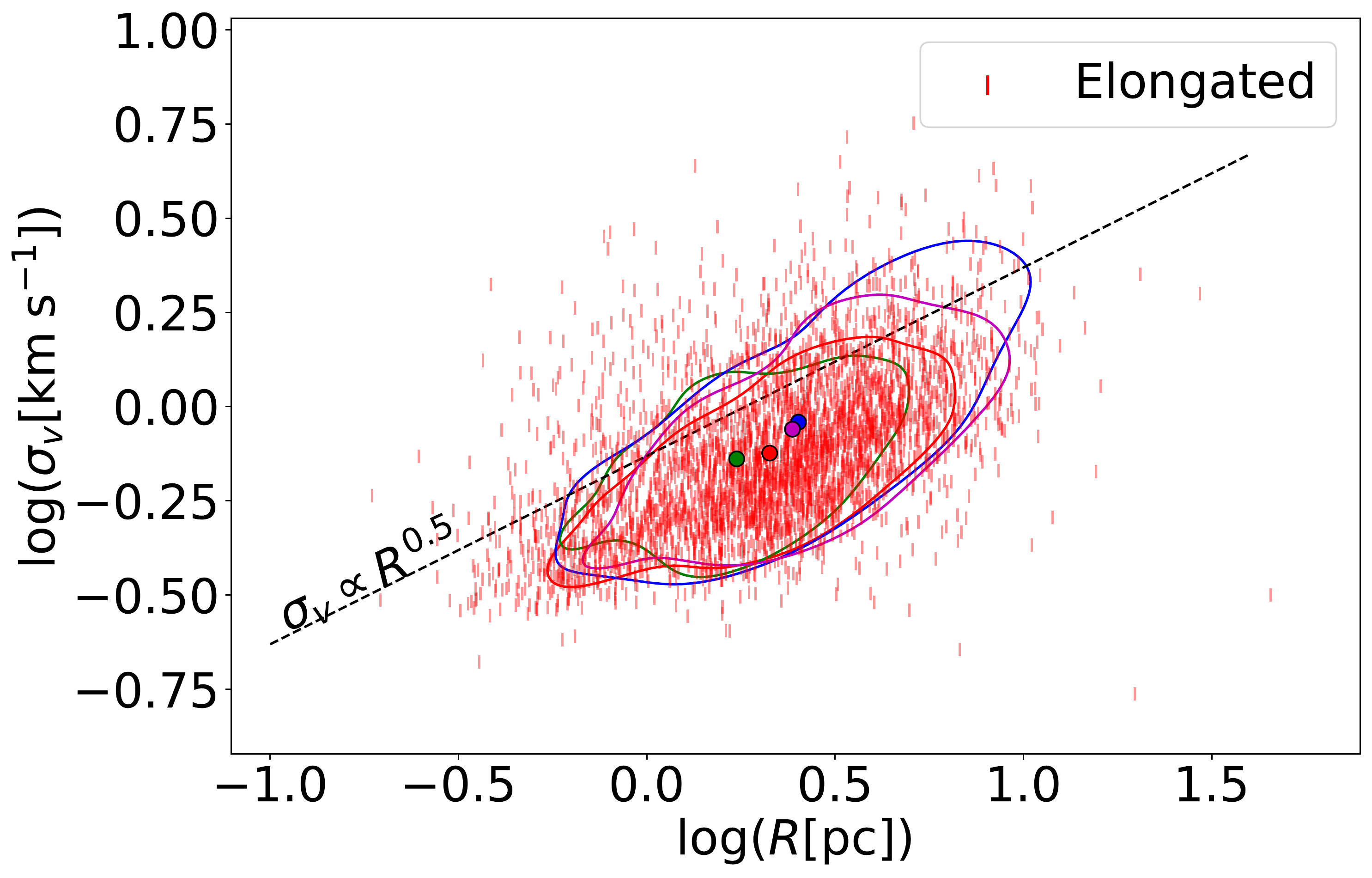}
    \includegraphics[width = .47 \textwidth, keepaspectratio]{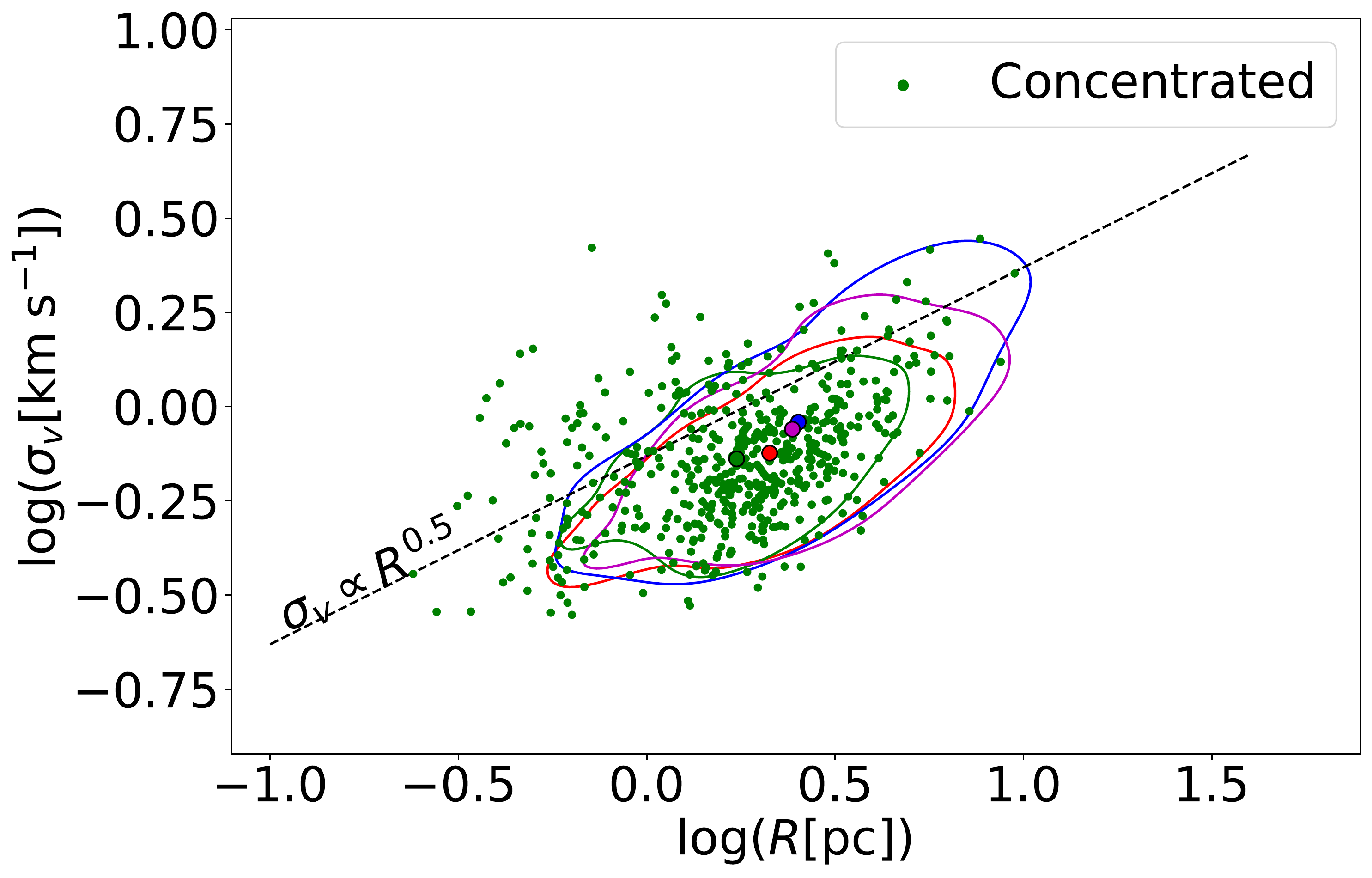}
    \includegraphics[width = .47 \textwidth, keepaspectratio]{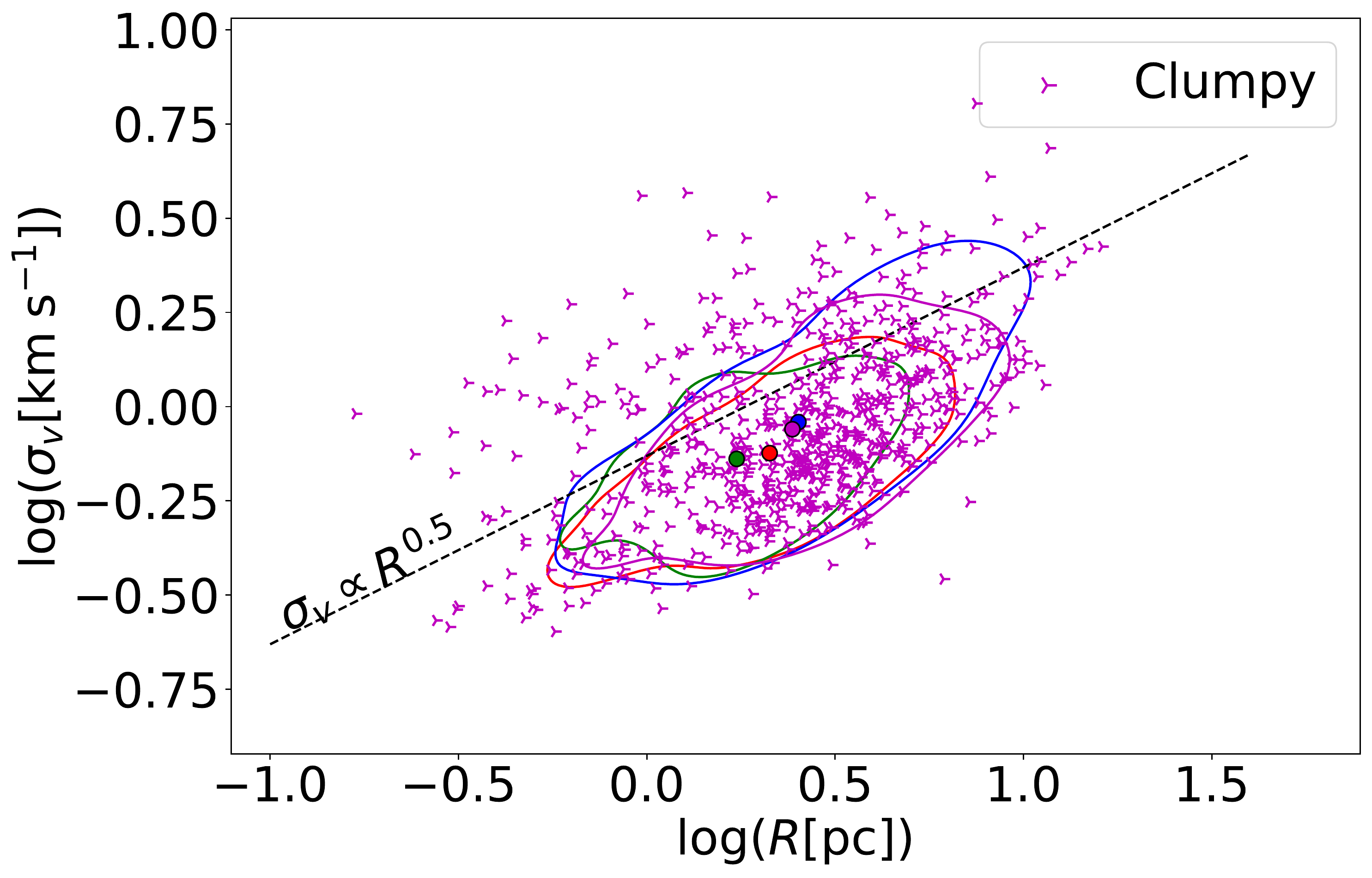}
    \caption{\add{Size-linewidth relation ($\sigma_\varv$ versus $R$) using the MR sample. The dashed line represents Larson's first relation} \citep{larson_realtion, solomon_1987}. \add{The different colours represent four cloud classes (Sec. \ref{sec: morphological cloud classification}); ring-like clouds in blue, elongated clouds are represented in red, concentrated clouds in green,  and clumpy clouds are represented in magenta. The contours include 68\% of the data (1-sigma level) and the small circles represent centroids (means) of the distributions. The expected slope for the relation is 0.5 (dashed black line)} \citep{larson_realtion}.}
    \label{fig: scaling relation larson}
\end{figure*}

\add{We present the size-linewidth relation for the various cloud morphologies in Fig. \ref{fig: scaling relation larson}.} \add{All cloud morphologies follow the general trend of high velocity dispersions for large clouds. However, it should be noted that the `size' in terms of an equivalent radius is a better approximation for some clouds than others.} The high values of velocity dispersion for ring-like clouds \add{with large radii (log(R) > 0.5)} might be attributed to the momentum imparted due to stellar feedback. A larger radius means that more of the gas (shell) has travelled farther away from the centre, resulting in a higher velocity difference between the centre and the edge. \add{Although we expect the ring-like clouds to have a vertical offset given their higher velocity dispersions, this is not the case and they still lie along Larson's relation (Fig. \ref{fig: scaling relation larson}).}
\add{We perform principal component analysis (PCA) for different cloud morphologies on the Larson's plot (Fig. \ref{fig: scaling relation larson}) to obtain the confidence ellipses (1-sigma) and compare the cloud distributions with the Larson's relation.} The slopes of the confidence ellipses for the clouds (Table \ref{table: larson slopes}) may indicate that elongated clouds and clumpy clouds \add{(slope = 0.48)} follow Larson's first law \add{(slope = 0.50)} more closely than other morphologies. However, the large scatter in the data (Fig. \ref{fig: scaling relation larson}) prevents us from confirming this.

\begin{table}
\caption{Slopes and their standard deviations (SD) recovered from PCA on Larson's scaling relation ($\sigma_\varv \propto \mathrm{R}^\mathrm{slope}$) for different cloud morphologies.}
\label{table: larson slopes}
\centering
\begin{tabular}{ccccc}
\hline
Cloud   & Ring & Elongated & Concentrated & Clumpy \\ \hline
Slope   & 0.63 & 0.48      & 0.41         & 0.48   \\ 
SD & 0.01 & < 0.01      & 0.01         & 0.01   \\ \hline
\end{tabular}
\tablefoot{The values are obtained using the bootstrap technique, where different samples are generated by varying the values of $x$ and $y$ parameters of the scaling relation within their errors. We calculate the slope for each sample and report the mean and standard deviation of these slopes in the table. The results are obtained using 10000 random samples generated using random seed 0. The expected value of the slope is 0.5\tablefootmark{a}.}
\tablebib{(a)~\citet{larson_realtion}}
\end{table}

\begin{figure*}
    \centering
    \includegraphics[width = .47 \textwidth, keepaspectratio]{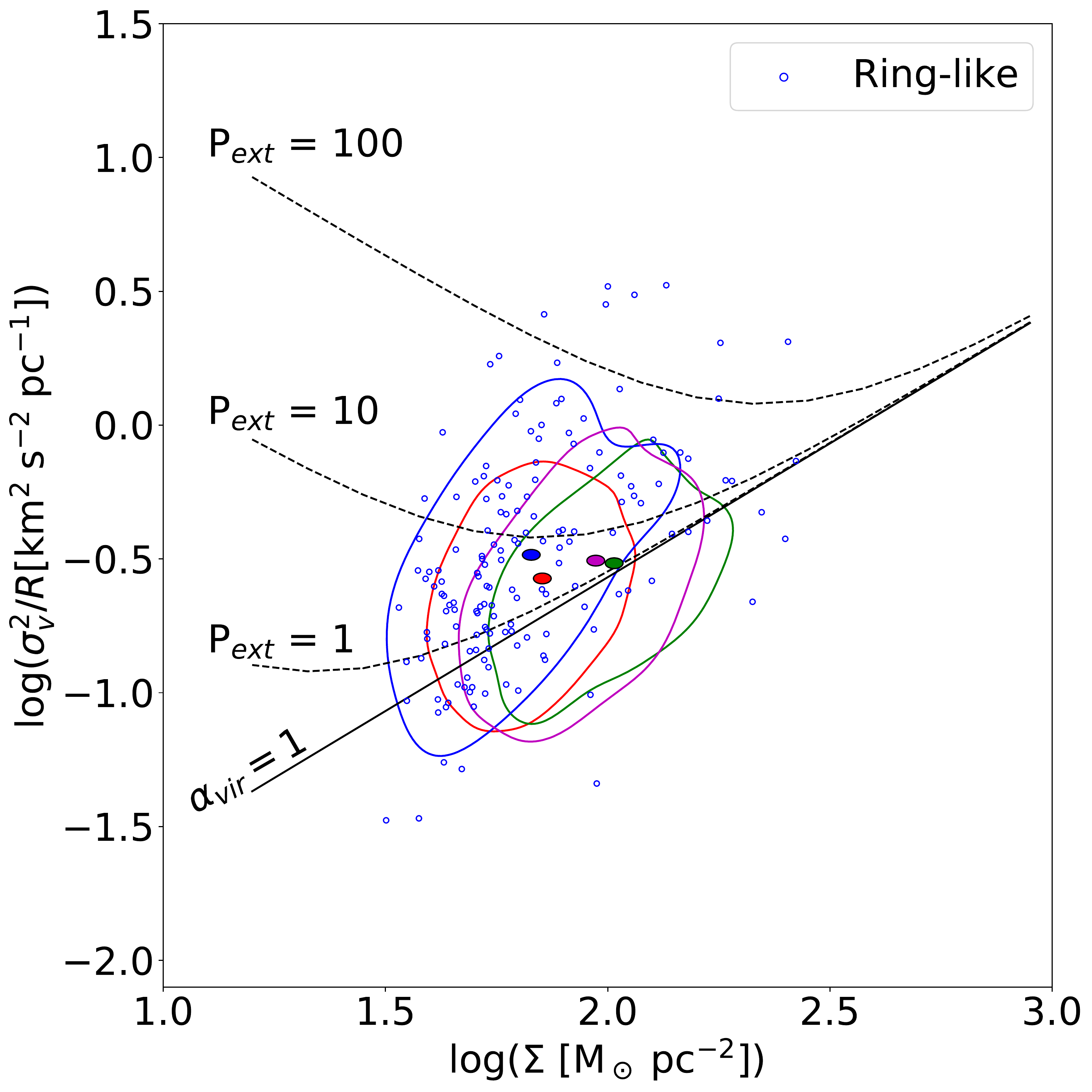}
    \includegraphics[width = .47 \textwidth, keepaspectratio]{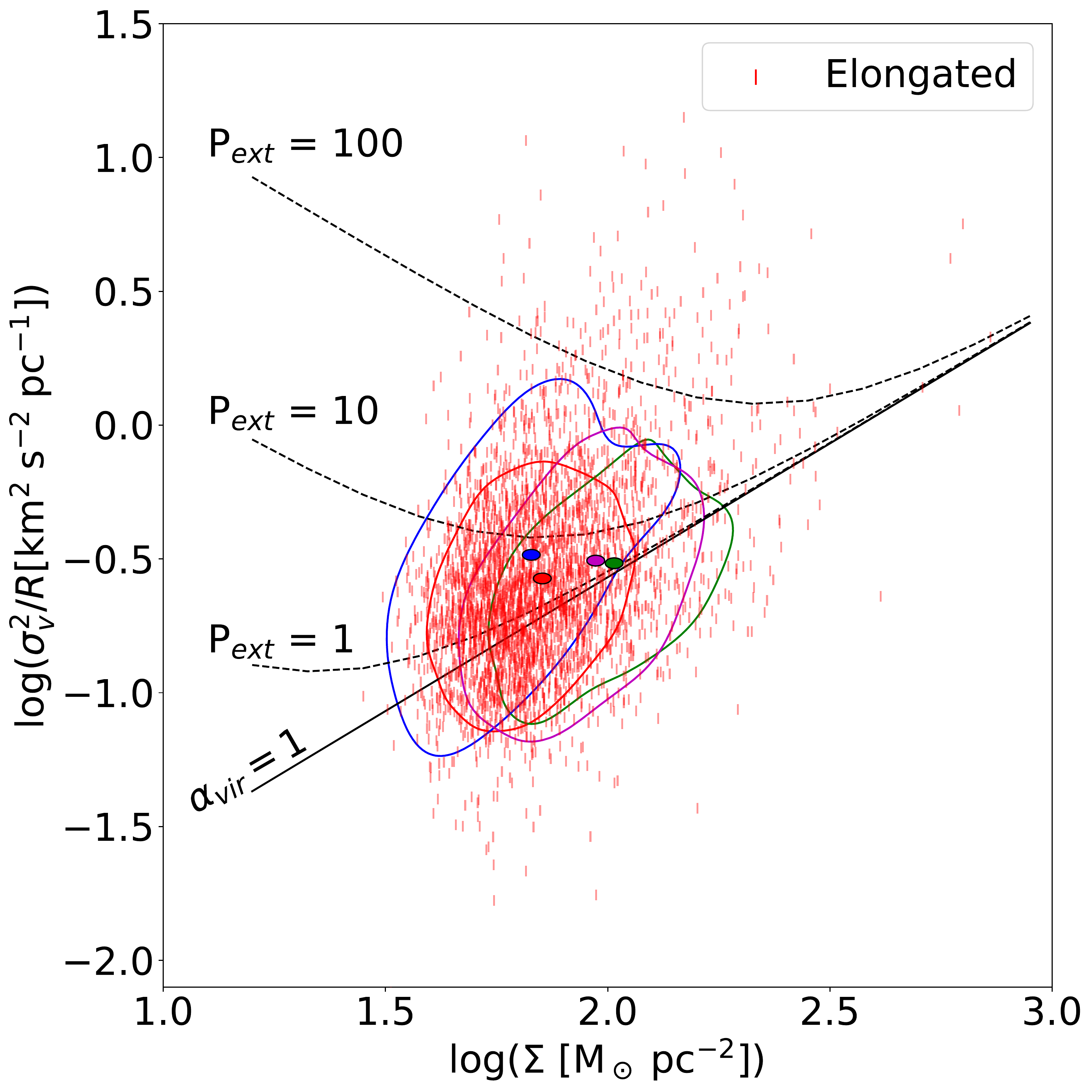}
    \includegraphics[width = .47 \textwidth, keepaspectratio]{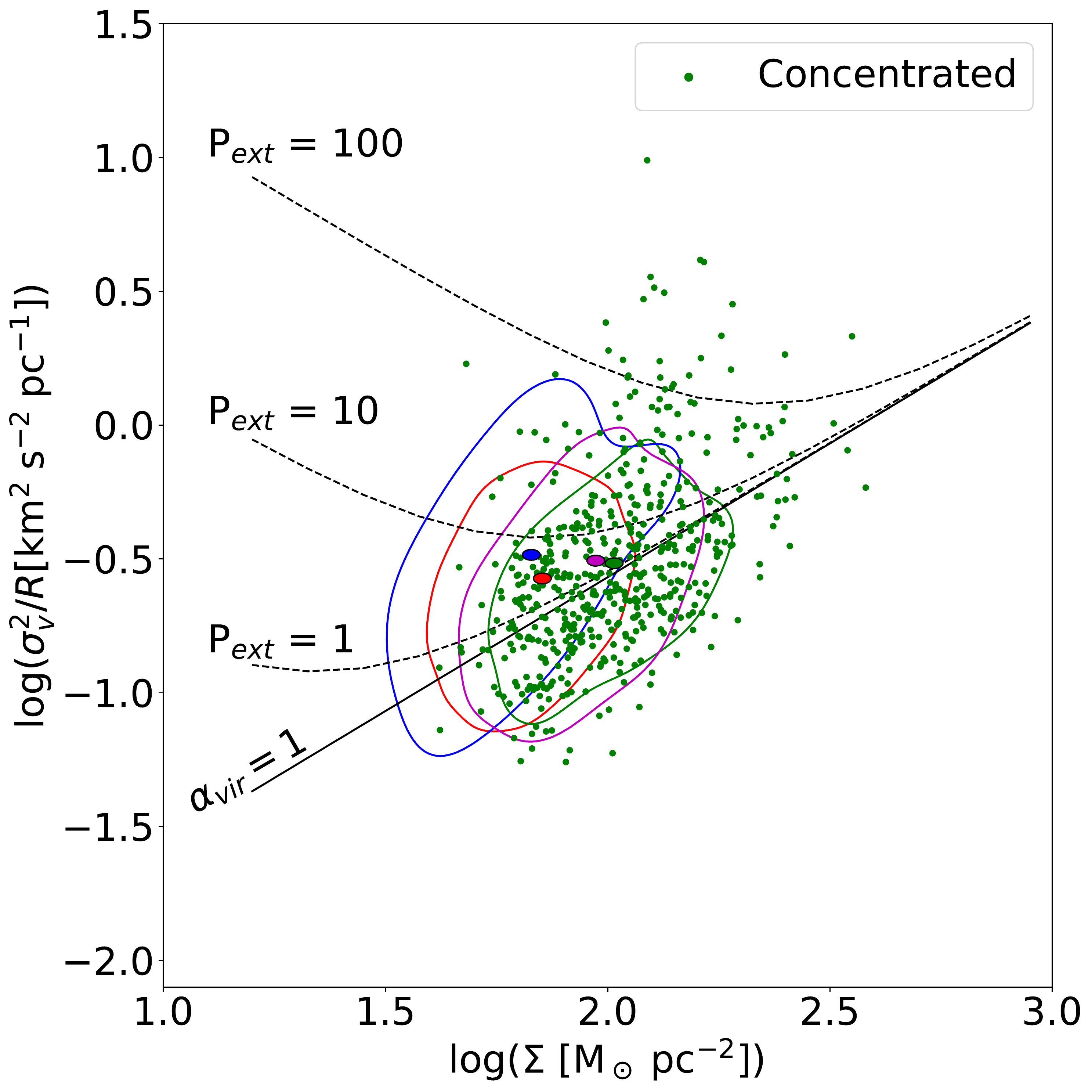}
    \includegraphics[width = .47 \textwidth, keepaspectratio]{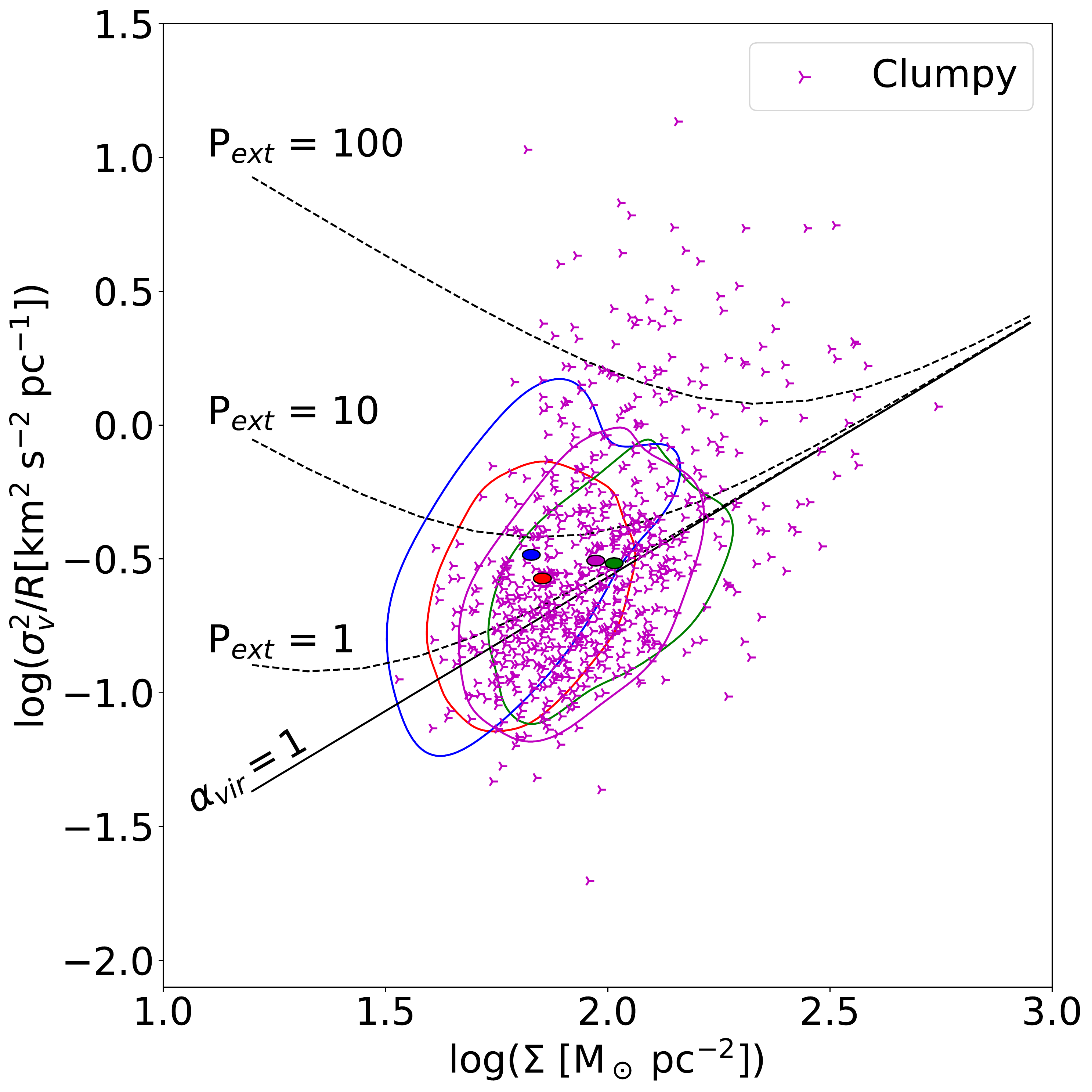}
    \caption{\add{Scaling relation between $\sigma_\varv^2$/$R$ and gas surface density ($\Sigma$) using the MR sample. The black solid lines and dashed lines (from bottom to top) represent $\alpha_\mathrm{vir} = 1$ in absence of external pressure and at an external pressure P$_\mathrm{ext}$ = 1, 10, 100 $\mathrm{M_\odot \; pc ^{-3} \; km^2 \; s^{-2}}$ respectively. The symbols and conventions follow Fig. \ref{fig: scaling relation larson}.}}
    \label{fig: scaling relation heyer}
\end{figure*}

Larson's second relation correlates the velocity dispersion to cloud mass implying that clouds are approximately in virial equilibrium, while the third relation shows the anticorrelation between cloud mean density and size. These relations together with the analysis of \citet{solomon_1987} assert that clouds have a constant surface mass density. \citet{heyer_2009} reanalysed the clouds in the first quadrant of the Galaxy and found that these clouds do not have a constant surface mass density and they are not necessarily gravitationally bound. We show Heyer's relation ($\sigma_\varv^2/R \propto \Sigma$) for different cloud morphologies in Fig. \ref{fig: scaling relation heyer} and report the slopes of the \add{corresponding 1-sigma PCA} ellipses in Table \ref{table: heyer slopes}. 
The different cloud morphologies show a trend with respect to the virial parameter, with ring-like clouds present towards a region of higher virial parameter. This is understood as ring-like clouds are effectively expanding and thus not gravitationally bound. Towards higher surface density and lower velocity dispersion (i.e. lower virial parameter), the regions are correspondingly occupied by elongated, clumpy and then concentrated clouds. This is in agreement with our previous analysis of concentrated clouds being dense, \add{gravitationally bound} compact objects. We also observe that none of the cloud distributions follow $\alpha_\mathrm{vir} = 1$ \citep{heyer_2009}, but the bulk of clouds could be virialised if they are under a relatively mild external pressure \add{or are contracting at some fraction of free-fall} (Fig. \ref{fig: scaling relation heyer}).

\begin{table}
\caption{Slopes recovered from PCA on the Heyer scaling relation ($\sigma_\varv^2/\mathrm{R} \propto \Sigma^\mathrm{slope}$). The terminology follows Table \ref{table: larson slopes}. The expected value of the slope is 1\tablefootmark{a}.}
\label{table: heyer slopes}
\centering
\begin{tabular}{lrrrr}
\hline
Cloud   & \multicolumn{1}{l}{Ring} & \multicolumn{1}{l}{Elongated} & \multicolumn{1}{l}{Concentrated} & \multicolumn{1}{l}{Clumpy} \\ \hline
Slope   & 3.17                     & 4.50                          & 3.46                             & 3.87                       \\ 
SD & 0.21                      & 0.12                           & 0.20                              & 0.21                        \\ \hline
\end{tabular}
\tablebib{(a)~\citet{heyer_2009}}
\end{table}

\section{Discussion}\label{sec: discussions}


The differences in properties of different morphologies can be associated with various physical processes that play a role in the formation and evolution of clouds. The large size (as measured by length and radius \add{in Fig. \ref{fig: prop rad vel dis} and \ref{fig: prop len asp}}) and aspect ratio \add{(Fig. \ref{fig: prop len asp})} for ring-like clouds can be attributed to their formation process. These clouds could be associated with infrared bubbles, which form due to stellar winds and radiation feedback. We find in Paper I that 605 ring-like clouds show association with infrared bubbles from the Milky Way Project (MWP). Stellar feedback could impart high velocity and turbulence to the shells, which then travel large distances to form large rings with high velocity dispersion \citep{snell_1980, Liu_2019}. 
Ring-like clouds have the highest mass (on average). 
\add{However,} the large sizes of these clouds result in them having lower surface densities than other structures.
Overall, ring-like clouds show the most extreme properties compared to the global distribution of clouds as well as other morphologies.

The high mass and surface density \add{(Fig. \ref{fig: prop mass surf dens})} of the clumpy clouds can be explained with two possibilities. The first is that these are multiple overlapping clouds, considered to be a single cloud by $\sc{scimes}$. The other hypothesis is that these clouds evolve similar to globular filaments \citep{globular_filament_ref_nessie_Stru}, having multiple dense regions in the filamentary structure. The second hypothesis is supported by higher star formation efficiencies (refer Paper I) for these clouds as compared to elongated clouds. Clumpy clouds have length distributions similar to elongated clouds, which could be due to a common formation/evolution mechanism, leading to filamentary structures of comparable sizes. It could also be an artefact of the cloud selection criteria used by the $\sc{scimes}$ algorithm. $\sc{scimes}$ is oriented towards selecting clouds with similar sizes and could thus present a single molecular cloud as smaller individual clouds or combine small molecular clouds and present them as a single cloud.

A major observation across all the data samples and throughout the different Galactic environments is that most of the molecular clouds are elongated \add{(Table \ref{tab: quantitative vc and mr samples}; histograms in Fig. \ref{fig: SCIENCE mass ridge R bins}, \ref{fig: morph mass ridge R bins}, \ref{fig: SCIENCE mass ridge z bins}, \ref{fig: morph mass ridge z bins}, \ref{fig: SCIENCE mass ridge distance} and \ref{fig: morph mass ridge distance})}. The average properties of elongated clouds lie between those of ring-like clouds and concentrated clouds. The large sample size for elongated clouds might be the reason for their values being non-extreme. We also find that the Galactic environment does not influence the trends in the integrated properties for most of the morphologies significantly (\add{see the ridge plots in }appendix \ref{app: telescope limitations}).

\section{Summary}\label{sec: summary}

In this paper, we have studied the connection between the morphology of clouds and their integrated properties. We also analyse, how the different cloud morphologies are distributed in the Galaxy.
Our analysis is based on the catalogue of molecular clouds from the SEDIGISM survey (presented in DC21; along with the integrated properties). Clouds were classified into four types based on their morphology (in Paper I).
Our main findings are as follows:

\begin{itemize}
    \item Elongated clouds are the most abundant type of clouds in the Galaxy, with their average properties lying between ring-like and concentrated clouds.
    \item Concentrated clouds have the smallest median sizes and highest median surface densities of all cloud categories.
    \item Ring-like clouds show the most discrepant properties compared with the global population of clouds, with higher average mass, radius, length, aspect ratio and velocity dispersion than other morphologies.
    \item The different morphology classes are not distributed differently in terms of Galactocentric distance (R$_gal$) and Galactic height ($z_\mathrm{gal}$).
    \item \add{The distributions of clouds do not show significant differences on Larson's size-linewidth plot for the different morphologies.}
    \item Heyer's scaling relation shows the evolution of morphologies with virial parameter; with ring-like, elongated, clumpy and concentrated clouds lying in order of decreasing virial parameter.
\end{itemize}

In conclusion, we found some connections between cloud integrated properties and their morphology. The ring-like clouds show an extreme behaviour compared to other cloud morphologies, which \add{may} be a consequence of their nature, i.e. formation due to stellar feedback. Moreover, the global trends in the properties for different morphologies are followed even in separate smaller regions of the Galaxy (i.e. throughout the distance bins (appendix \ref{app 1})), irrespective of the observational biases (appendix \ref{app: telescope limitations}). 
\add{Even though we did not find any particular trend in the distribution of different types of clouds as a function of Galactocentric distances or Galactic height, we have not looked into other environments within the Galaxy, possibly more critical to the evolution of clouds, such as spiral arms or the Galactic centre. The categorisation of clouds into different morphologies could thus give way to various potential follow up studies to help understand the role of the Galaxy in shaping the formation and evolution of molecular clouds.}

\begin{acknowledgement}
\add{The authors thank the referee for valuable comments on the draft which helped improve the quality of the paper.}
This publication is based on data acquired with the Atacama Pathfinder Experiment (APEX) under programmes 092.F-9315 and 193.C-0584. APEX is a collaboration among the Max-Planck-Institut f\"ur Radioastronomie, the European Southern Observatory, and the Onsala Space Observatory. 
This publication uses data generated via the Zooniverse.org platform, development of which is funded by generous support, including a Global Impact Award from Google, and by a grant from the Alfred P. Sloan Foundation. A part of this work is based on observations made with the Spitzer Space Telescope, which is operated by the Jet Propulsion Laboratory, California Institute of Technology under a contract with NASA.
KRN thanks Henrik Beuther for the valuable suggestions and feedback on the paper. KRN would also like to thank Antonio Hernandez for careful reading of the manuscript. 
DC acknowledges support by the German \emph{Deut\-sche For\-schungs\-ge\-mein\-schaft, DFG\/} project number SFB956A.
ADC acknowledges the support from the Royal Society University Research Fellowship (URF/R1/191609). HB acknowledges support from the European Research Council under the Horizon 2020 Framework Programme via the ERC Consol-idator Grant CSF-648505. HB also acknowledges support from the Deutsche Forschungsgemeinschaft in the Collaborative Research Center (SFB 881) “The Milky Way System” (subproject B1). CLD acknowledges funding from the European Research Council for the Horizon 2020 ERC consolidator grant project ICYBOB, grant number 818940.
\end{acknowledgement}

\footnotesize{
\bibliographystyle{aa}
\bibliography{reference}

\begin{thebibliography}{63}
\expandafter\ifx\csname natexlab\endcsname\relax\def\natexlab#1{#1}\fi

\bibitem[{{Arzoumanian} {et~al.}(2011){Arzoumanian}, {Andr{\'e}}, {Didelon},
  {K{\"o}nyves}, {Schneider}, {Men'shchikov}, {Sousbie}, {Zavagno}, {Bontemps},
  {di Francesco}, {Griffin}, {Hennemann}, {Hill}, {Kirk}, {Martin}, {Minier},
  {Molinari}, {Motte}, {Peretto}, {Pezzuto}, {Spinoglio}, {Ward-Thompson},
  {White}, \& {Wilson}}]{globular_filaments_2}
{Arzoumanian}, D., {Andr{\'e}}, P., {Didelon}, P., {et~al.} 2011,
  \href{http://dx.doi.org/10.1051/0004-6361/201116596}{\color{magenta}\aap},
  \href{https://ui.adsabs.harvard.edu/abs/2011A&A...529L...6A}{529, L6}

\bibitem[{{Arzoumanian} {et~al.}(2022){Arzoumanian}, {Russeil}, {Zavagno},
  {Chun-Yuan Chen}, {Andr{\'e}}, {Inutsuka}, {Misugi}, {S{\'a}nchez-Monge},
  {Schilke}, {Men'shchikov}, \& {Kohno}}]{arzoumanian_2022}
{Arzoumanian}, D., {Russeil}, D., {Zavagno}, A., {et~al.} 2022,
  \href{http://dx.doi.org/10.1051/0004-6361/202141699}{\color{magenta}\aap},
  \href{https://ui.adsabs.harvard.edu/abs/2022A&A...660A..56A}{660, A56}

\bibitem[{{Ballesteros-Paredes} {et~al.}(2020){Ballesteros-Paredes},
  {Andr{\'e}}, {Hennebelle}, {Klessen}, {Kruijssen}, {Chevance}, {Nakamura},
  {Adamo}, \& {V{\'a}zquez-Semadeni}}]{ballesteros_2020}
{Ballesteros-Paredes}, J., {Andr{\'e}}, P., {Hennebelle}, P., {et~al.} 2020,
  \href{http://dx.doi.org/10.1007/s11214-020-00698-3}{\color{magenta}\ssr},
  \href{https://ui.adsabs.harvard.edu/abs/2020SSRv..216...76B}{216, 76}

\bibitem[{{Ballesteros-Paredes} {et~al.}(2011){Ballesteros-Paredes},
  {Hartmann}, {V{\'a}zquez-Semadeni}, {Heitsch}, \&
  {Zamora-Avil{\'e}s}}]{ballesteros_paredes_2011}
{Ballesteros-Paredes}, J., {Hartmann}, L.~W., {V{\'a}zquez-Semadeni}, E.,
  {Heitsch}, F., \& {Zamora-Avil{\'e}s}, M.~A. 2011,
  \href{http://dx.doi.org/10.1111/j.1365-2966.2010.17657.x}{\color{magenta}\mnras},
  \href{https://ui.adsabs.harvard.edu/abs/2011MNRAS.411...65B}{411, 65}

\bibitem[{{Barnes} {et~al.}(2015){Barnes}, {Li}, {Telesco}, {Tanakul},
  {Mari{\~n}as}, {Wright}, {Packham}, {Pantin}, {Roche}, \&
  {Hough}}]{barnes_2015}
{Barnes}, P., {Li}, D., {Telesco}, C., {et~al.} 2015,
  \href{http://dx.doi.org/10.1093/mnras/stv1272}{\color{magenta}\mnras},
  \href{https://ui.adsabs.harvard.edu/abs/2015MNRAS.453.2622B}{453, 2622}

\bibitem[{{Barnes} {et~al.}(2018){Barnes}, {Hernandez}, {Muller}, \&
  {Pitts}}]{barnes_2018}
{Barnes}, P.~J., {Hernandez}, A.~K., {Muller}, E., \& {Pitts}, R.~L. 2018,
  \href{http://dx.doi.org/10.3847/1538-4357/aad4ab}{\color{magenta}\apj},
  \href{https://ui.adsabs.harvard.edu/abs/2018ApJ...866...19B}{866, 19}

\bibitem[{{Beuther} {et~al.}(2022){Beuther}, {Schneider}, {Simon}, {Suri},
  {Ossenkopf-Okada}, {Kabanovic}, {R{\"o}llig}, {Guevara}, {Tielens},
  {Sandell}, {Buchbender}, {Ricken}, \& {G{\"u}sten}}]{beuther_2022}
{Beuther}, H., {Schneider}, N., {Simon}, R., {et~al.} 2022,
  \href{http://dx.doi.org/10.1051/0004-6361/202142689}{\color{magenta}\aap},
  \href{https://ui.adsabs.harvard.edu/abs/2022A&A...659A..77B}{659, A77}

\bibitem[{{Blitz} \& {Stark}(1986)}]{blitz_1986}
{Blitz}, L. \& {Stark}, A.~A. 1986,
  \href{http://dx.doi.org/10.1086/184609}{\color{magenta}\apjl},
  \href{https://ui.adsabs.harvard.edu/abs/1986ApJ...300L..89B}{300, L89}

\bibitem[{{Bresnahan} {et~al.}(2018){Bresnahan}, {Ward-Thompson}, {Kirk},
  {Pattle}, {Eyres}, {White}, {K{\"o}nyves}, {Men'shchikov}, {Andr{\'e}},
  {Schneider}, {Di Francesco}, {Arzoumanian}, {Benedettini}, {Ladjelate},
  {Palmeirim}, {Bracco}, {Molinari}, {Pezzuto}, \&
  {Spinoglio}}]{bresnahan_2018}
{Bresnahan}, D., {Ward-Thompson}, D., {Kirk}, J.~M., {et~al.} 2018,
  \href{http://dx.doi.org/10.1051/0004-6361/201730515}{\color{magenta}\aap},
  \href{https://ui.adsabs.harvard.edu/abs/2018A&A...615A.125B}{615, A125}

\bibitem[{{Chapman} {et~al.}(2011){Chapman}, {Goldsmith}, {Pineda}, {Clemens},
  {Li}, \& {Kr{\v{c}}o}}]{chapman_2011}
{Chapman}, N.~L., {Goldsmith}, P.~F., {Pineda}, J.~L., {et~al.} 2011,
  \href{http://dx.doi.org/10.1088/0004-637X/741/1/21}{\color{magenta}\apj},
  \href{https://ui.adsabs.harvard.edu/abs/2011ApJ...741...21C}{741, 21}

\bibitem[{{Chen} {et~al.}(2020){Chen}, {Mundy}, {Ostriker}, {Storm}, \&
  {Dhabal}}]{chen_2020}
{Chen}, C.-Y., {Mundy}, L.~G., {Ostriker}, E.~C., {Storm}, S., \& {Dhabal}, A.
  2020, \href{http://dx.doi.org/10.1093/mnras/staa960}{\color{magenta}\mnras},
  \href{https://ui.adsabs.harvard.edu/abs/2020MNRAS.494.3675C}{494, 3675}

\bibitem[{{Colombo} {et~al.}(2021){Colombo}, {K{\"o}nig}, {Urquhart},
  {Wyrowski}, {Mattern}, {Menten}, {Lee}, {Brand}, {Wienen}, {Mazumdar},
  {Schuller}, \& {Leurini}}]{colombo_2021}
{Colombo}, D., {K{\"o}nig}, C., {Urquhart}, J.~S., {et~al.} 2021,
  \href{http://dx.doi.org/10.1051/0004-6361/202142182}{\color{magenta}\aap},
  \href{https://ui.adsabs.harvard.edu/abs/2021A&A...655L...2C}{655, L2}

\bibitem[{{Colombo} {et~al.}(2019){Colombo}, {Rosolowsky}, {Duarte-Cabral},
  {Ginsburg}, {Glenn}, {Zetterlund}, {Hernand ez}, {Dempsey}, \&
  {Currie}}]{scimes_2}
{Colombo}, D., {Rosolowsky}, E., {Duarte-Cabral}, A., {et~al.} 2019,
  \href{http://dx.doi.org/10.1093/mnras/sty3283}{\color{magenta}\mnras},
  \href{https://ui.adsabs.harvard.edu/abs/2019MNRAS.483.4291C}{483, 4291}

\bibitem[{{Colombo} {et~al.}(2015){Colombo}, {Rosolowsky}, {Ginsburg},
  {Duarte-Cabral}, \& {Hughes}}]{SCIMES}
{Colombo}, D., {Rosolowsky}, E., {Ginsburg}, A., {Duarte-Cabral}, A., \&
  {Hughes}, A. 2015,
  \href{http://dx.doi.org/10.1093/mnras/stv2063}{\color{magenta}\mnras},
  \href{https://ui.adsabs.harvard.edu/abs/2015MNRAS.454.2067C}{454, 2067}

\bibitem[{{Colombo} {et~al.}(2013){Colombo}, {Schinnerer}, {Hughes}, {Meidt},
  {Leroy}, {Pety}, {Dobbs}, {Garcia Burillo}, {Dumas}, {Thompson}, {Schuster},
  \& {Kramer}}]{colombo_2013}
{Colombo}, D., {Schinnerer}, E., {Hughes}, A., {et~al.} 2013, in American
  Astronomical Society Meeting Abstracts, Vol. 221, American Astronomical
  Society Meeting Abstracts \#221,
  \href{https://ui.adsabs.harvard.edu/abs/2013AAS...22134916C}{349.16}

\bibitem[{{Crutcher}(2012)}]{crutcher_2012}
{Crutcher}, R.~M. 2012,
  \href{http://dx.doi.org/10.1146/annurev-astro-081811-125514}{\color{magenta}\araa},
  \href{https://ui.adsabs.harvard.edu/abs/2012ARA&A..50...29C}{50, 29}

\bibitem[{{Crutcher} {et~al.}(2010){Crutcher}, {Wandelt}, {Heiles},
  {Falgarone}, \& {Troland}}]{crutcher_2010}
{Crutcher}, R.~M., {Wandelt}, B., {Heiles}, C., {Falgarone}, E., \& {Troland},
  T.~H. 2010,
  \href{http://dx.doi.org/10.1088/0004-637X/725/1/466}{\color{magenta}\apj},
  \href{https://ui.adsabs.harvard.edu/abs/2010ApJ...725..466C}{725, 466}

\bibitem[{{Deharveng} {et~al.}(2010){Deharveng}, {Schuller}, {Anderson},
  {Zavagno}, {Wyrowski}, {Menten}, {Bronfman}, {Testi}, {Walmsley}, \&
  {Wienen}}]{deharveng_2010}
{Deharveng}, L., {Schuller}, F., {Anderson}, L.~D., {et~al.} 2010,
  \href{http://dx.doi.org/10.1051/0004-6361/201014422}{\color{magenta}\aap},
  \href{https://ui.adsabs.harvard.edu/abs/2010A&A...523A...6D}{523, A6}

\bibitem[{{Dobbs} {et~al.}(2014){Dobbs}, {Krumholz}, {Ballesteros-Paredes},
  {Bolatto}, {Fukui}, {Heyer}, {Low}, {Ostriker}, \&
  {V{\'a}zquez-Semadeni}}]{cloud_review_krumholz}
{Dobbs}, C.~L., {Krumholz}, M.~R., {Ballesteros-Paredes}, J., {et~al.} 2014, in
  Protostars and Planets VI, ed. H.~{Beuther}, R.~S. {Klessen}, C.~P.
  {Dullemond}, \& T.~{Henning},
  \href{https://ui.adsabs.harvard.edu/abs/2014prpl.conf....3D}{3}

\bibitem[{{Duarte-Cabral} {et~al.}(2021){Duarte-Cabral}, {Colombo}, {Urquhart},
  {Ginsburg}, {Russeil}, {Schuller}, {Anderson}, {Barnes}, {Beltr{\'a}n},
  {Beuther}, {Bontemps}, {Bronfman}, {Csengeri}, {Dobbs}, {Eden}, {Giannetti},
  {Kauffmann}, {Mattern}, {Medina}, {Menten}, {Lee}, {Pettitt}, {Riener},
  {Rigby}, {Traficante}, {Veena}, {Wienen}, {Wyrowski}, {Agurto}, {Azagra},
  {Cesaroni}, {Finger}, {Gonzalez}, {Henning}, {Hernandez}, {Kainulainen},
  {Leurini}, {Lopez}, {Mac-Auliffe}, {Mazumdar}, {Molinari}, {Motte}, {Muller},
  {Nguyen-Luong}, {Parra}, {Perez-Beaupuits}, {Montenegro-Montes}, {Moore},
  {Ragan}, {S{\'a}nchez-Monge}, {Sanna}, {Schilke}, {Schisano}, {Schneider},
  {Suri}, {Testi}, {Torstensson}, {Venegas}, {Wang}, \& {Zavagno}}]{ana_paper}
{Duarte-Cabral}, A., {Colombo}, D., {Urquhart}, J.~S., {et~al.} 2021,
  \href{http://dx.doi.org/10.1093/mnras/staa2480}{\color{magenta}\mnras},
  \href{https://ui.adsabs.harvard.edu/abs/2021MNRAS.500.3027D}{500, 3027}

\bibitem[{{Duarte-Cabral} \& {Dobbs}(2016)}]{filaments_simulation_ana_16}
{Duarte-Cabral}, A. \& {Dobbs}, C.~L. 2016,
  \href{http://dx.doi.org/10.1093/mnras/stw469}{\color{magenta}\mnras},
  \href{https://ui.adsabs.harvard.edu/abs/2016MNRAS.458.3667D}{458, 3667}

\bibitem[{{Elmegreen} \& {Lada}(1977)}]{elmegreen_1977}
{Elmegreen}, B.~G. \& {Lada}, C.~J. 1977,
  \href{http://dx.doi.org/10.1086/155302}{\color{magenta}\apj},
  \href{https://ui.adsabs.harvard.edu/abs/1977ApJ...214..725E}{214, 725}

\bibitem[{{Field} {et~al.}(2011){Field}, {Blackman}, \& {Keto}}]{field_2011}
{Field}, G.~B., {Blackman}, E.~G., \& {Keto}, E.~R. 2011,
  \href{http://dx.doi.org/10.1111/j.1365-2966.2011.19091.x}{\color{magenta}\mnras},
  \href{https://ui.adsabs.harvard.edu/abs/2011MNRAS.416..710F}{416, 710}

\bibitem[{{Fukui} {et~al.}(2008){Fukui}, {Kawamura}, {Minamidani}, {Mizuno},
  {Kanai}, {Mizuno}, {Onishi}, {Yonekura}, {Mizuno}, {Ogawa}, \&
  {Rubio}}]{fukui_2008}
{Fukui}, Y., {Kawamura}, A., {Minamidani}, T., {et~al.} 2008,
  \href{http://dx.doi.org/10.1086/589833}{\color{magenta}\apjs},
  \href{https://ui.adsabs.harvard.edu/abs/2008ApJS..178...56F}{178, 56}

\bibitem[{{Geen} {et~al.}(2016){Geen}, {Hennebelle}, {Tremblin}, \&
  {Rosdahl}}]{geen_2016}
{Geen}, S., {Hennebelle}, P., {Tremblin}, P., \& {Rosdahl}, J. 2016,
  \href{http://dx.doi.org/10.1093/mnras/stw2235}{\color{magenta}\mnras},
  \href{https://ui.adsabs.harvard.edu/abs/2016MNRAS.463.3129G}{463, 3129}

\bibitem[{{G{\"u}sten} {et~al.}(2006){G{\"u}sten}, {Nyman}, {Schilke},
  {Menten}, {Cesarsky}, \& {Booth}}]{gusten_2006}
{G{\"u}sten}, R., {Nyman}, L.~{\r{A}}., {Schilke}, P., {et~al.} 2006,
  \href{http://dx.doi.org/10.1051/0004-6361:20065420}{\color{magenta}\aap},
  \href{https://ui.adsabs.harvard.edu/abs/2006A&A...454L..13G}{454, L13}

\bibitem[{{Hacar} {et~al.}(2013){Hacar}, {Tafalla}, {Kauffmann}, \&
  {Kov{\'a}cs}}]{hacar_2013}
{Hacar}, A., {Tafalla}, M., {Kauffmann}, J., \& {Kov{\'a}cs}, A. 2013,
  \href{http://dx.doi.org/10.1051/0004-6361/201220090}{\color{magenta}\aap},
  \href{https://ui.adsabs.harvard.edu/abs/2013A&A...554A..55H}{554, A55}

\bibitem[{{Heyer} \& {Dame}(2015)}]{intro_refer_1}
{Heyer}, M. \& {Dame}, T.~M. 2015,
  \href{http://dx.doi.org/10.1146/annurev-astro-082214-122324}{\color{magenta}\araa},
  \href{https://ui.adsabs.harvard.edu/abs/2015ARA&A..53..583H}{53, 583}

\bibitem[{{Heyer} {et~al.}(2009){Heyer}, {Krawczyk}, {Duval}, \&
  {Jackson}}]{heyer_2009}
{Heyer}, M., {Krawczyk}, C., {Duval}, J., \& {Jackson}, J.~M. 2009,
  \href{http://dx.doi.org/10.1088/0004-637X/699/2/1092}{\color{magenta}\apj},
  \href{https://ui.adsabs.harvard.edu/abs/2009ApJ...699.1092H}{699, 1092}

\bibitem[{{Jaffa} {et~al.}(2018){Jaffa}, {Whitworth}, {Clarke}, \&
  {Howard}}]{jaffa_2018}
{Jaffa}, S.~E., {Whitworth}, A.~P., {Clarke}, S.~D., \& {Howard}, A.~D.~P.
  2018, \href{http://dx.doi.org/10.1093/mnras/sty696}{\color{magenta}\mnras},
  \href{https://ui.adsabs.harvard.edu/abs/2018MNRAS.477.1940J}{477, 1940}

\bibitem[{{Jayasinghe} {et~al.}(2019){Jayasinghe}, {Dixon}, {Povich}, {Binder},
  {Velasco}, {Lepore}, {Xu}, {Offner}, {Kobulnicky}, {Anderson}, {Kendrew}, \&
  {Simpson}}]{jayasingghe_2019}
{Jayasinghe}, T., {Dixon}, D., {Povich}, M.~S., {et~al.} 2019,
  \href{http://dx.doi.org/10.1093/mnras/stz1738}{\color{magenta}\mnras},
  \href{https://ui.adsabs.harvard.edu/abs/2019MNRAS.488.1141J}{488, 1141}

\bibitem[{{Kainulainen} {et~al.}(2017){Kainulainen}, {Stutz}, {Stanke},
  {Abreu-Vicente}, {Beuther}, {Henning}, {Johnston}, \&
  {Megeath}}]{kainulainen_2017}
{Kainulainen}, J., {Stutz}, A.~M., {Stanke}, T., {et~al.} 2017,
  \href{http://dx.doi.org/10.1051/0004-6361/201628481}{\color{magenta}\aap},
  \href{https://ui.adsabs.harvard.edu/abs/2017A&A...600A.141K}{600, A141}

\bibitem[{{Koda} {et~al.}(2006){Koda}, {Sawada}, {Hasegawa}, \&
  {Scoville}}]{GMF_formation_1}
{Koda}, J., {Sawada}, T., {Hasegawa}, T., \& {Scoville}, N.~Z. 2006,
  \href{http://dx.doi.org/10.1086/498640}{\color{magenta}\apj},
  \href{https://ui.adsabs.harvard.edu/abs/2006ApJ...638..191K}{638, 191}

\bibitem[{{Larson}(1981)}]{larson_realtion}
{Larson}, R.~B. 1981,
  \href{http://dx.doi.org/10.1093/mnras/194.4.809}{\color{magenta}\mnras},
  \href{https://ui.adsabs.harvard.edu/abs/1981MNRAS.194..809L}{194, 809}

\bibitem[{{Li} {et~al.}(2016){Li}, {Urquhart}, {Leurini}, {Csengeri},
  {Wyrowski}, {Menten}, \& {Schuller}}]{ATLASGAL_filaments}
{Li}, G.-X., {Urquhart}, J.~S., {Leurini}, S., {et~al.} 2016,
  \href{http://dx.doi.org/10.1051/0004-6361/201527468}{\color{magenta}\aap},
  \href{https://ui.adsabs.harvard.edu/abs/2016A&A...591A...5L}{591, A5}

\bibitem[{{Liu} {et~al.}(2019){Liu}, {Li}, {Kr{\v{c}}o}, {Ho}, {Xu}, \&
  {Li}}]{Liu_2019}
{Liu}, M., {Li}, D., {Kr{\v{c}}o}, M., {et~al.} 2019,
  \href{http://dx.doi.org/10.3847/1538-4357/ab4880}{\color{magenta}\apj},
  \href{https://ui.adsabs.harvard.edu/abs/2019ApJ...885..124L}{885, 124}

\bibitem[{{Luisi} {et~al.}(2021){Luisi}, {Anderson}, {Schneider}, {Simon},
  {Kabanovic}, {G{\"u}sten}, {Zavagno}, {Broos}, {Buchbender}, {Guevara},
  {Jacobs}, {Justen}, {Klein}, {Linville}, {R{\"o}llig}, {Russeil}, {Stutzki},
  {Tiwari}, {Townsley}, \& {Tielens}}]{luisi_2021}
{Luisi}, M., {Anderson}, L.~D., {Schneider}, N., {et~al.} 2021,
  \href{http://dx.doi.org/10.1126/sciadv.abe9511}{\color{magenta}Science
  Advances}, \href{https://ui.adsabs.harvard.edu/abs/2021SciA....7.9511L}{7,
  eabe9511}

\bibitem[{{Mattern} {et~al.}(2018){Mattern}, {Kauffmann}, {Csengeri},
  {Urquhart}, {Leurini}, {Wyrowski}, {Giannetti}, {Barnes}, {Beuther},
  {Bronfman}, {Duarte-Cabral}, {Henning}, {Kainulainen}, {Menten}, {Schisano},
  \& {Schuller}}]{mass_vel_scaling_relation}
{Mattern}, M., {Kauffmann}, J., {Csengeri}, T., {et~al.} 2018,
  \href{http://dx.doi.org/10.1051/0004-6361/201833406}{\color{magenta}\aap},
  \href{https://ui.adsabs.harvard.edu/abs/2018A&A...619A.166M}{619, A166}

\bibitem[{{Miville-Desch{\^e}nes} {et~al.}(2017){Miville-Desch{\^e}nes},
  {Murray}, \& {Lee}}]{miville_2017}
{Miville-Desch{\^e}nes}, M.-A., {Murray}, N., \& {Lee}, E.~J. 2017,
  \href{http://dx.doi.org/10.3847/1538-4357/834/1/57}{\color{magenta}\apj},
  \href{https://ui.adsabs.harvard.edu/abs/2017ApJ...834...57M}{834, 57}

\bibitem[{{Neralwar} {et~al.}(2022){Neralwar}, {Colombo}, {Duarte-Cabral},
  {Urquhart}, {Mattern}, {Wyrowski}, {Menten}, {Barnes}, {Sanchez-Monge},
  {Beuther}, {Rigby}, {Mazumdar}, {Eden}, {Csengeri}, {Dobbs}, {Veena},
  {Neupane}, {Henning}, {Schuller}, {Leurini}, {Wienen}, {Yang}, {Ragan},
  {Medina}, \& {Nguyen-Luong}}]{neralwar_2022}
{Neralwar}, K.~R., {Colombo}, D., {Duarte-Cabral}, A., {et~al.} 2022,
  \href{https://ui.adsabs.harvard.edu/abs/2022arXiv220302504N}{arXiv e-prints,
  arXiv:2203.02504}

\bibitem[{{Olmi} {et~al.}(2016){Olmi}, {Cunningham}, {Elia}, \&
  {Jones}}]{olmi_2016}
{Olmi}, L., {Cunningham}, M., {Elia}, D., \& {Jones}, P. 2016,
  \href{http://dx.doi.org/10.1051/0004-6361/201628519}{\color{magenta}\aap},
  \href{https://ui.adsabs.harvard.edu/abs/2016A&A...594A..58O}{594, A58}

\bibitem[{{Polychroni} {et~al.}(2013){Polychroni}, {Schisano}, {Elia}, {Roy},
  {Molinari}, {Martin}, {Andr{\'e}}, {Turrini}, {Rygl}, {Di Francesco},
  {Benedettini}, {Busquet}, {di Giorgio}, {Pestalozzi}, {Pezzuto},
  {Arzoumanian}, {Bontemps}, {Hennemann}, {Hill}, {K{\"o}nyves},
  {Men'shchikov}, {Motte}, {Nguyen-Luong}, {Peretto}, {Schneider}, \&
  {White}}]{polychroni_2013}
{Polychroni}, D., {Schisano}, E., {Elia}, D., {et~al.} 2013,
  \href{http://dx.doi.org/10.1088/2041-8205/777/2/L33}{\color{magenta}\apjl},
  \href{https://ui.adsabs.harvard.edu/abs/2013ApJ...777L..33P}{777, L33}

\bibitem[{{Rebolledo} {et~al.}(2012){Rebolledo}, {Wong}, {Leroy}, {Koda}, \&
  {Donovan Meyer}}]{rebolledo_2012}
{Rebolledo}, D., {Wong}, T., {Leroy}, A., {Koda}, J., \& {Donovan Meyer}, J.
  2012,
  \href{http://dx.doi.org/10.1088/0004-637X/757/2/155}{\color{magenta}\apj},
  \href{https://ui.adsabs.harvard.edu/abs/2012ApJ...757..155R}{757, 155}

\bibitem[{{Rodgers} {et~al.}(1960){Rodgers}, {Campbell}, \&
  {Whiteoak}}]{rodgers_1960}
{Rodgers}, A.~W., {Campbell}, C.~T., \& {Whiteoak}, J.~B. 1960,
  \href{http://dx.doi.org/10.1093/mnras/121.1.103}{\color{magenta}\mnras},
  \href{https://ui.adsabs.harvard.edu/abs/1960MNRAS.121..103R}{121, 103}

\bibitem[{{Roman-Duval} {et~al.}(2010){Roman-Duval}, {Jackson}, {Heyer},
  {Rathborne}, \& {Simon}}]{roman_duval_2010}
{Roman-Duval}, J., {Jackson}, J.~M., {Heyer}, M., {Rathborne}, J., \& {Simon},
  R. 2010,
  \href{http://dx.doi.org/10.1088/0004-637X/723/1/492}{\color{magenta}\apj},
  \href{https://ui.adsabs.harvard.edu/abs/2010ApJ...723..492R}{723, 492}

\bibitem[{{Rosolowsky} \& {Leroy}(2006)}]{crops}
{Rosolowsky}, E. \& {Leroy}, A. 2006,
  \href{http://dx.doi.org/10.1086/502982}{\color{magenta}Publications of the
  ASP}, \href{https://ui.adsabs.harvard.edu/abs/2006PASP..118..590R}{118, 590}

\bibitem[{{Rosolowsky} {et~al.}(2008){Rosolowsky}, {Pineda}, {Kauffmann}, \&
  {Goodman}}]{dendrograms}
{Rosolowsky}, E.~W., {Pineda}, J.~E., {Kauffmann}, J., \& {Goodman}, A.~A.
  2008, \href{http://dx.doi.org/10.1086/587685}{\color{magenta}\apj},
  \href{https://ui.adsabs.harvard.edu/abs/2008ApJ...679.1338R}{679, 1338}

\bibitem[{{Schneider} {et~al.}(2020){Schneider}, {Simon}, {Guevara},
  {Buchbender}, {Higgins}, {Okada}, {Stutzki}, {G{\"u}sten}, {Anderson},
  {Bally}, {Beuther}, {Bonne}, {Bontemps}, {Chambers}, {Csengeri}, {Graf},
  {Gusdorf}, {Jacobs}, {Justen}, {Kabanovic}, {Karim}, {Luisi}, {Menten},
  {Mertens}, {Mookerjea}, {Ossenkopf-Okada}, {Pabst}, {Pound}, {Richter},
  {Reyes}, {Ricken}, {R{\"o}llig}, {Russeil}, {S{\'a}nchez-Monge}, {Sandell},
  {Tiwari}, {Wiesemeyer}, {Wolfire}, {Wyrowski}, {Zavagno}, \&
  {Tielens}}]{schneider_2020}
{Schneider}, N., {Simon}, R., {Guevara}, C., {et~al.} 2020,
  \href{http://dx.doi.org/10.1088/1538-3873/aba840}{\color{magenta}Publications
  of the ASP},
  \href{https://ui.adsabs.harvard.edu/abs/2020PASP..132j4301S}{132, 104301}

\bibitem[{{Schneider} \& {Elmegreen}(1979)}]{globular_filament_ref_nessie_Stru}
{Schneider}, S. \& {Elmegreen}, B.~G. 1979,
  \href{http://dx.doi.org/10.1086/190609}{\color{magenta}\apjs},
  \href{https://ui.adsabs.harvard.edu/abs/1979ApJS...41...87S}{41, 87}

\bibitem[{{Schuller} {et~al.}(2017){Schuller}, {Csengeri}, {Urquhart},
  {Duarte-Cabral}, {Barnes}, {Giannetti}, {Hernandez}, {Leurini}, {Mattern},
  {Medina}, {Agurto}, {Azagra}, {Anderson}, {Beltr{\'a}n}, {Beuther},
  {Bontemps}, {Bronfman}, {Dobbs}, {Dumke}, {Finger}, {Ginsburg}, {Gonzalez},
  {Henning}, {Kauffmann}, {Mac-Auliffe}, {Menten}, {Montenegro-Montes},
  {Moore}, {Muller}, {Parra}, {Perez-Beaupuits}, {Pettitt}, {Russeil},
  {S{\'a}nchez-Monge}, {Schilke}, {Schisano}, {Suri}, {Testi}, {Torstensson},
  {Venegas}, {Wang}, {Wienen}, {Wyrowski}, \& {Zavagno}}]{schuller_2017}
{Schuller}, F., {Csengeri}, T., {Urquhart}, J.~S., {et~al.} 2017,
  \href{http://dx.doi.org/10.1051/0004-6361/201628933}{\color{magenta}\aap},
  \href{https://ui.adsabs.harvard.edu/abs/2017A&A...601A.124S}{601, A124}

\bibitem[{{Schuller} {et~al.}(2021){Schuller}, {Urquhart}, {Csengeri},
  {Colombo}, {Duarte-Cabral}, {Mattern}, {Ginsburg}, {Pettitt}, {Wyrowski},
  {Anderson}, {Azagra}, {Barnes}, {Beltran}, {Beuther}, {Billington},
  {Bronfman}, {Cesaroni}, {Dobbs}, {Eden}, {Lee}, {Medina}, {Menten}, {Moore},
  {Montenegro-Montes}, {Ragan}, {Rigby}, {Riener}, {Russeil}, {Schisano},
  {Sanchez-Monge}, {Traficante}, {Zavagno}, {Agurto}, {Bontemps}, {Finger},
  {Giannetti}, {Gonzalez}, {Hernandez}, {Henning}, {Kainulainen}, {Kauffmann},
  {Leurini}, {Lopez}, {Mac-Auliffe}, {Mazumdar}, {Molinari}, {Motte}, {Muller},
  {Nguyen-Luong}, {Parra}, {Perez-Beaupuits}, {Schilke}, {Schneider}, {Suri},
  {Testi}, {Torstensson}, {Veena}, {Venegas}, {Wang}, \& {Wienen}}]{SEDIGISM_1}
{Schuller}, F., {Urquhart}, J.~S., {Csengeri}, T., {et~al.} 2021,
  \href{http://dx.doi.org/10.1093/mnras/staa2369}{\color{magenta}\mnras},
  \href{https://ui.adsabs.harvard.edu/abs/2021MNRAS.500.3064S}{500, 3064}

\bibitem[{{Snell} {et~al.}(1980){Snell}, {Loren}, \& {Plambeck}}]{snell_1980}
{Snell}, R.~L., {Loren}, R.~B., \& {Plambeck}, R.~L. 1980,
  \href{http://dx.doi.org/10.1086/183283}{\color{magenta}\apjl},
  \href{https://ui.adsabs.harvard.edu/abs/1980ApJ...239L..17S}{239, L17}

\bibitem[{{Solomon} {et~al.}(1987){Solomon}, {Rivolo}, {Barrett}, \&
  {Yahil}}]{solomon_1987}
{Solomon}, P.~M., {Rivolo}, A.~R., {Barrett}, J., \& {Yahil}, A. 1987,
  \href{http://dx.doi.org/10.1086/165493}{\color{magenta}\apj},
  \href{https://ui.adsabs.harvard.edu/abs/1987ApJ...319..730S}{319, 730}

\bibitem[{{Sun} {et~al.}(2018){Sun}, {Leroy}, {Schruba}, {Rosolowsky},
  {Hughes}, {Kruijssen}, {Meidt}, {Schinnerer}, {Blanc}, {Bigiel}, {Bolatto},
  {Chevance}, {Groves}, {Herrera}, {Hygate}, {Pety}, {Querejeta}, {Usero}, \&
  {Utomo}}]{sun_2018}
{Sun}, J., {Leroy}, A.~K., {Schruba}, A., {et~al.} 2018,
  \href{http://dx.doi.org/10.3847/1538-4357/aac326}{\color{magenta}\apj},
  \href{https://ui.adsabs.harvard.edu/abs/2018ApJ...860..172S}{860, 172}

\bibitem[{{Tiwari} {et~al.}(2021){Tiwari}, {Karim}, {Pound}, {Wolfire},
  {Jacob}, {Buchbender}, {G{\"u}sten}, {Guevara}, {Higgins}, {Kabanovic},
  {Pabst}, {Ricken}, {Schneider}, {Simon}, {Stutzki}, \&
  {Tielens}}]{tiwari_2021}
{Tiwari}, M., {Karim}, R., {Pound}, M.~W., {et~al.} 2021,
  \href{http://dx.doi.org/10.3847/1538-4357/abf6ce}{\color{magenta}\apj},
  \href{https://ui.adsabs.harvard.edu/abs/2021ApJ...914..117T}{914, 117}

\bibitem[{{Urquhart} {et~al.}(2021){Urquhart}, {Figura}, {Cross}, {Wells},
  {Moore}, {Eden}, {Ragan}, {Pettitt}, {Duarte-Cabral}, {Colombo}, {Schuller},
  {Csengeri}, {Mattern}, {Beuther}, {Menten}, {Wyrowski}, {Anderson}, {Barnes},
  {Beltr{\'a}n}, {Billington}, {Bronfman}, {Giannetti}, {Kainulainen},
  {Kauffmann}, {Lee}, {Leurini}, {Medina}, {Montenegro-Montes}, {Riener},
  {Rigby}, {S{\'a}nchez-Monge}, {Schilke}, {Schisano}, {Traficante}, \&
  {Wienen}}]{james_paper}
{Urquhart}, J.~S., {Figura}, C., {Cross}, J.~R., {et~al.} 2021,
  \href{http://dx.doi.org/10.1093/mnras/staa2512}{\color{magenta}\mnras},
  \href{https://ui.adsabs.harvard.edu/abs/2021MNRAS.500.3050U}{500, 3050}

\bibitem[{{V{\'a}zquez-Semadeni} {et~al.}(2019){V{\'a}zquez-Semadeni}, {Palau},
  {Ballesteros-Paredes}, {G{\'o}mez}, \& {Zamora-Avil{\'e}s}}]{vazquez_2019}
{V{\'a}zquez-Semadeni}, E., {Palau}, A., {Ballesteros-Paredes}, J.,
  {G{\'o}mez}, G.~C., \& {Zamora-Avil{\'e}s}, M. 2019,
  \href{http://dx.doi.org/10.1093/mnras/stz2736}{\color{magenta}\mnras},
  \href{https://ui.adsabs.harvard.edu/abs/2019MNRAS.490.3061V}{490, 3061}

\bibitem[{{Yuan} {et~al.}(2021){Yuan}, {Yang}, {Du}, {Liu}, {Zhang}, {Lin},
  {Sun}, {Yan}, {Ma}, {Su}, {Sun}, \& {Zhou}}]{yuan_2021}
{Yuan}, L., {Yang}, J., {Du}, F., {et~al.} 2021,
  \href{http://dx.doi.org/10.3847/1538-4365/ac242a}{\color{magenta}\apjs},
  \href{https://ui.adsabs.harvard.edu/abs/2021ApJS..257...51Y}{257, 51}

\bibitem[{{Zavagno} {et~al.}(2006){Zavagno}, {Deharveng}, {Comer{\'o}n},
  {Brand}, {Massi}, {Caplan}, \& {Russeil}}]{zavagno_2006}
{Zavagno}, A., {Deharveng}, L., {Comer{\'o}n}, F., {et~al.} 2006,
  \href{http://dx.doi.org/10.1051/0004-6361:20053952}{\color{magenta}\aap},
  \href{https://ui.adsabs.harvard.edu/abs/2006A&A...446..171Z}{446, 171}

\bibitem[{{Zavagno} {et~al.}(2007){Zavagno}, {Pomar{\`e}s}, {Deharveng},
  {Hosokawa}, {Russeil}, \& {Caplan}}]{zavagno_2007}
{Zavagno}, A., {Pomar{\`e}s}, M., {Deharveng}, L., {et~al.} 2007,
  \href{http://dx.doi.org/10.1051/0004-6361:20077474}{\color{magenta}\aap},
  \href{https://ui.adsabs.harvard.edu/abs/2007A&A...472..835Z}{472, 835}

\bibitem[{{Zhou} {et~al.}(2020){Zhou}, {Zhou}, {Esimbek}, {Baan}, {Wu}, {Ji},
  {He}, {Li}, {Sailanbek}, {Komesh}, \& {Tang}}]{jianjun_2020}
{Zhou}, J., {Zhou}, D., {Esimbek}, J., {et~al.} 2020,
  \href{http://dx.doi.org/10.3847/1538-4357/ab94c0}{\color{magenta}\apj},
  \href{https://ui.adsabs.harvard.edu/abs/2020ApJ...897...74Z}{897, 74}

\bibitem[{{Zucker} {et~al.}(2018){Zucker}, {Battersby}, \&
  {Goodman}}]{filaments_catherine_zucker}
{Zucker}, C., {Battersby}, C., \& {Goodman}, A. 2018,
  \href{http://dx.doi.org/10.3847/1538-4357/aacc66}{\color{magenta}\apj},
  \href{https://ui.adsabs.harvard.edu/abs/2018ApJ...864..153Z}{864, 153}

\bibitem[{{Zucker} {et~al.}(2021){Zucker}, {Goodman}, {Alves}, {Bialy}, {Koch},
  {Speagle}, {Foley}, {Finkbeiner}, {Leike}, {En{\ss}lin}, {Peek}, \&
  {Edenhofer}}]{zucker_2021}
{Zucker}, C., {Goodman}, A., {Alves}, J., {et~al.} 2021,
  \href{http://dx.doi.org/10.3847/1538-4357/ac1f96}{\color{magenta}\apj},
  \href{https://ui.adsabs.harvard.edu/abs/2021ApJ...919...35Z}{919, 35}

\end{thebibliography}
}

\begin{appendix}

\section{Kolmogorov–Smirnov test and p-value}\label{app: ks test}

The two-sample Kolmogorov–Smirnov (KS) test is a non-parametric goodness of fit test that compares the cumulative distribution of two datasets by calculating p-value and the maximum difference between them. We use the null hypothesis that the two distributions are obtained from the same distribution. Hence, a low p-value (typically below 0.01) leads to rejection of the null hypothesis, and is an indication that the two samples belong to different distributions. All the p-values throughout this paper are obtained from the KS test. We use the ks\_2samp module from python scipy package to implement KS test on our distributions.

We use the KS test for two major tasks. The first is to confirm whether the distribution of property for a particular morphology is \add{the }same for VC and MR samples. We confirm that VC and MR samples are consistent with each other  in Table \ref{tab: KS test VC vs MR}. The second task is to check if different morphologies show different distribution for a given property. The differences in the properties for different morphologies can be visually distinguished in the violin plots (Fig. \ref{fig: prop len asp}--\ref{fig: prop vir}) and ridge plots (appendix \ref{app: telescope limitations} and \ref{app 1}) and statistically confirmed through Table \ref{tab: KS test diff morphologies}.

\begin{table*}
\centering
\caption{P-values for the two sample KS test conducted on the VC and MR samples for each property and morphology.}
\label{tab: KS test VC vs MR}
\begin{tabular}{llllllll}
\hline
Cloud & M    & $\Sigma$ & R  & $\sigma_\varv$ & $AR$ & $\alpha_\mathrm{vir}$ & Length  \\ \hline
Total  & 7.7e-02 & 1.1e-01         & 1.1e-01 & 7.5e-02             & 2.7e-01      & 1.0e+00          & 8.8e-02 \\
Ring-like   & 4.5e-03 & 3.8e-05         & 3.5e-02 & 4.4e-02             & 6.8e-03      & 1.5e-01          & 3.9e-02 \\
Elongated   & 1.0e+00 & 1.0e+00         & 1.0e+00 & 1.0e+00             & 1.0e+00      & 1.0e+00          & 1.0e+00 \\
Concentrated   & 9.0e-01 & 1.8e-03         & 2.9e-01 & 6.3e-01             & 3.5e-03      & 1.0e+00          & 4.2e-02 \\
Clumpy & 1.0e+00 & 1.0e+00         & 1.0e+00 & 1.0e+00             & 1.0e+00      & 1.0e+00          & 1.0e+00 \\ \hline
\end{tabular}
\tablefoot{ The rows represent the cloud morphologies whereas the columns represent the cloud properties. The symbols follow Table \ref{Table: Properties}.}
\end{table*}

\begin{table*}
\centering
\caption{P-values for the two sample KS test conducted on different morphologies for each cloud property (VC sample).}
\label{tab: KS test diff morphologies}
\begin{tabular}{llllllll}
\hline
Cloud & M    & $\Sigma$ & R  & $\sigma_\varv$ & $AR$ & $\alpha_\mathrm{vir}$ & Length   \\ \hline
Tot-Ring          & 5.9e-15 & 3.3e-02         & 5.9e-15 & 5.9e-15             & 5.9e-15      & 5.3e-10          & 5.9e-15  \\
Tot-Elon          & 2.4e-04 & 2.9e-15         & 1.1e-01 & 1.2e-03             & 8.5e-09      & 1.2e-03          & 4.6e-01  \\
Tot-Conc          & 1.6e-07 & 1.8e-15         & 1.8e-15 & 4.1e-10             & 1.8e-15      & 1.8e-15          & 1.8e-15  \\
Tot-Clumpy        & 5.4e-10 & 8.8e-29         & 5.7e-06 & 3.3e-05             & 2.1e-02      & 3.3e-05          & 1.7e-02  \\
Ring-Elon         & 2.1e-15 & 1.5e-13         & 2.1e-15 & 2.1e-15             & 2.1e-15      & 8.6e-06          & 2.1e-15  \\
Ring-Conc         & 6.9e-32 & 1.3e-30         & 2.7e-54 & 1.1e-43             & 1.8e-191     & 2.7e-28          & 5.0e-103 \\
Ring-Clumpy       & 1.4e-03 & 1.7e-15         & 2.1e-07 & 8.2e-07             & 1.7e-15      & 1.1e-14          & 1.7e-15  \\
Elon-Conc         & 5.6e-03 & 9.6e-90         & 3.5e-12 & 1.3e-04             & 1.7e-141     & 1.6e-24          & 6.8e-50  \\
Elon-Clumpy       & 3.0e-16 & 4.1e-59         & 3.7e-08 & 3.4e-09             & 5.8e-07      & 6.5e-09          & 1.9e-02  \\
Conc-Clumpy       & 1.2e-15 & 1.4e-01         & 1.2e-15 & 5.3e-15             & 1.2e-15      & 9.8e-03          & 1.2e-15  \\ \hline
\end{tabular}
\tablefoot{The rows represent the two morphologies for which the KS test was conducted. Tot, Ring, Elon, Conc and Clumpy refer to the total, ring-like, elongated,  concentrated and clumpy clouds respectively. The columns represent the property for which the p-value is obtained. The symbols follow Table \ref{Table: Properties}.}
\end{table*}
\FloatBarrier

\section{Testing the observational biases affecting cloud properties and morphological trends}\label{app: telescope limitations} 

\begin{figure*}[ht]
    \centering
    \includegraphics[width = 1\textwidth, keepaspectratio]{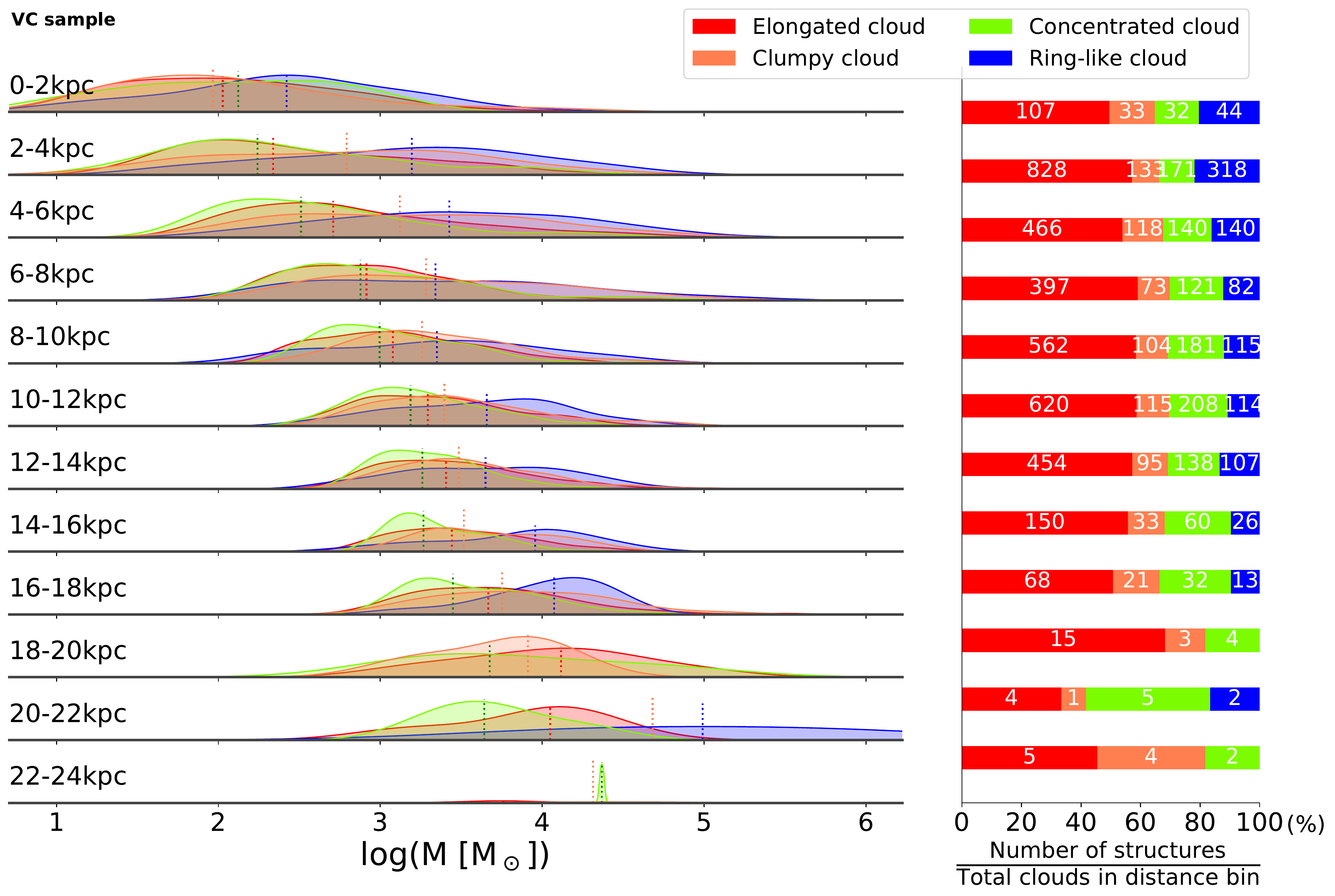}
    \caption{Mass ridge plot for the VC sample with $R_\mathrm{d}$ bins.}
    \label{fig: SCIENCE mass ridge distance}
\end{figure*}

The sensitivity and resolution limitations of the telescope may result in incorrect classification of cloud morphology at large distances ($R_\mathrm{d}$). We discuss the effects of these technical limitations in details in this section using the ridge plots for heliocentric distance ($R_\mathrm{d}$). In general, ridge plots are a visualisation tool that show the distribution of a numerical value for various groups. In our case, they can be considered violin plots for cloud properties in separate distance bins. 
The clouds are separated into 9 distance bins of width 2 kpc each, covering 0--18 kpc, based on their proximity from the Sun. Fig. \ref{fig: SCIENCE mass ridge distance}  shows the number of structures in each distance bin from 0--24 kpc.

\begin{figure*}
    \centering
    \includegraphics[width = 1\textwidth, keepaspectratio]{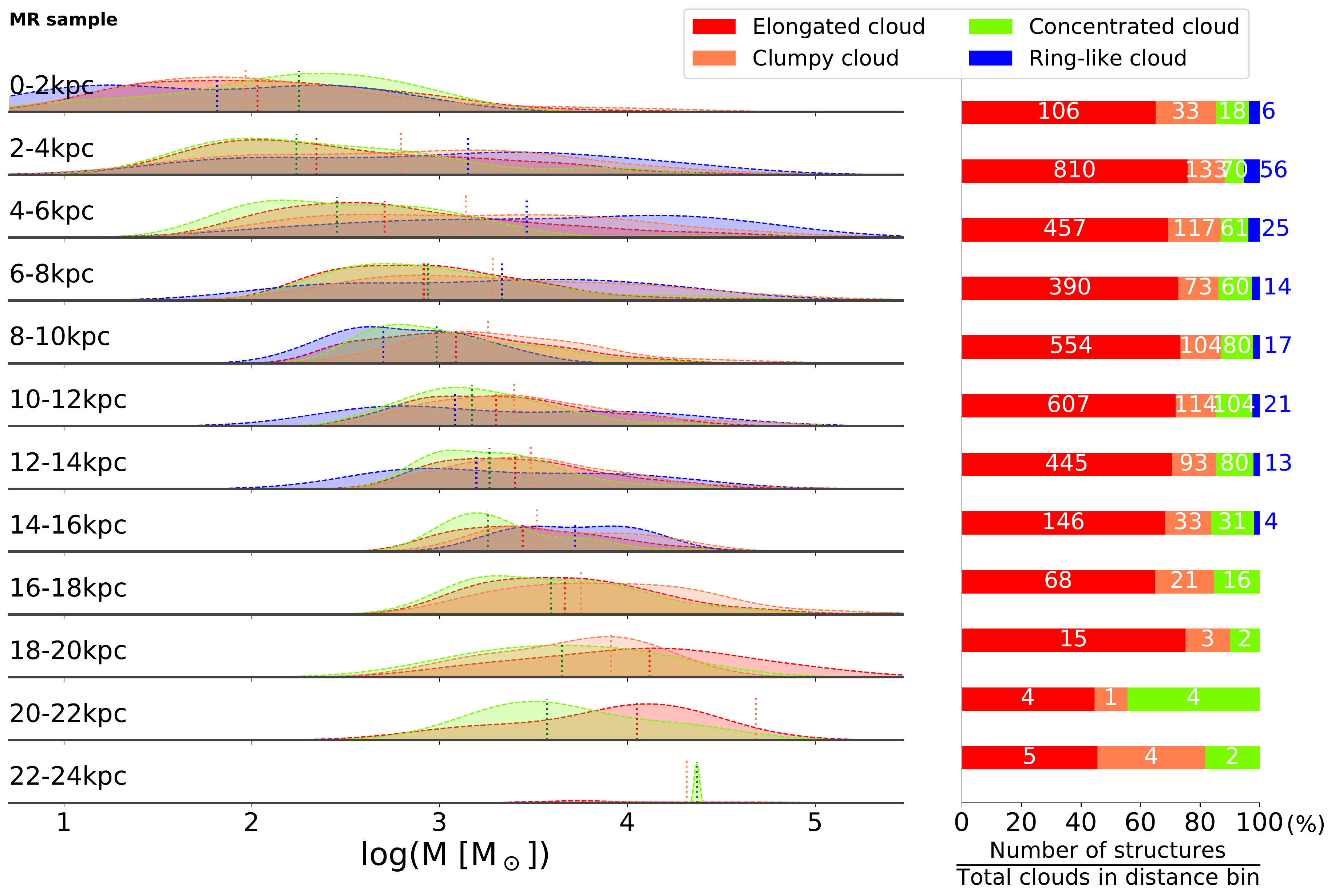}
    \caption{Mass ridge plot for the MR sample with $R_\mathrm{d}$ bins.}
    \label{fig: morph mass ridge distance}
\end{figure*}

\begin{figure*}
    \centering
    \begin{minipage}{\textwidth}
    \includegraphics[width = .5\textwidth, keepaspectratio]{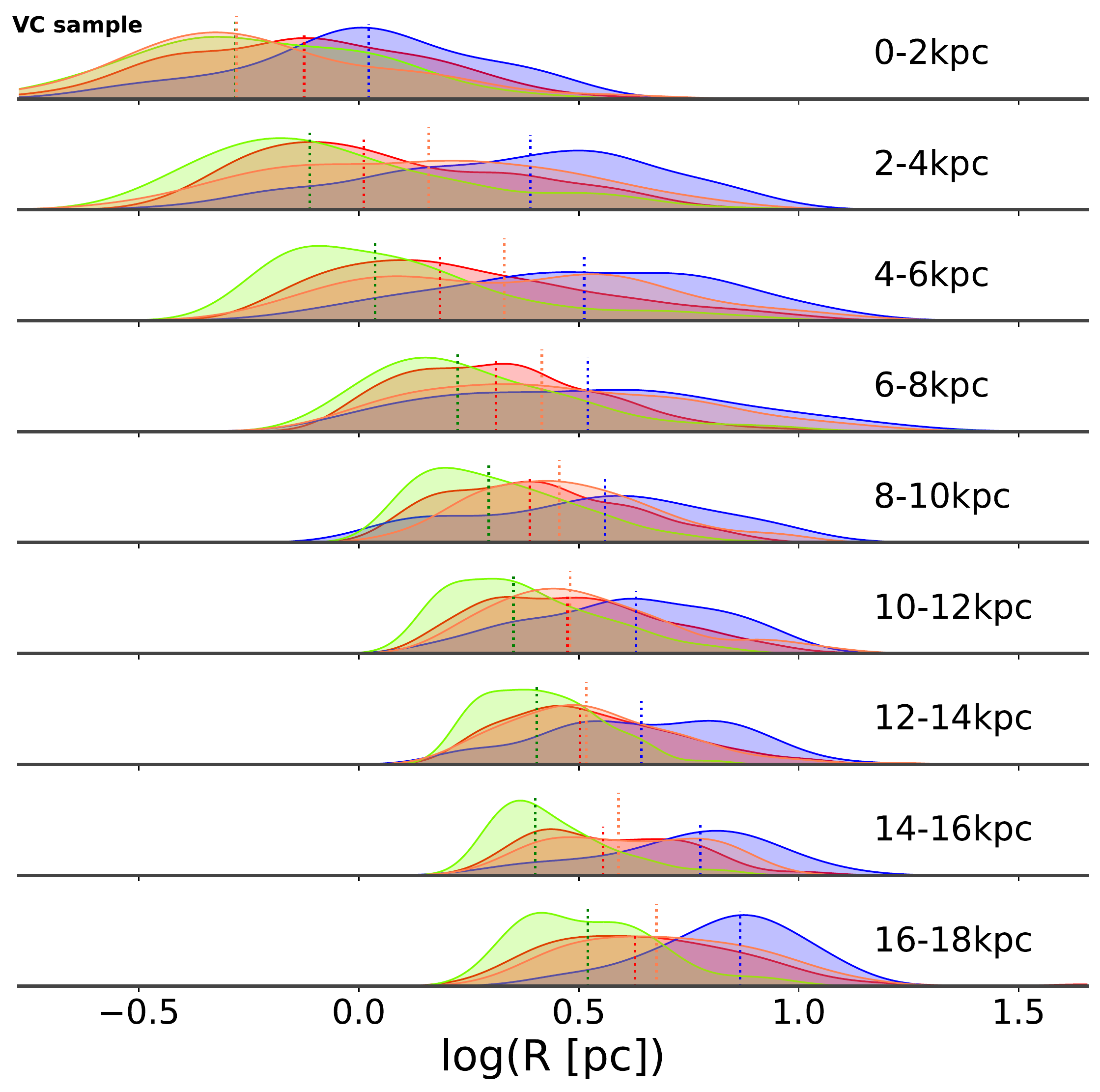}
    \includegraphics[width = .5\textwidth, keepaspectratio]{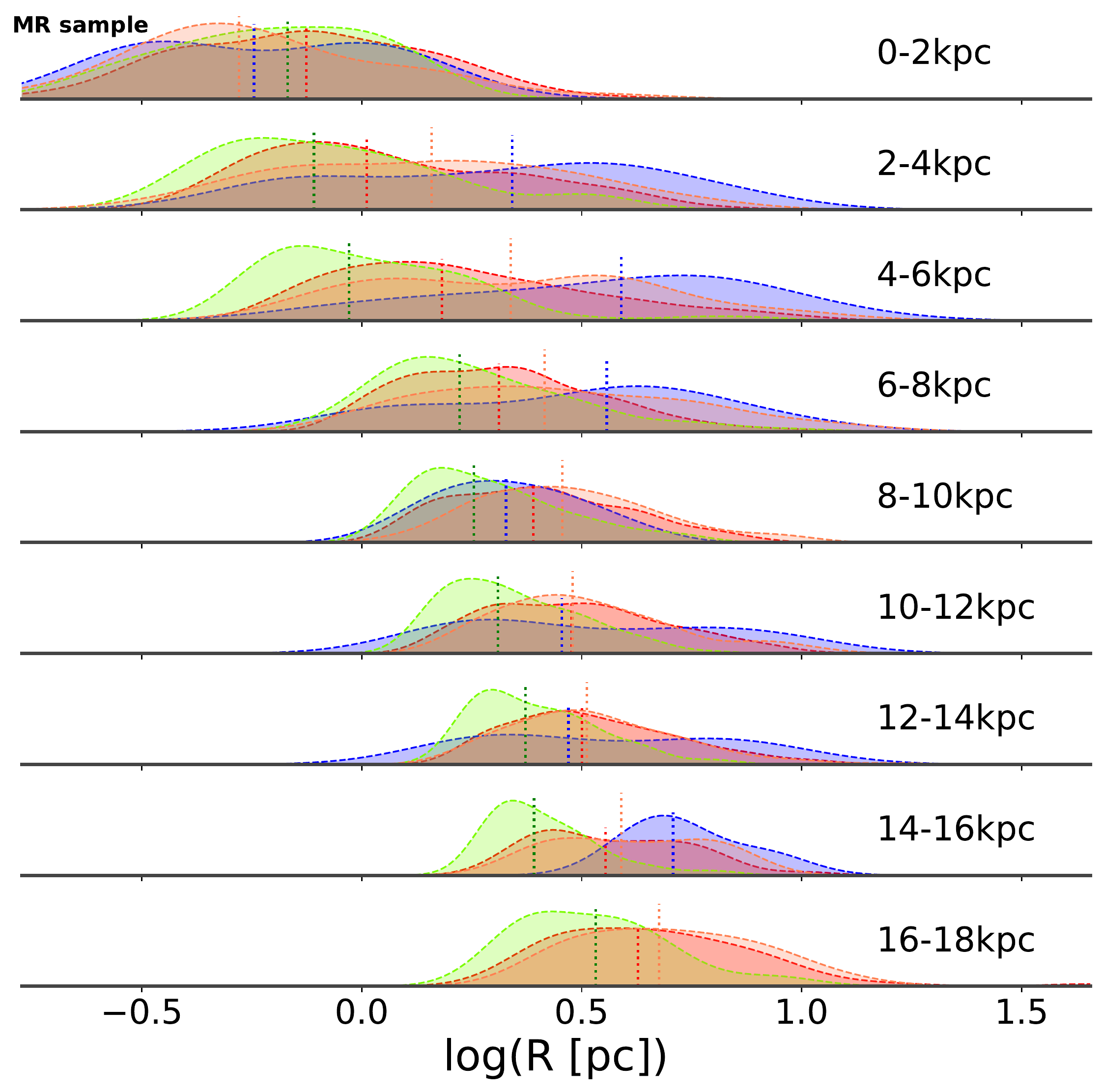}
    \caption{Radius ridge plots with $R_\mathrm{d}$ bins. \textit{Left} (solid): VC sample, \textit{Right} (dashed): MR sample.}
    \label{fig: radius ridge plot distance}
    \end{minipage}\hfill
\end{figure*}

\begin{figure*}
    \centering
    \begin{minipage}{\textwidth}
    \includegraphics[width = .5\textwidth, keepaspectratio]{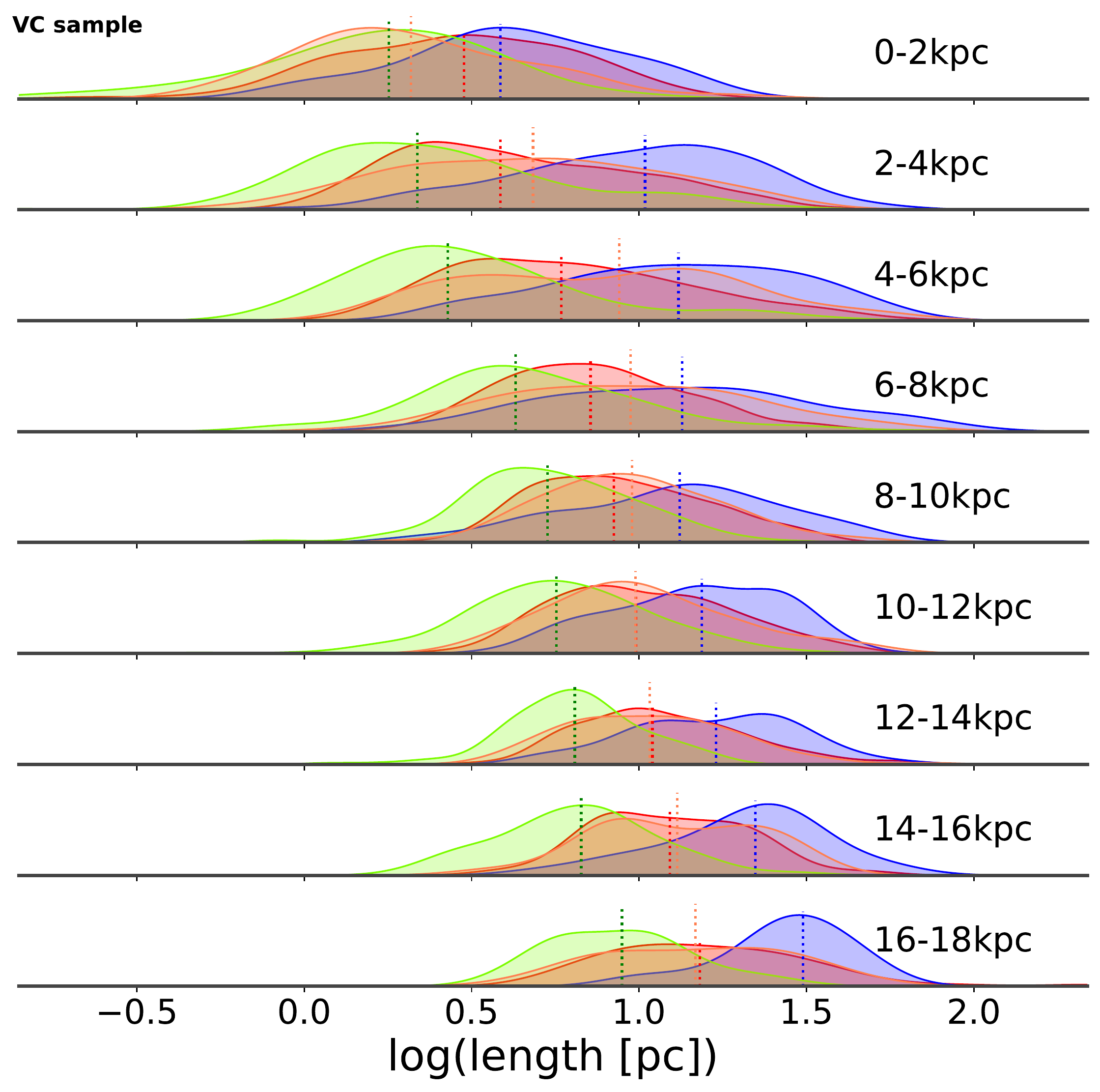}
    \includegraphics[width = .5\textwidth, keepaspectratio]{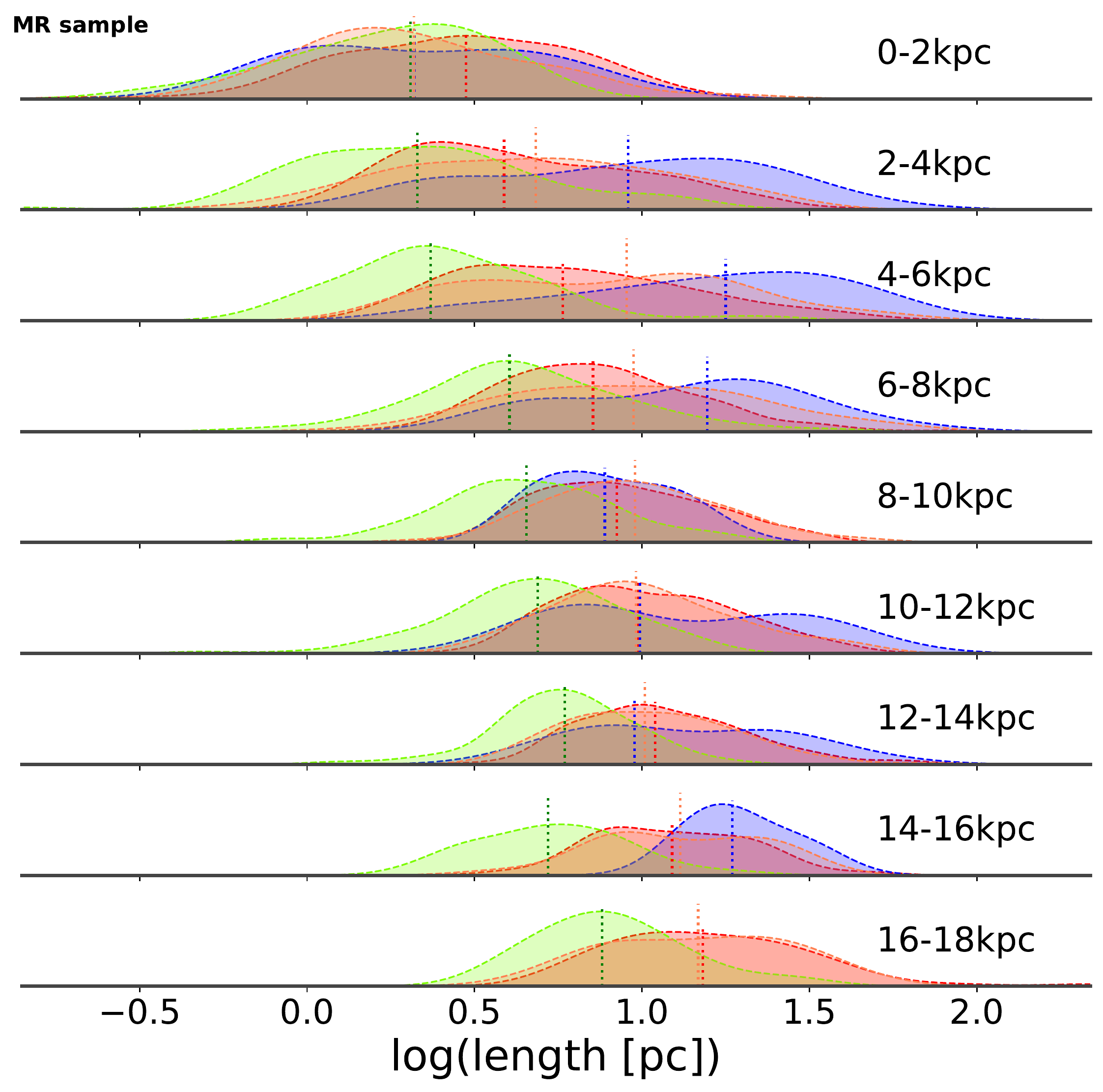}
    \caption{Length ridge plots with $R_\mathrm{d}$ bins. \textit{Left} (solid): VC sample, \textit{Right} (dashed): MR sample.}
    \label{fig: length ridge plot distance}
    \end{minipage}\hfill
\end{figure*}

The cloud properties that are most affected by the technical limitations are mass and size (radius and length). They show visible changes in the median as well as the distribution range across the distance bins.
Fig. \ref{fig: morph mass ridge distance} shows that the lower limit on mass increases with $R_\mathrm{d}$. At a smaller $R_\mathrm{d}$, low-mass clouds can be detected due to the high resolution and sensitivity of the telescope, but as $R_\mathrm{d}$ increases, the low mass clouds are perceived as background noise. 
Similarly, we also find that the lower limit on radius (Fig. \ref{fig: radius ridge plot distance}) and length (Fig. \ref{fig: length ridge plot distance}) increases with distance, following the same principle as mass distributions.

\begin{figure*}
    \centering
    \begin{minipage}{\textwidth}
    \includegraphics[width = .5\textwidth, keepaspectratio]{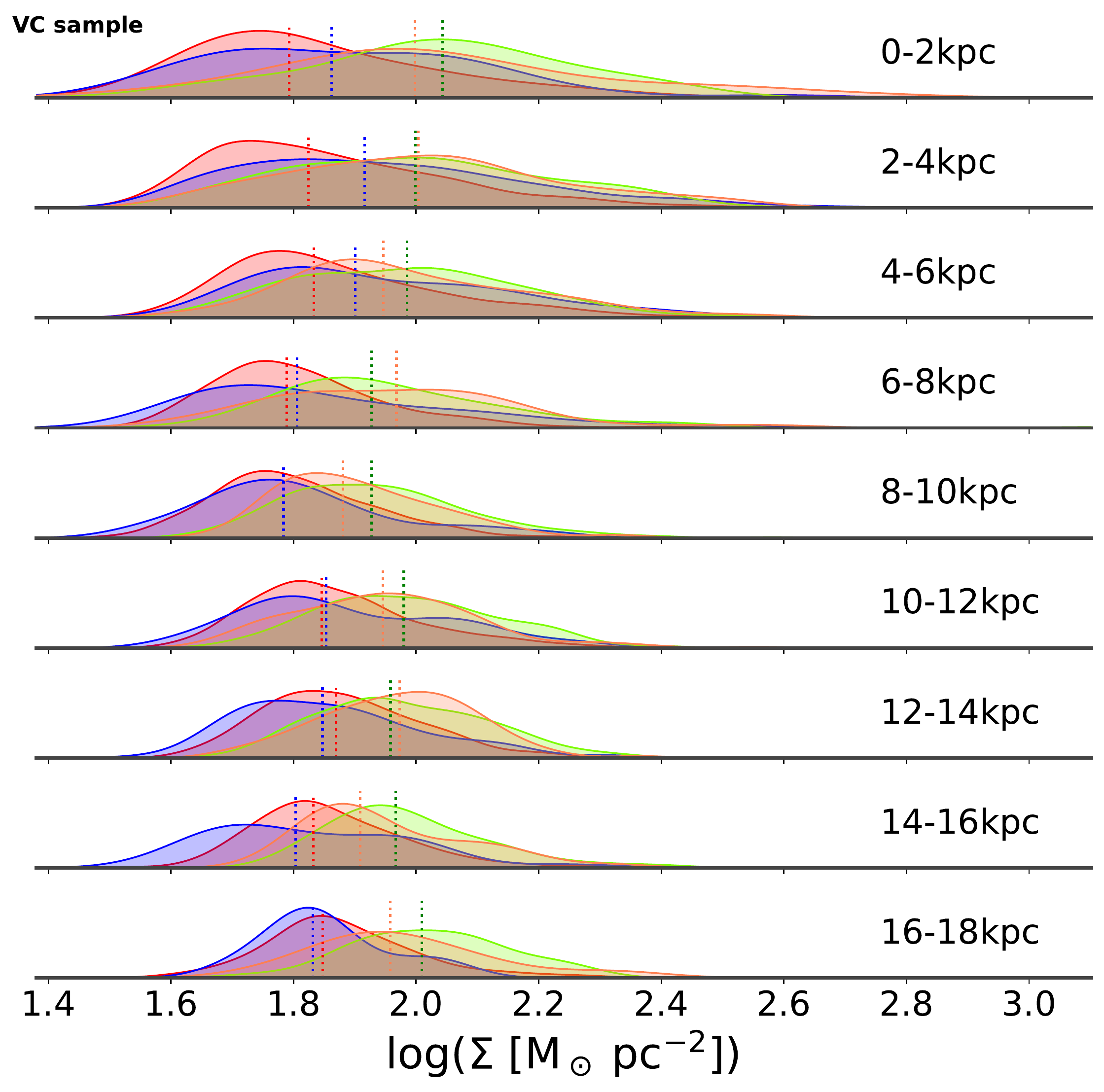}
    \includegraphics[width = .5\textwidth, keepaspectratio]{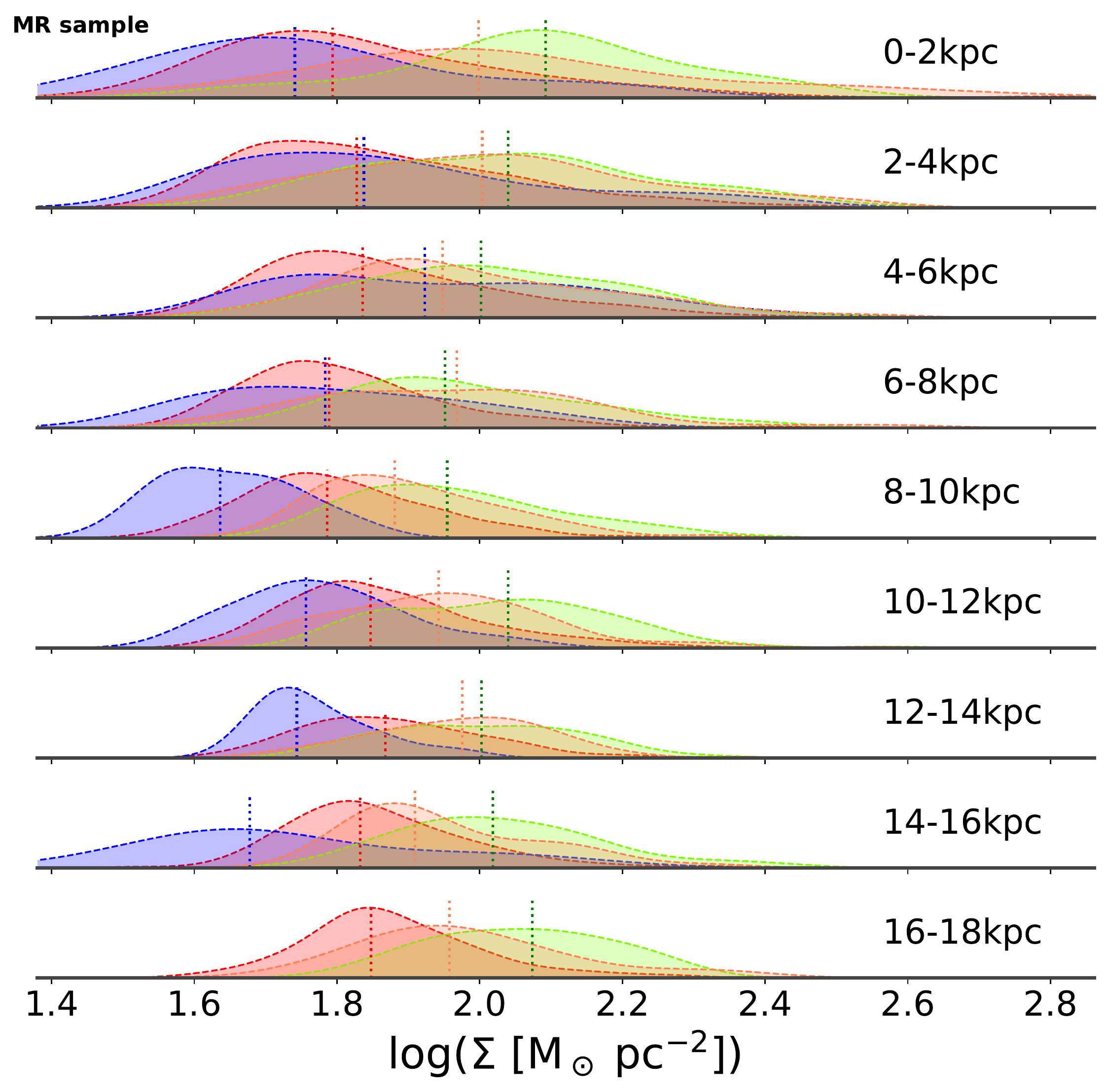}
    \caption{Surface density ridge plots with $R_\mathrm{d}$ bins. \textit{Left} (solid): VC sample, \textit{Right} (dashed): MR sample.}
    \label{fig: surface density ridge plot distance}
    \end{minipage}\hfill
\end{figure*}

\begin{figure*}
    \centering
    \begin{minipage}{\textwidth}
    \includegraphics[width = .5\textwidth, keepaspectratio]{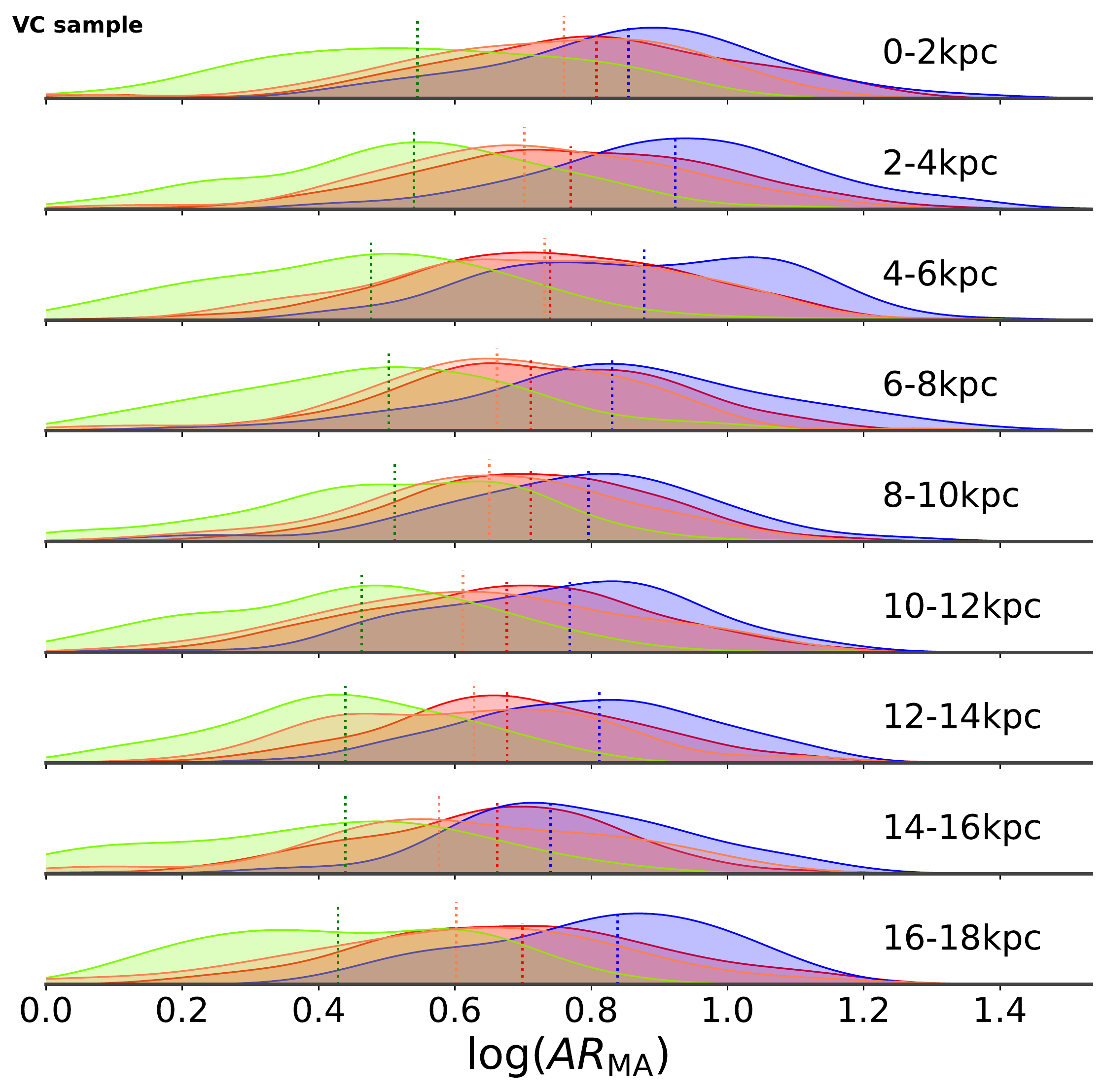}
    \includegraphics[width = .5\textwidth, keepaspectratio]{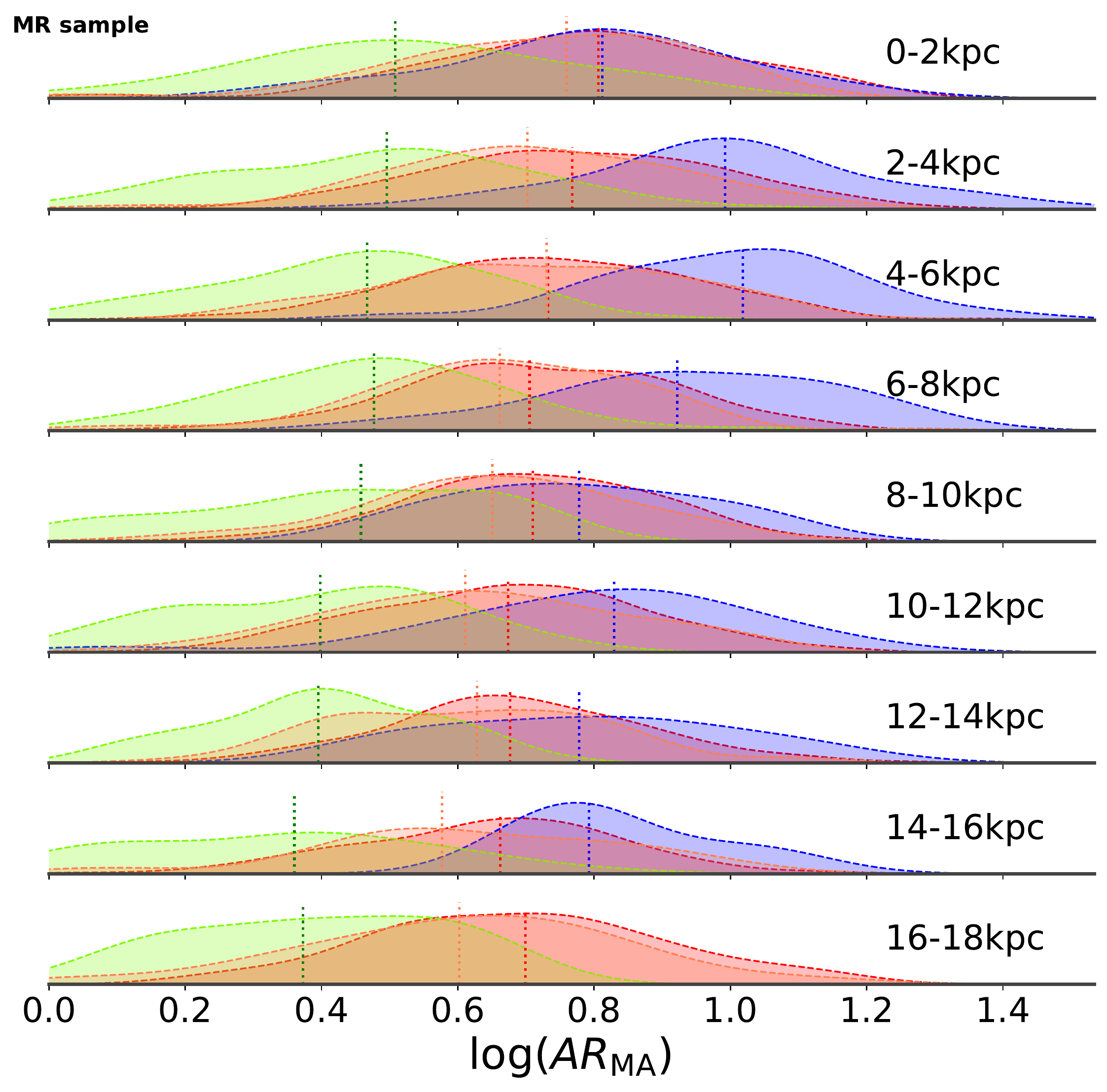}
    \caption{Aspect ratio ridge plots with $R_\mathrm{d}$ bins. \textit{Left} (solid): VC sample, \textit{Right} (dashed): MR sample.}
    \label{fig: aspect ratio ridge plot distance}
    \end{minipage}\hfill
\end{figure*}

\begin{figure*}
    \centering
    \begin{minipage}{\textwidth}
    \includegraphics[width = .5\textwidth, keepaspectratio]{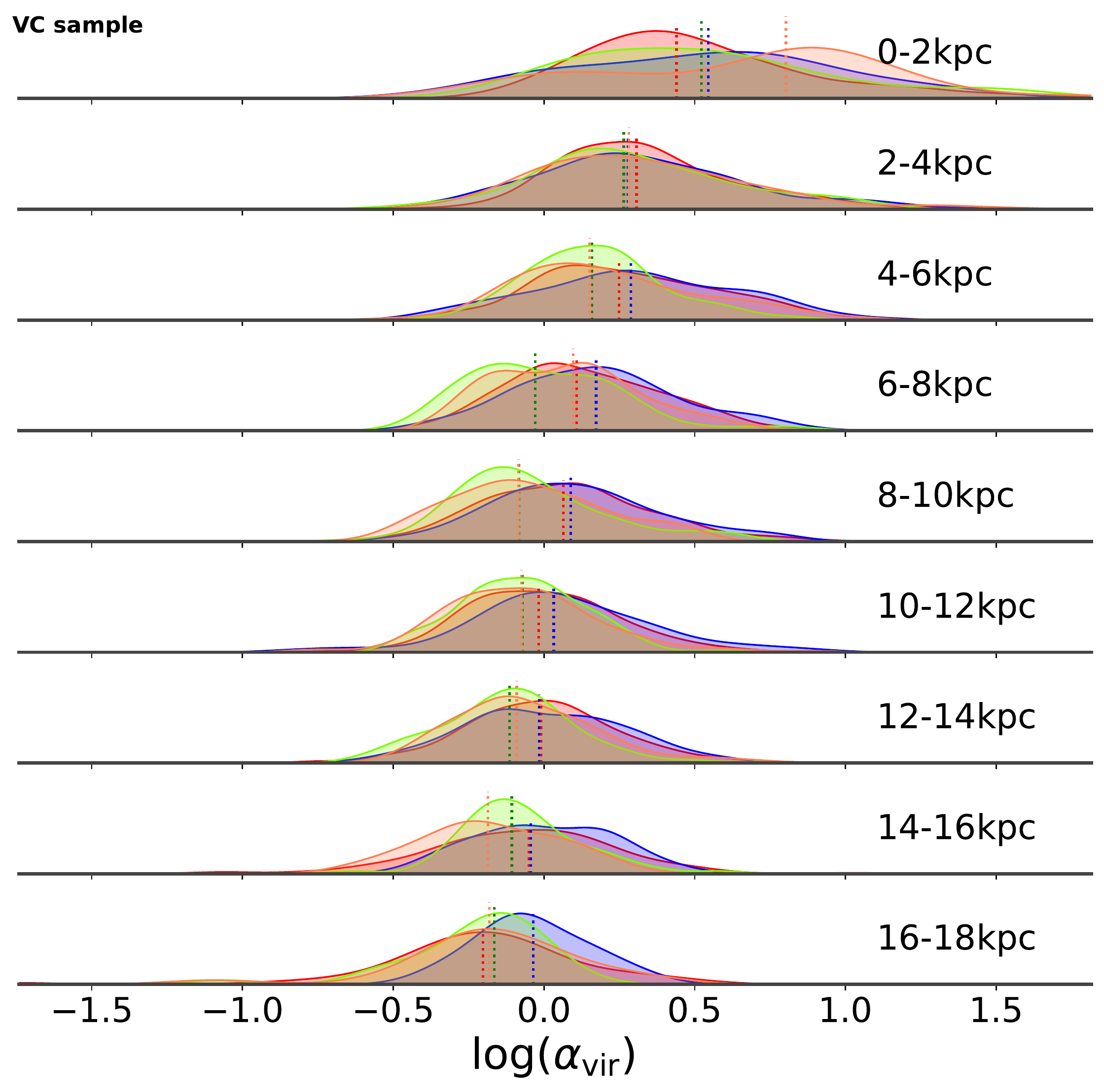}
    \includegraphics[width = .5\textwidth, keepaspectratio]{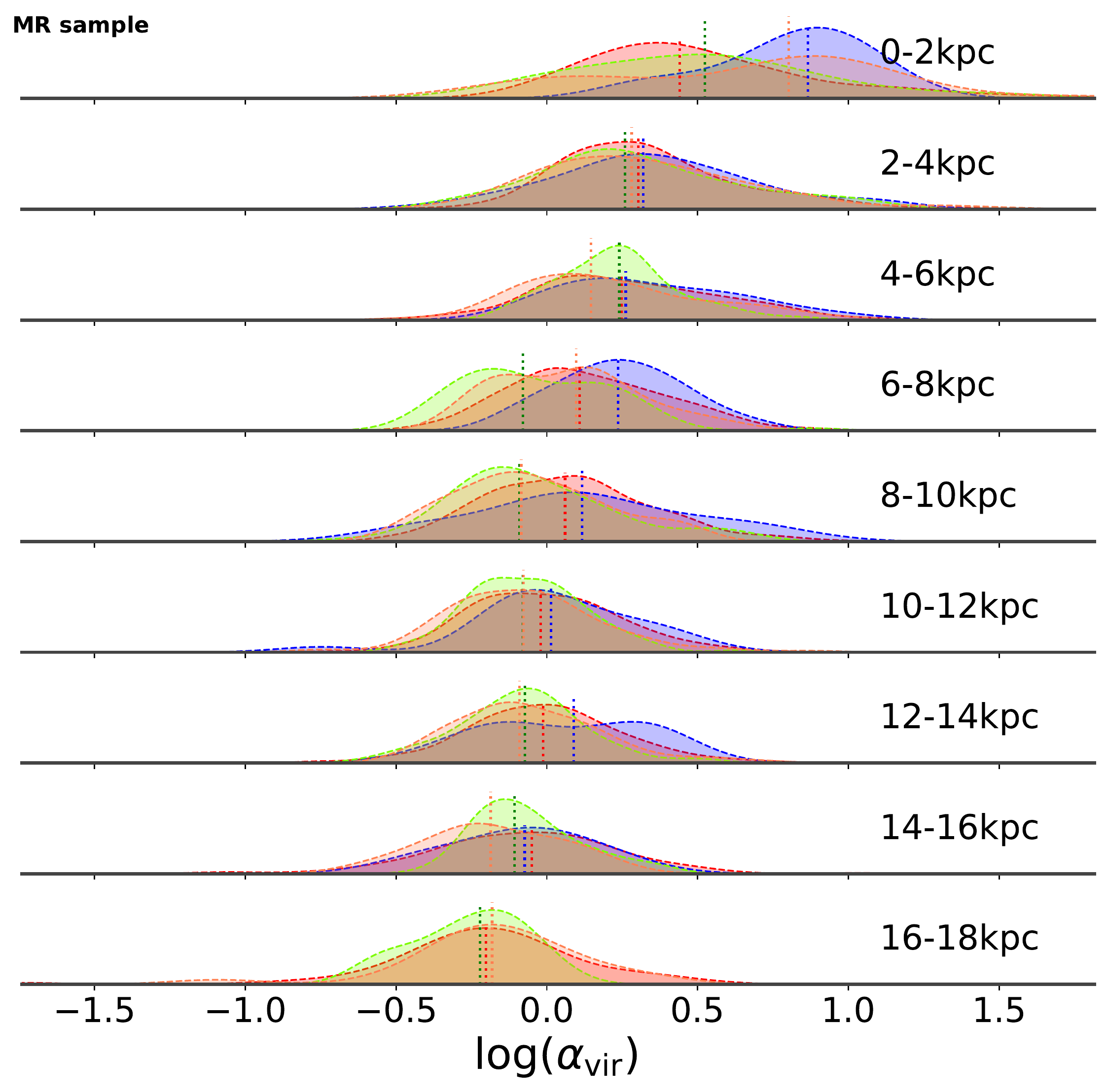}
    \caption{Virial parameter ridge plots with $R_\mathrm{d}$ bins. \textit{Left} (solid): VC sample, \textit{Right} (dashed): MR sample.}
    \label{fig: virial parameter ridge plot distance}
    \end{minipage}\hfill
\end{figure*}

The surface density (Fig. \ref{fig: surface density ridge plot distance}), aspect ratio (Fig. \ref{fig: aspect ratio ridge plot distance}) and velocity dispersion are not strongly affected by the technical limitations.
The distributions for virial parameter (Fig. \ref{fig: virial parameter ridge plot distance}) show the most overlap compared to those of other properties. The decrease in the virial parameter with increasing $R_\mathrm{d}$ can be attributed to observational biases at play (discussed in appendix C of DC21).

We study the cloud properties in two-dimensional position-position space, and it could lead to the projection effects affecting our analysis. For example, our cloud length and width do not represent the original dimensions of the molecular clouds, but rather a projected length and width. Similarly, the cloud morphology is also influenced due to the positioning of clouds with respect to the telescope's line of sight. However, given the large size of our data sample, the general trends in the results are unlikely to be affected and thus we ignore the projection effects in this work.

It is interesting to note that regardless of the changes in the properties with distance, the properties show similar trends across all the distance bins for the different morphologies validating our conclusions on the trends found for the full samples (see Sec. \ref{sec: results}). 

\begin{figure*}
    \centering
    \begin{minipage}{\textwidth}
    \includegraphics[width = .5\textwidth, keepaspectratio]{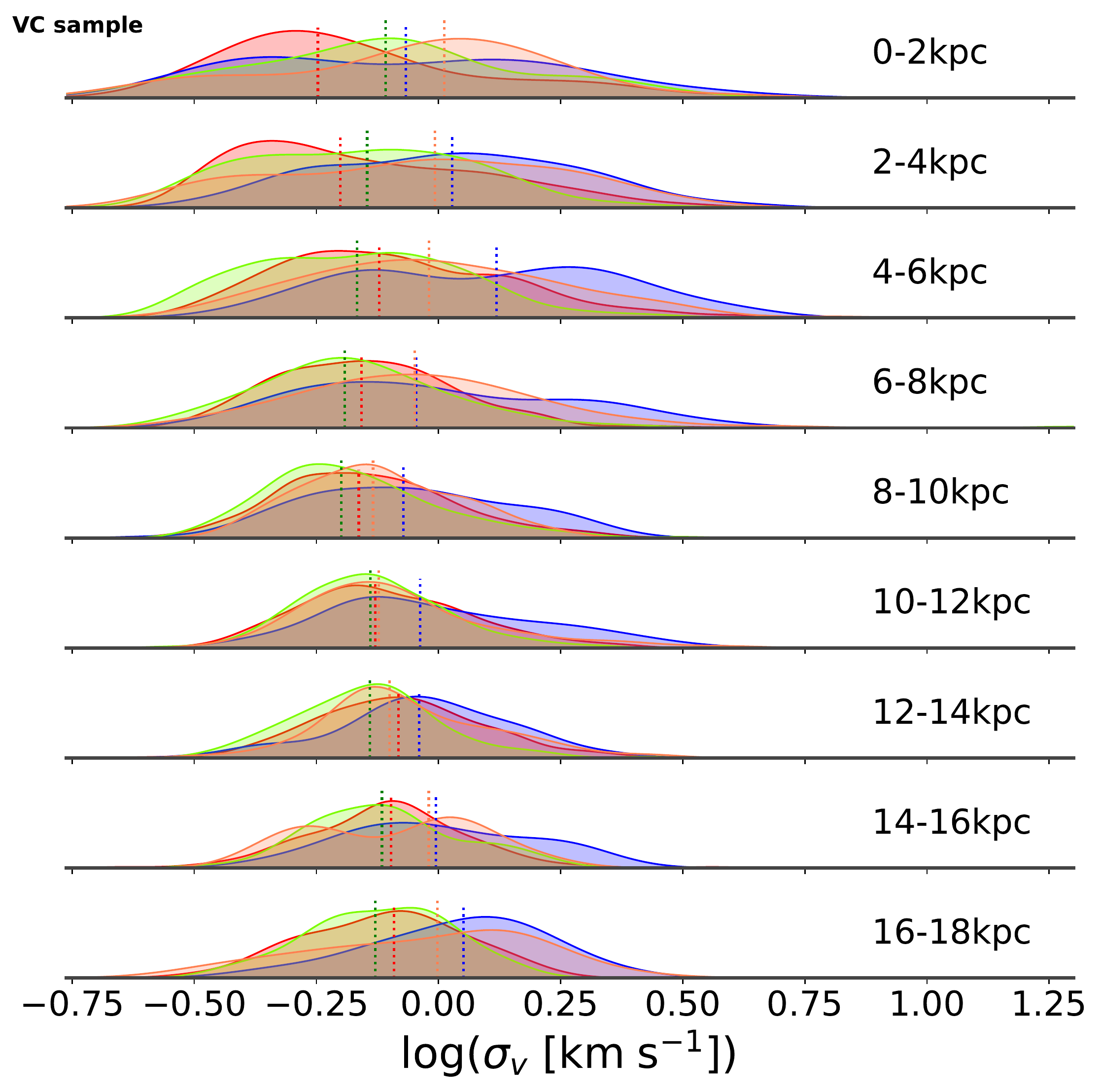}
    \includegraphics[width = .5\textwidth, keepaspectratio]{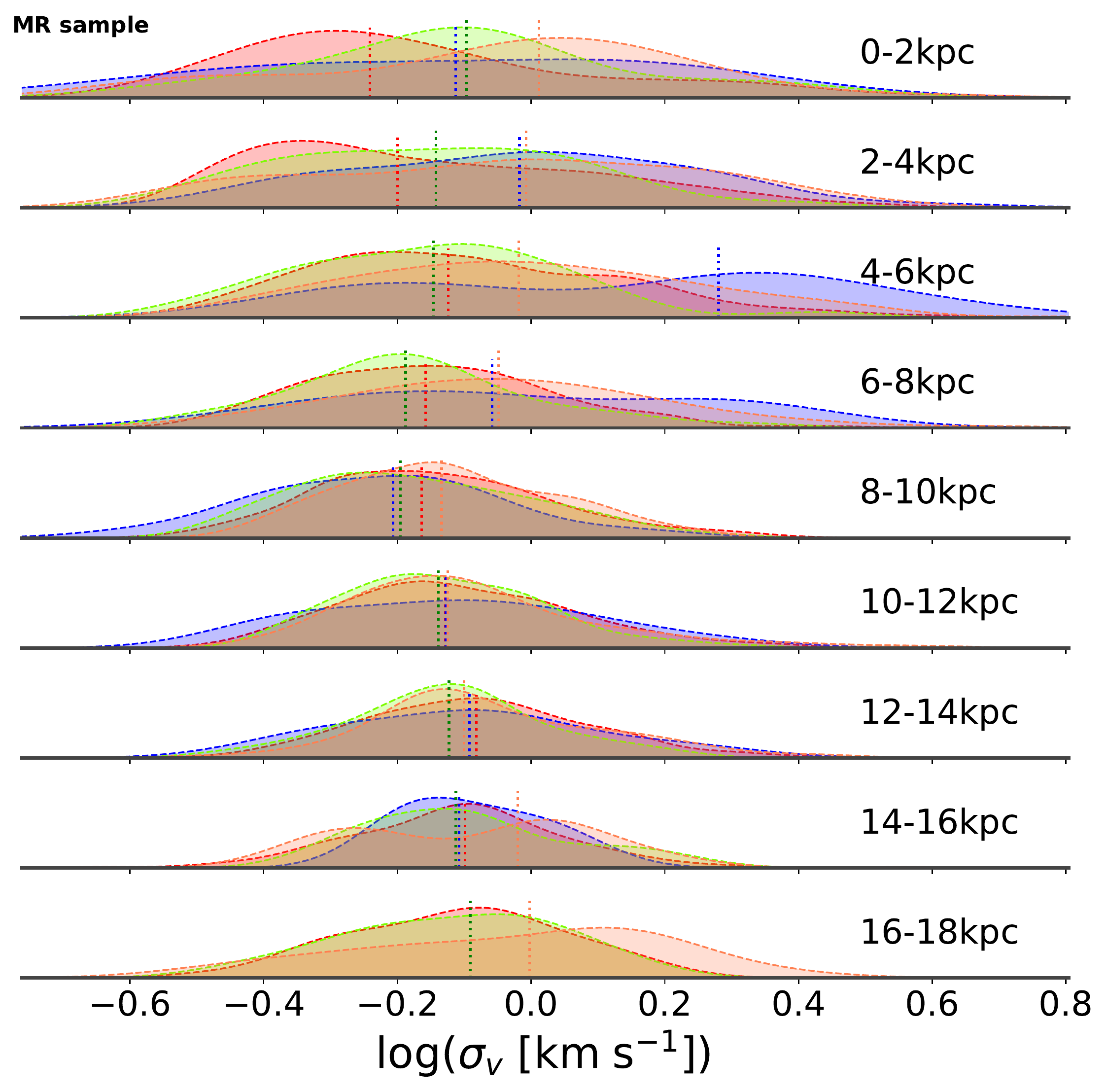}
    \caption{Velocity dispersion ridge plots with $R_\mathrm{d}$ bins. \textit{Left} (solid): VC sample, \textit{Right} (dashed): MR sample.}
    \label{fig: velocity dispersion ridge plot distance}
    \end{minipage}\hfill
\end{figure*}

\FloatBarrier

\section{Dependence of cloud properties on Galactic environment}\label{app 1}

The properties of molecular clouds might show changes within different Galactic environments and based on their position in the Galaxy. We study changes in the integrated properties of molecular clouds based on their location in the Galaxy for the VC and MR samples using ridge plots. The data in each bin is plotted separately for each morphology to understand the environmental influence on each cloud properties for the different morphologies. 
The distributions of elongated clouds and clumpy clouds for the two samples are nearly identical as only a few of these clouds from the VC sample are not a part of the MR sample. The clouds are separated into distance bins based on two distance\footnote{The distances are provided as $R\_\mathrm{gal}$ and $z\_\mathrm{gal\_kpc}$ in the cloud catalogue of DC21.} types. \add{The first is Galactocentric radius ($R_\mathrm{gal}$), i.e., the distance of a cloud from the centre of Milky Way. The second is the Galactic height ($z_\mathrm{gal}$), i.e., vertical height of a cloud from the Galactic plane.}
The cloud properties, i.e. mass, surface density, radius, velocity dispersion, aspect ratio, virial parameter and length are studied using ridge plots. The low sample size of morphologies in some bins leads to sporadic peaks in the plots and also prohibits an in-depth study.

\subsection{Galactocentric radius ($R_\mathrm{gal}$) bins}

\begin{figure*}[]
    \centering
    \includegraphics[width = \textwidth, keepaspectratio]{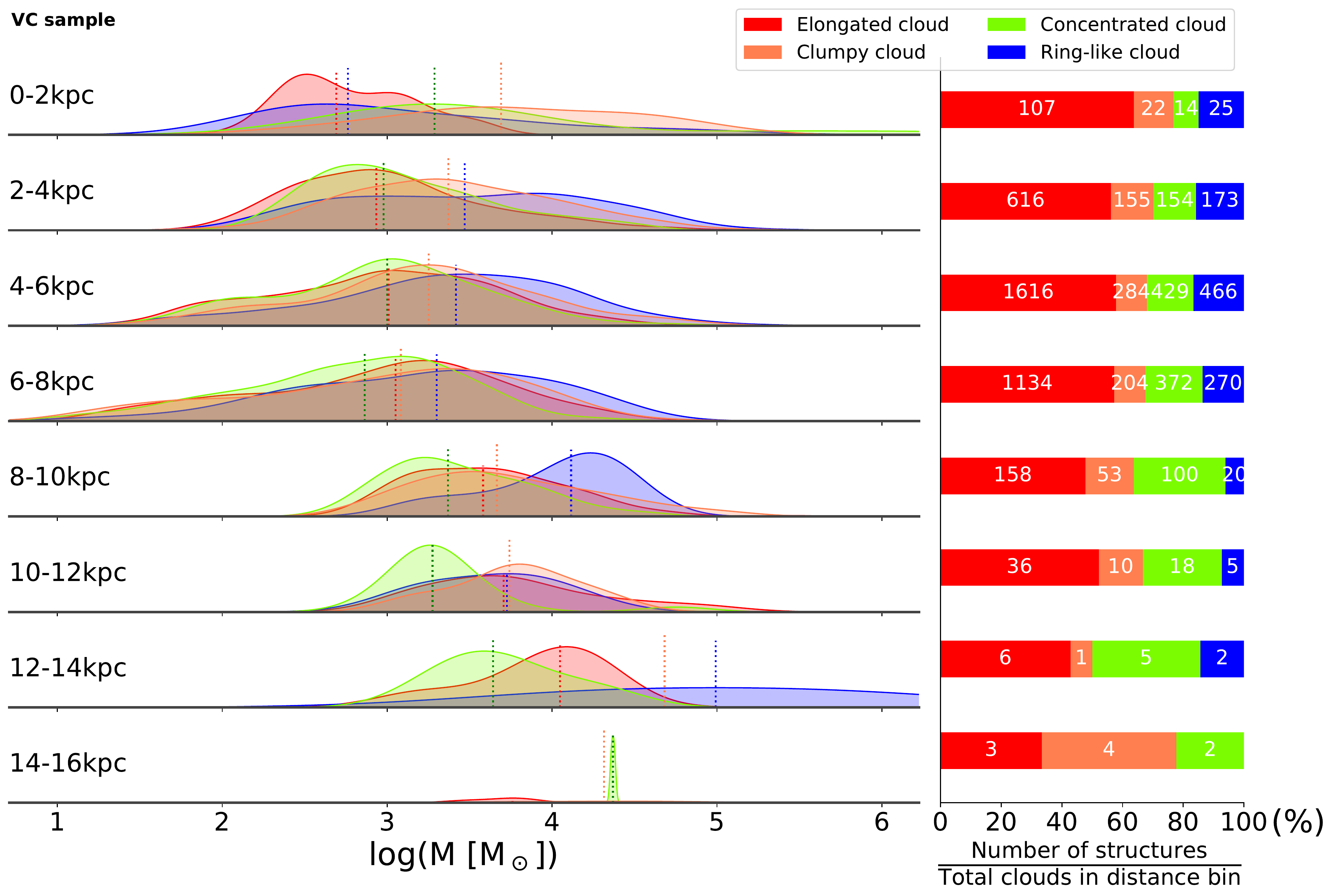}
    \caption{Mass ridge plot for the VC sample with $R_\mathrm{gal}$ bins. The vertical dashed lines represent the medians of the distributions.}
    \label{fig: SCIENCE mass ridge R bins}
\end{figure*}

\begin{figure*}[]
    \centering
    \includegraphics[width = \textwidth, keepaspectratio]{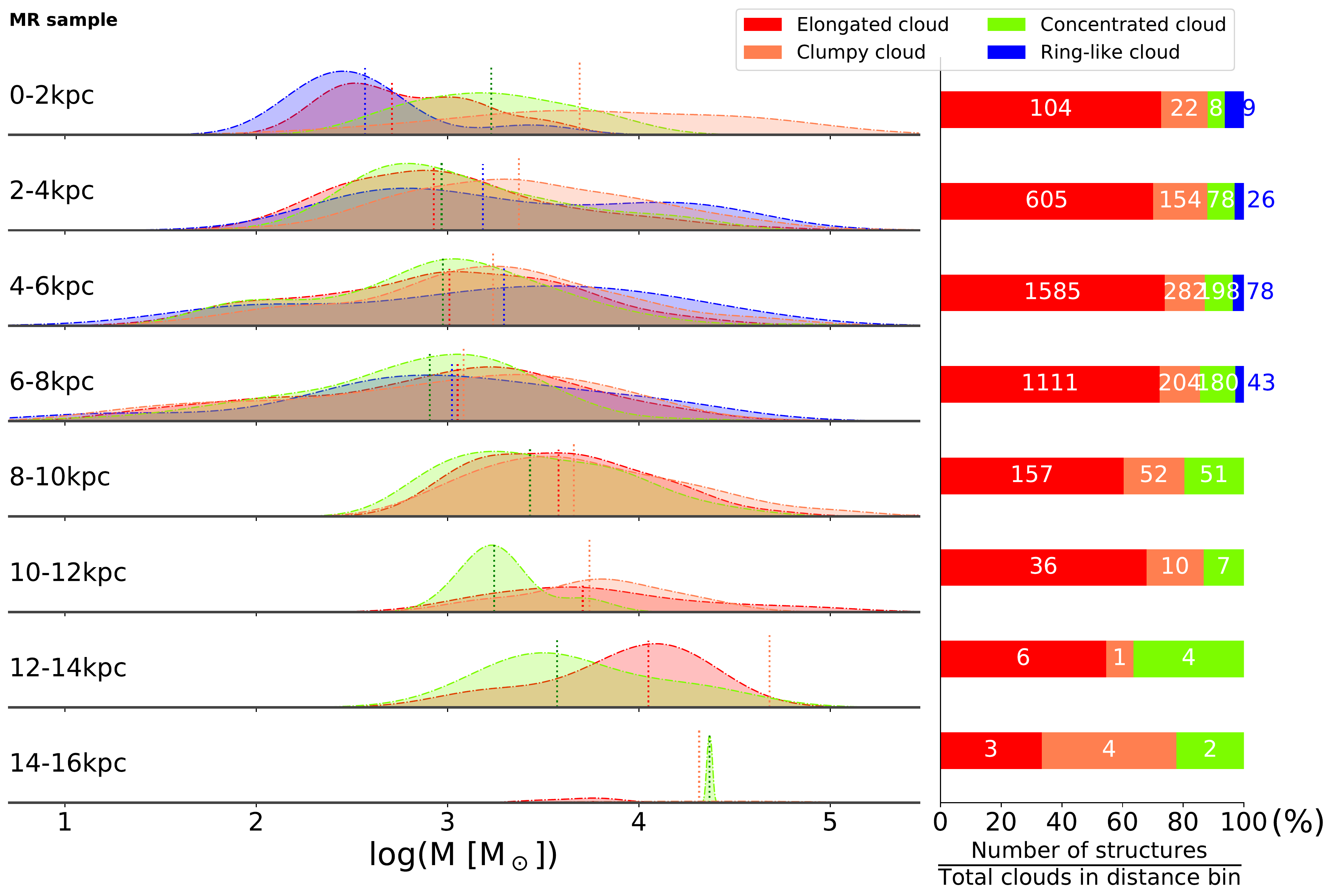}
    \caption{Mass ridge plot for the MR sample with $R_\mathrm{gal}$ bins. The vertical dashed lines represent the medians of the distributions.}
    \label{fig: morph mass ridge R bins}
\end{figure*}

This section presents the ridge plots with Galactocentric radius ($R_\mathrm{gal}$) bins. The data is divided into 8 bins of width 1 kpc from 2--10 kpc\footnote{The MR sample has ring-like clouds only upto $R_\mathrm{gal}$ = 8 kpc.}. The central part of the Galaxy ($<$ 1 kpc) is excluded as the local gas motions influence the overall properties, causing a change in the dynamics of the region. The decrease in sample size at large distance from the Galactic centre sets an upper limit of 10 kpc for $R_\mathrm{gal}$ bins (Fig. \ref{fig: SCIENCE mass ridge R bins} and \ref{fig: morph mass ridge R bins}).

\begin{figure*}[]
    \centering
    \begin{minipage}{\textwidth}
    \includegraphics[width = .5\textwidth, keepaspectratio]{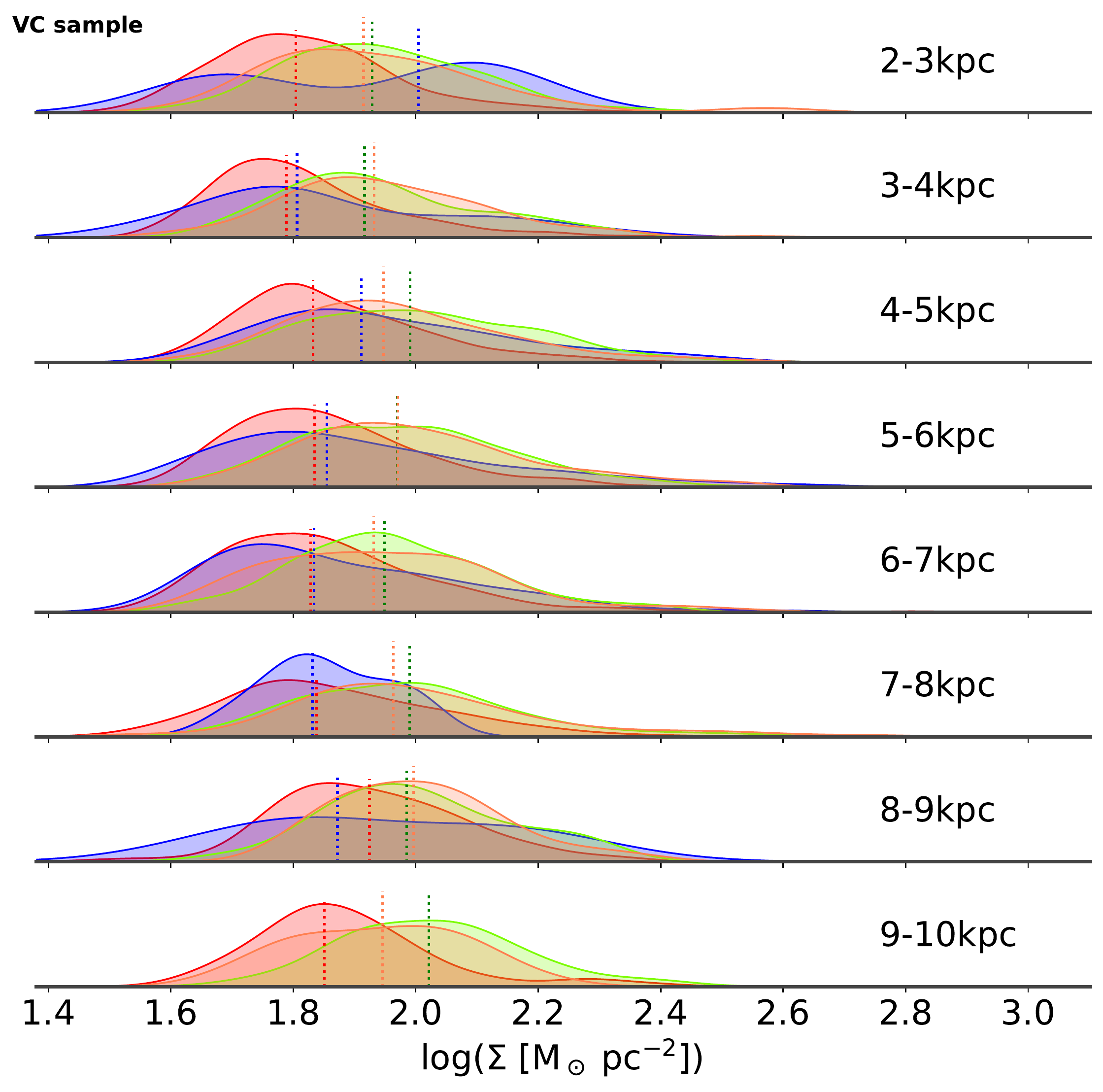}
    \includegraphics[width = .5\textwidth, keepaspectratio]{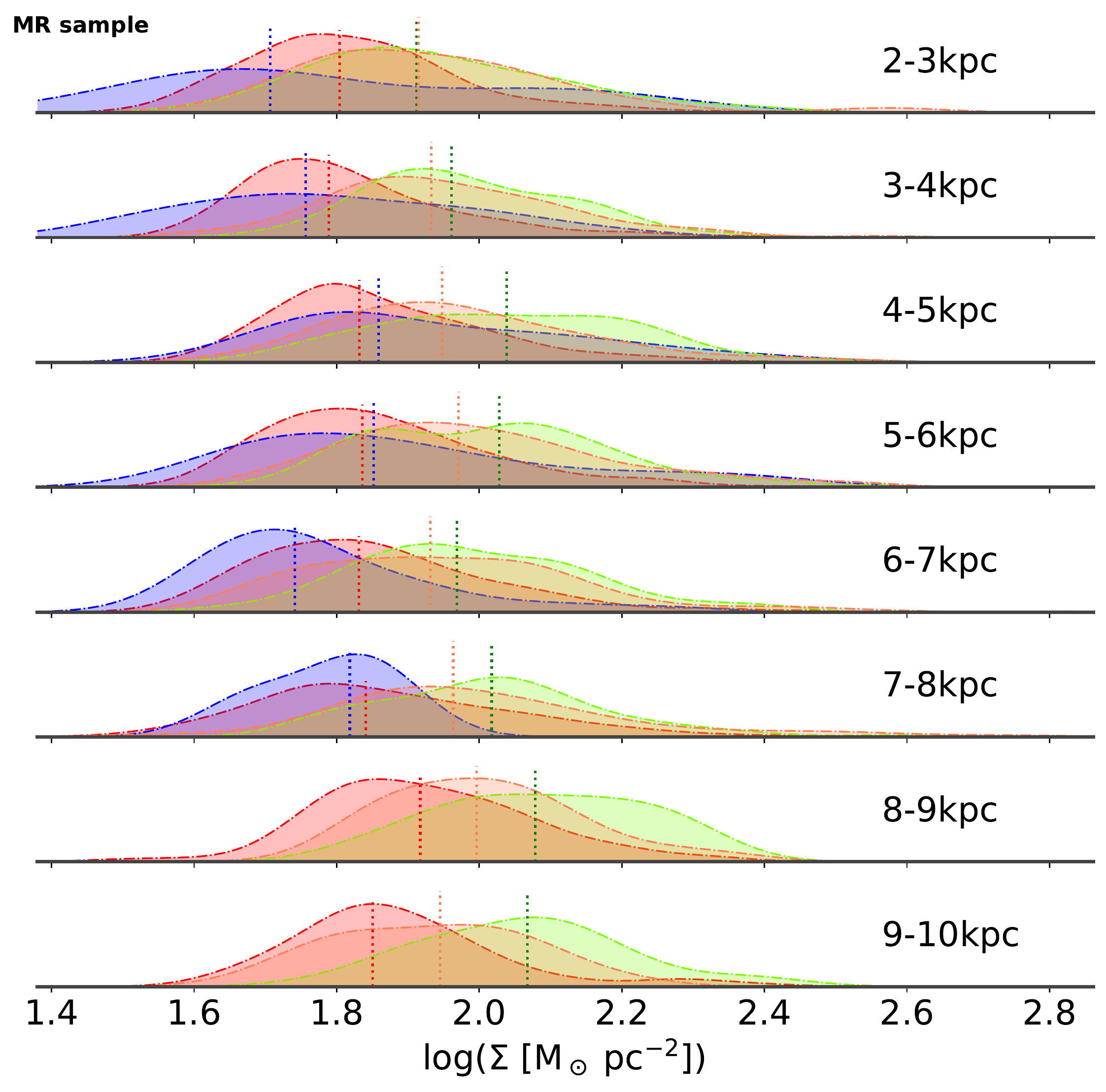}
    \caption{Surface density ridge plots with $R_\mathrm{gal}$ bins. \textit{Left} (solid): VC sample, \textit{Right} (dashed): MR sample.}
    \label{fig: surface density ridge plot R bins}
    \end{minipage}\hfill
\end{figure*}

\begin{figure*}[]
    \centering
    \begin{minipage}{\textwidth}
    \includegraphics[width = .5\textwidth, keepaspectratio]{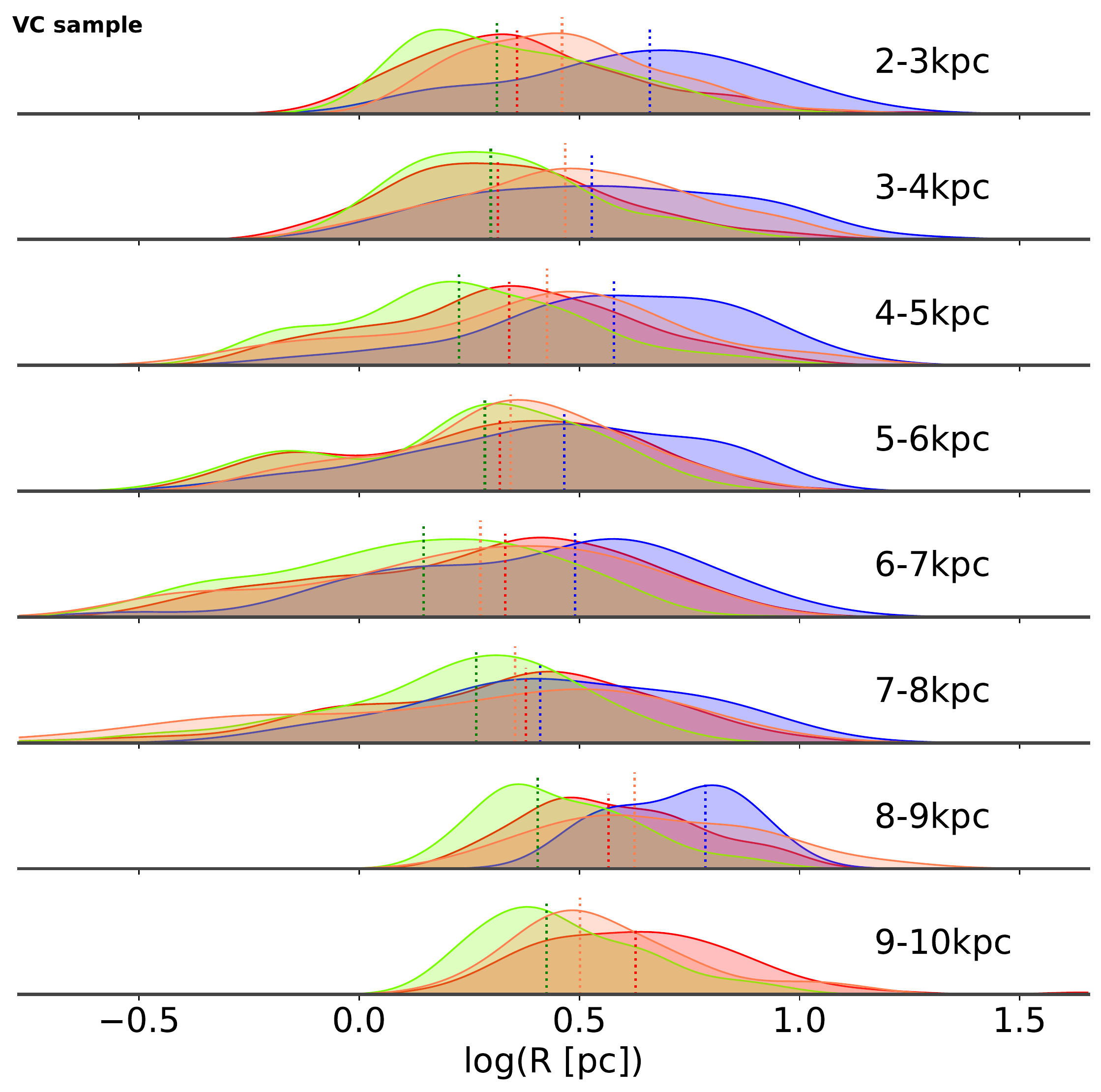}
    \includegraphics[width = .5\textwidth, keepaspectratio]{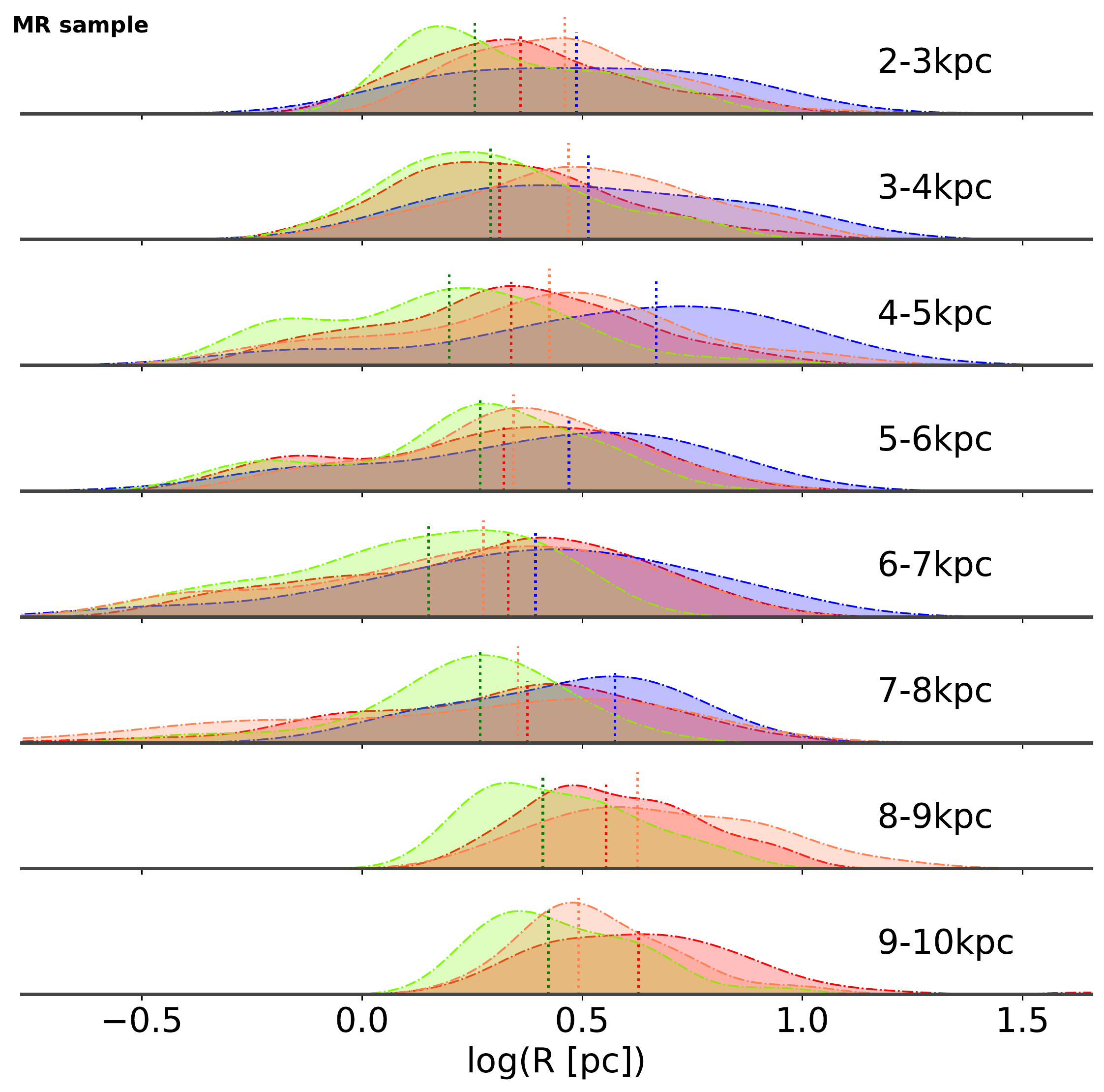}
    \caption{Radius ridge plots with $R_\mathrm{gal}$ bins. \textit{Left} (solid): VC sample, \textit{Right} (dashed): MR sample.}
    \label{fig: radius ridge plot R bins}
    \end{minipage}\hfill
\end{figure*}

\begin{figure*}
    \centering
    \begin{minipage}{\textwidth}
    \includegraphics[width = .5\textwidth, keepaspectratio]{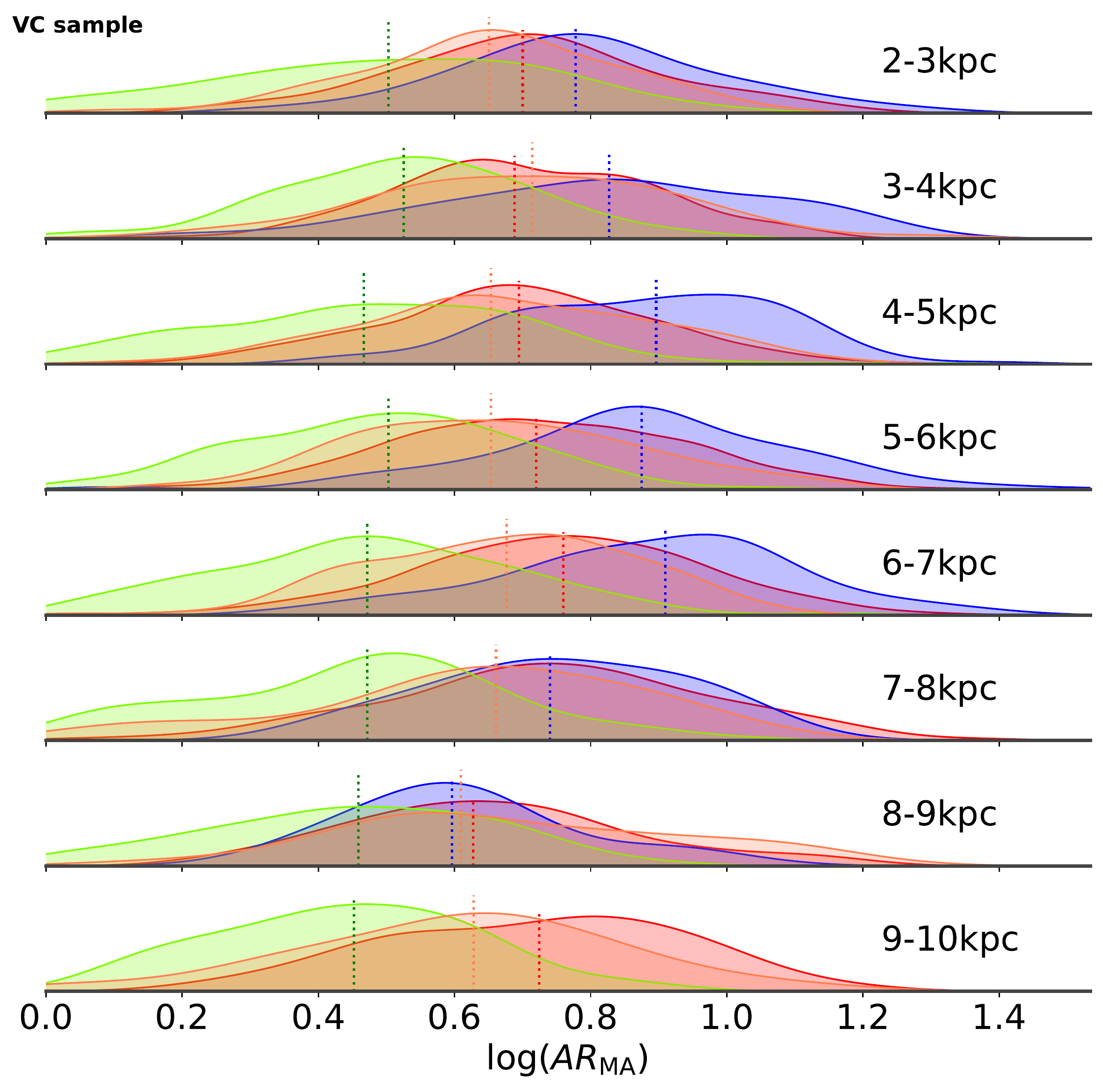}
    \includegraphics[width = .5\textwidth, keepaspectratio]{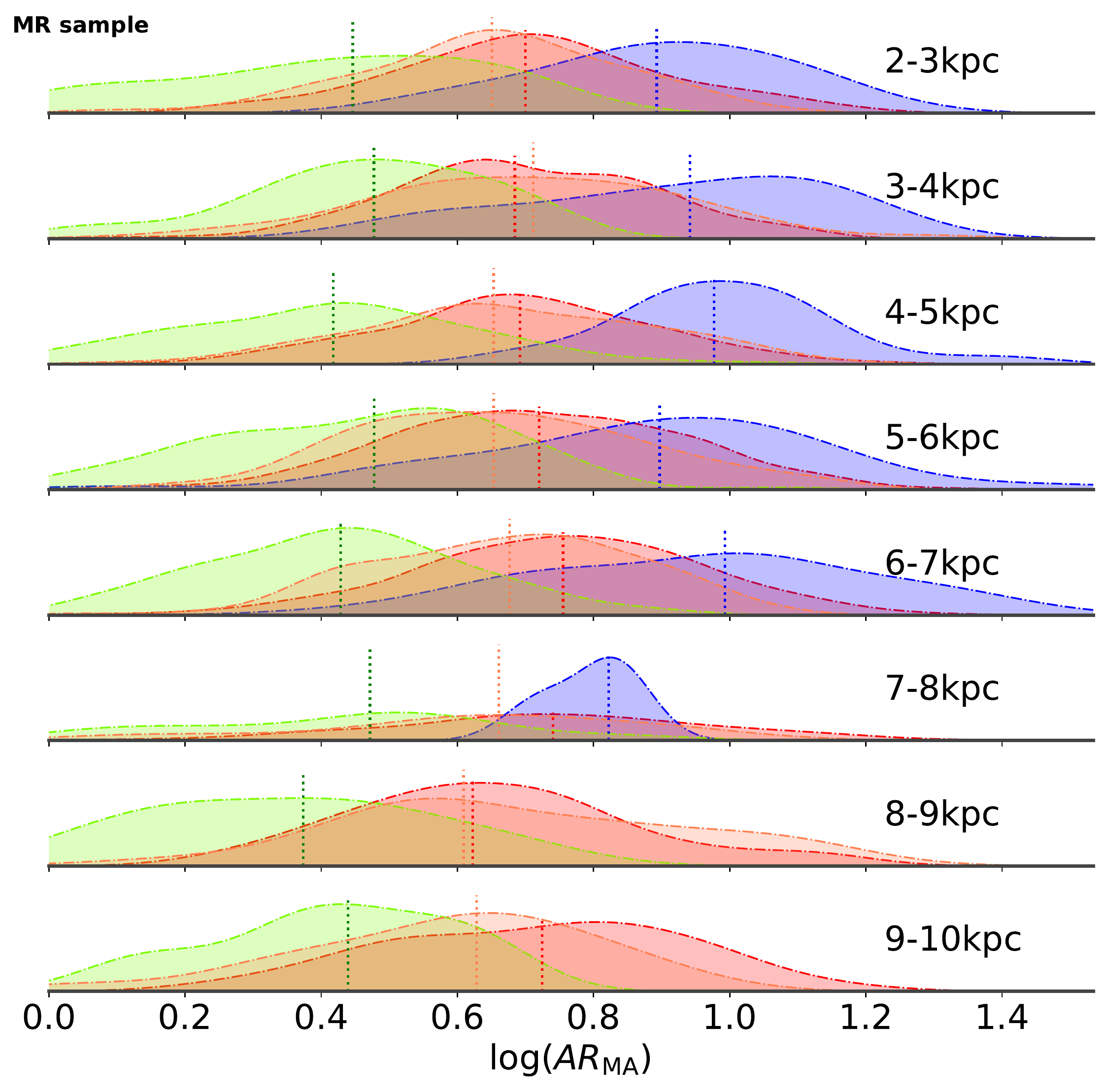}
    \caption{Aspect ratio ridge plots with $R_\mathrm{gal}$ bins. \textit{Left} (solid): VC sample, \textit{Right} (dashed): MR sample.}
    \label{fig: aspect ratio ridge plot R bins}
    \end{minipage}\hfill
\end{figure*}

\begin{figure*}
    \centering
    \begin{minipage}{\textwidth}
    \includegraphics[width = .5\textwidth, keepaspectratio]{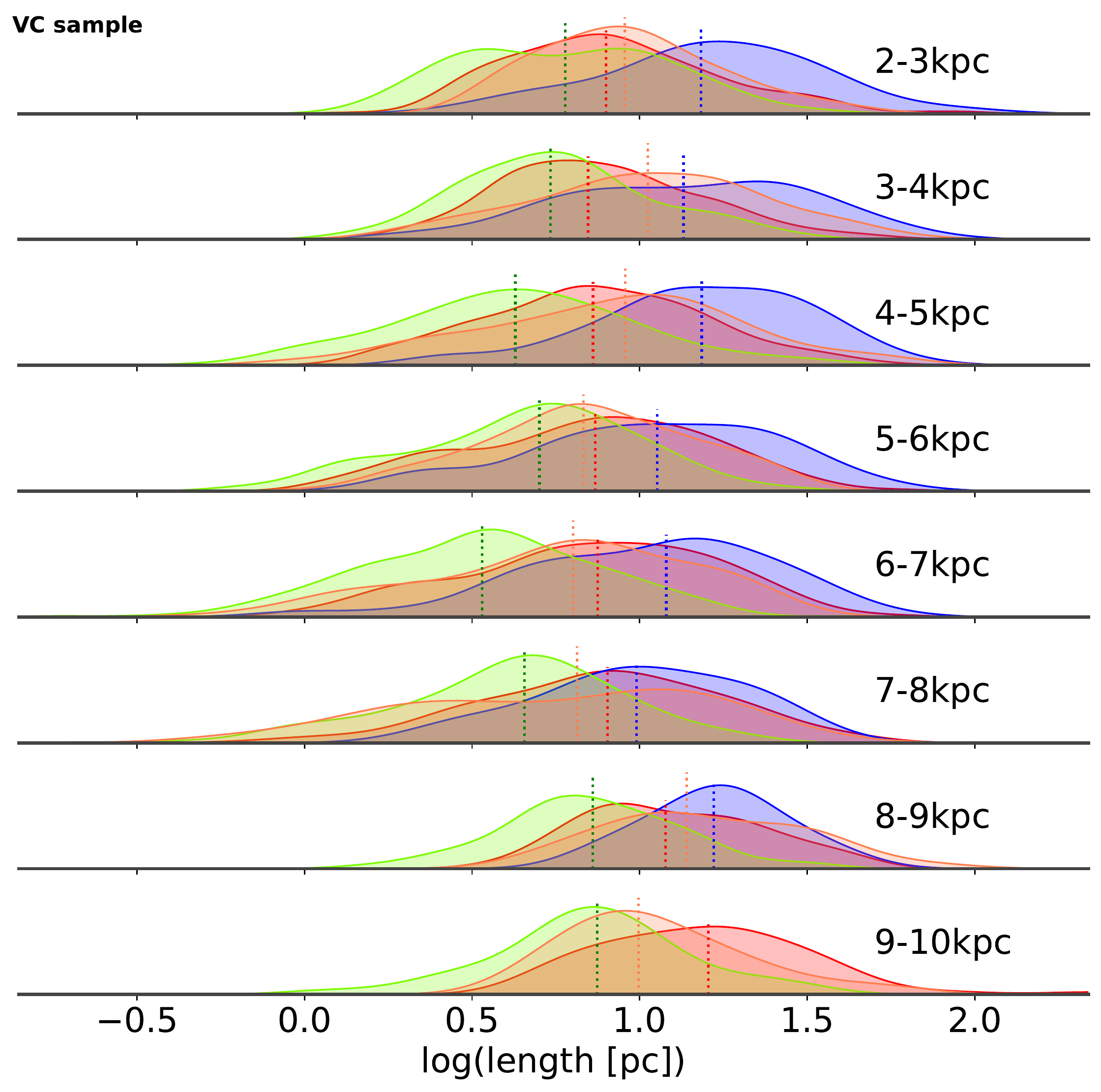}
    \includegraphics[width = .5\textwidth, keepaspectratio]{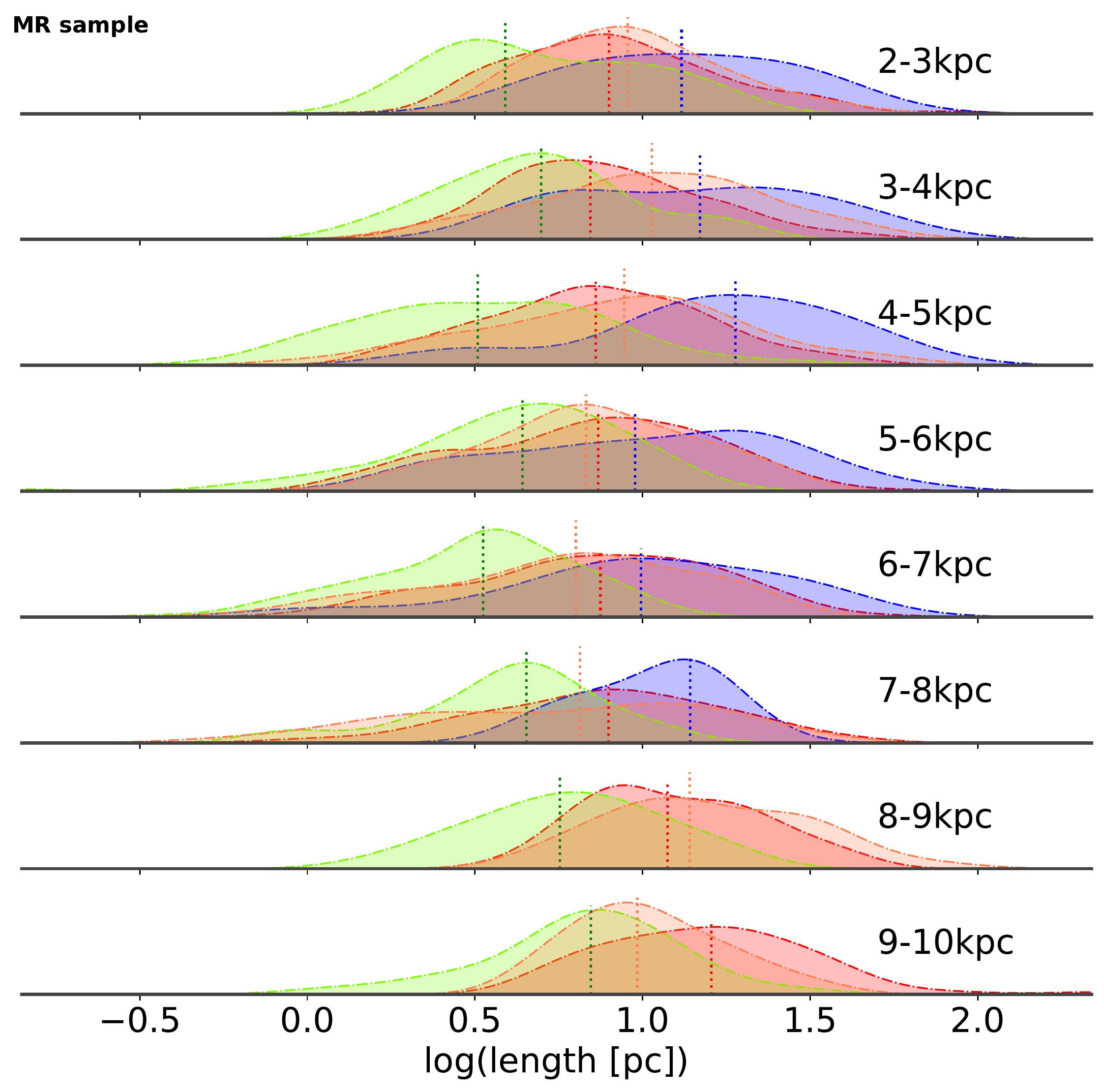}
    \caption{Length ridge plots with $R_\mathrm{gal}$ bins. \textit{Left} (solid): VC sample, \textit{Right} (dashed): MR sample.}
    \label{fig: length ridge plot R bins}
    \end{minipage}\hfill
\end{figure*}

The mass distribution for the two samples (Fig. \ref{fig: SCIENCE mass ridge R bins} and \ref{fig: morph mass ridge R bins}) show highest values for ring-like clouds and clumpy clouds. 
The surface density distributions (Fig. \ref{fig: surface density ridge plot R bins}) follow the global trends (Sec. \ref{sec: results}), showing high values for concentrated and clumpy clouds. The properties pertaining to the size of the clouds show similar features across most of the $R_\mathrm{gal}$ bins. These are the radius (Fig. \ref{fig: radius ridge plot R bins}) and length (Fig. \ref{fig: length ridge plot R bins}), which show highest values for ring-like clouds. The aspect ratio distribution (Fig. \ref{fig: aspect ratio ridge plot R bins}) also follows similar trends. These high values can be explained as a consequence of stellar feedback expanding the bubbles (see Sec. \ref{sec: discussions}). 

\begin{figure*}
    \centering
    \begin{minipage}{\textwidth}
    \includegraphics[width = .5\textwidth, keepaspectratio]{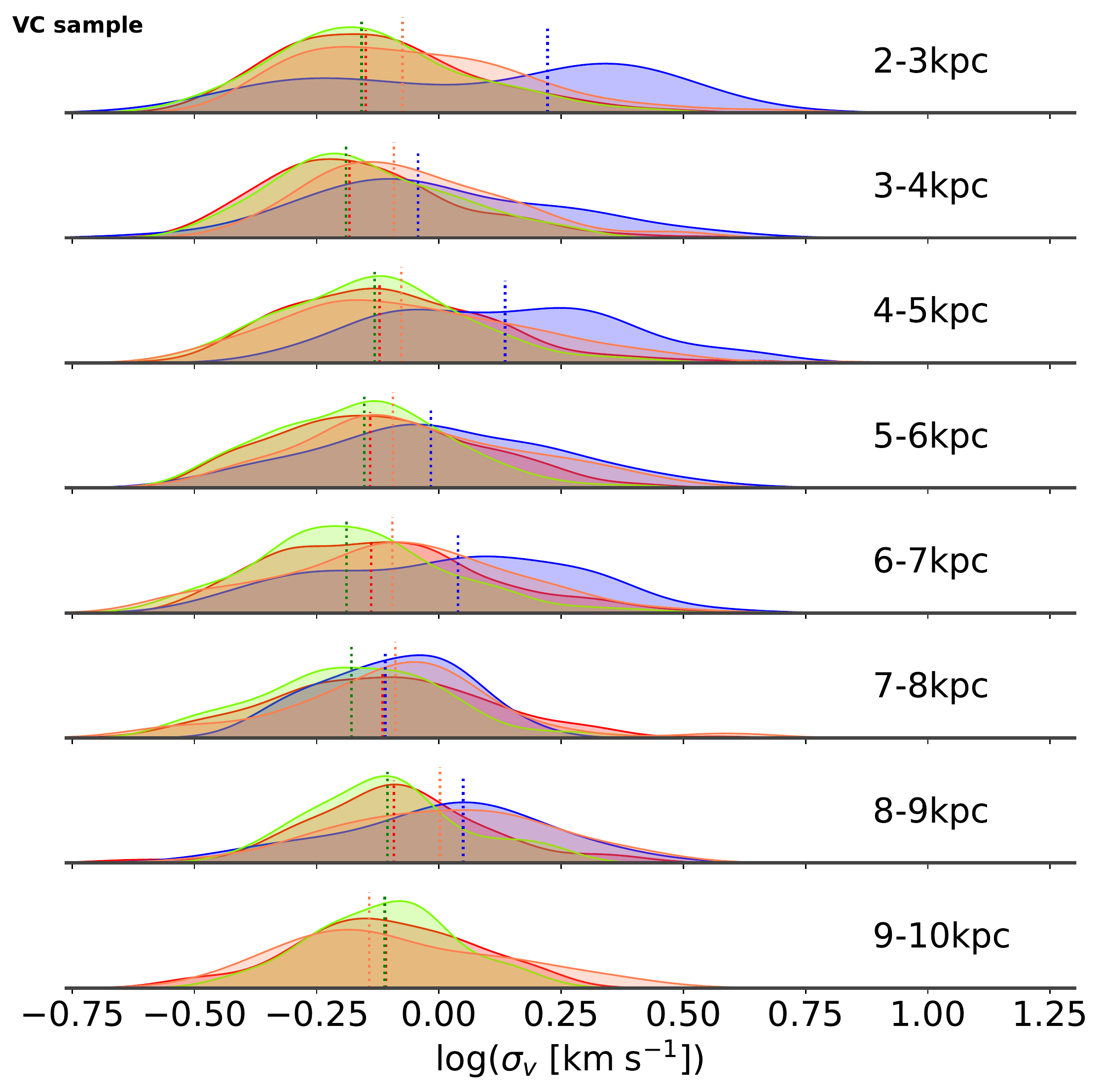}
    \includegraphics[width = .5\textwidth, keepaspectratio]{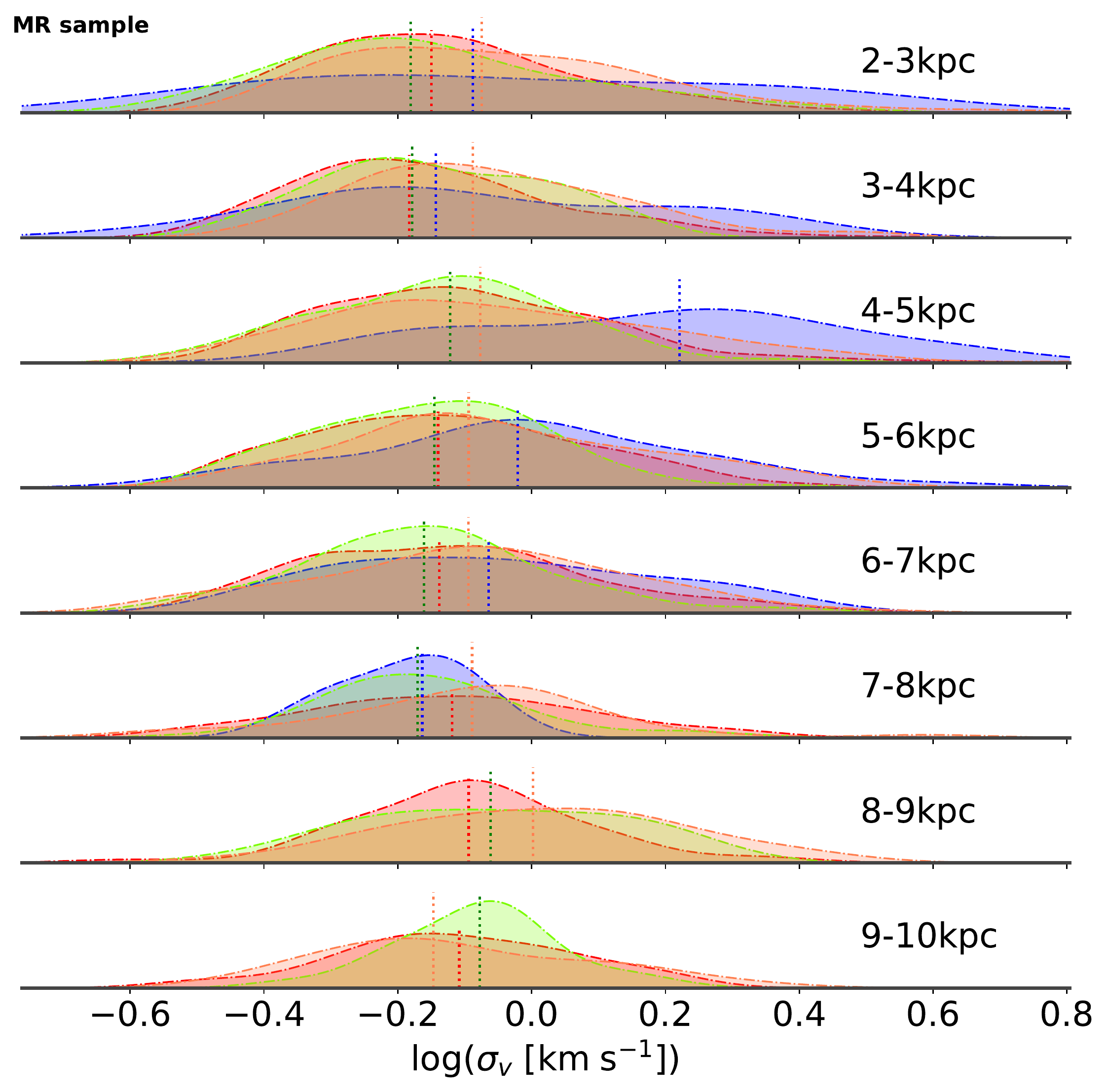}
    \caption{Velocity Dispersion ridge plots with $R_\mathrm{gal}$ bins. \textit{Left} (solid): VC sample, \textit{Right} (dashed): MR sample.}
    \label{fig: velocity dispersion ridge plot R bins}
    \end{minipage}\hfill
\end{figure*}

\begin{figure*}
    \centering
    \begin{minipage}{\textwidth}
    \includegraphics[width = .5\textwidth, keepaspectratio]{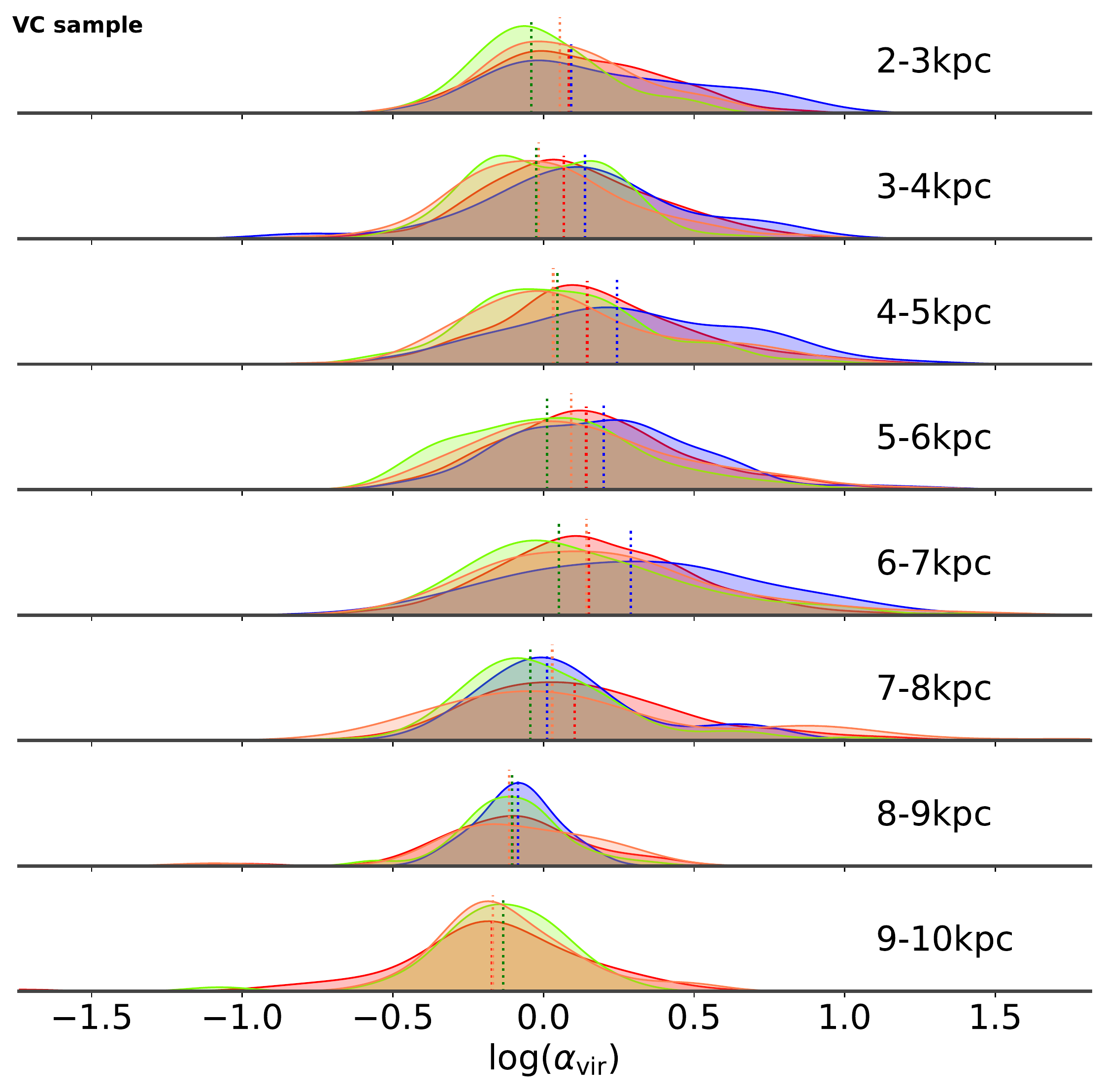}
    \includegraphics[width = .5\textwidth, keepaspectratio]{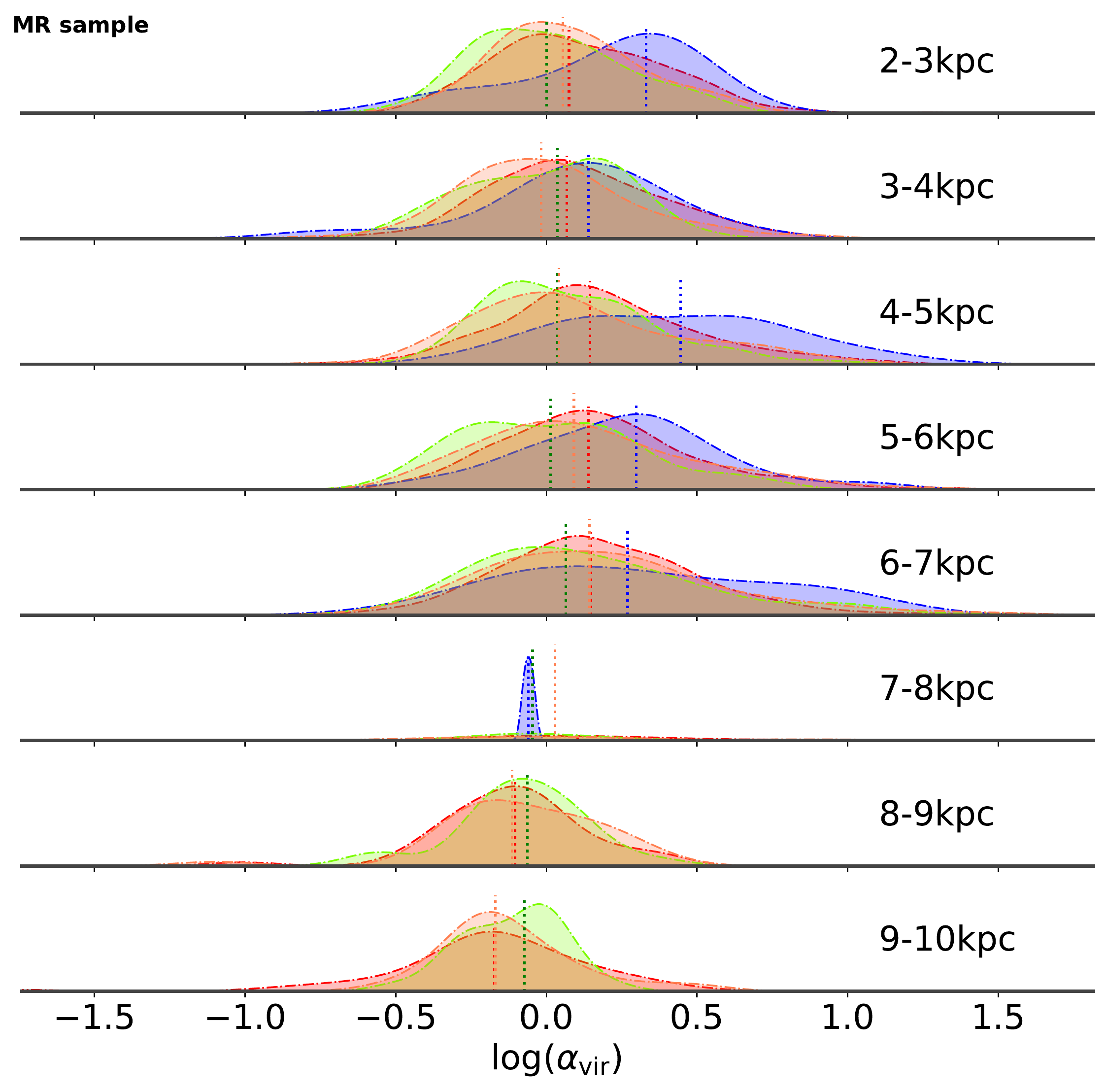}
    \caption{Virial Parameter ridge plots with $R_\mathrm{gal}$ bins. \textit{Left} (solid): VC sample, \textit{Right} (dashed): MR sample.}
    \label{fig: virial parameter ridge plot R bins}
    \end{minipage}\hfill
\end{figure*}

\subsection{Galactic height ($z_\mathrm{gal}$) bins}

\begin{figure*}[!h]
    \centering
    \includegraphics[width = 1\textwidth, keepaspectratio]{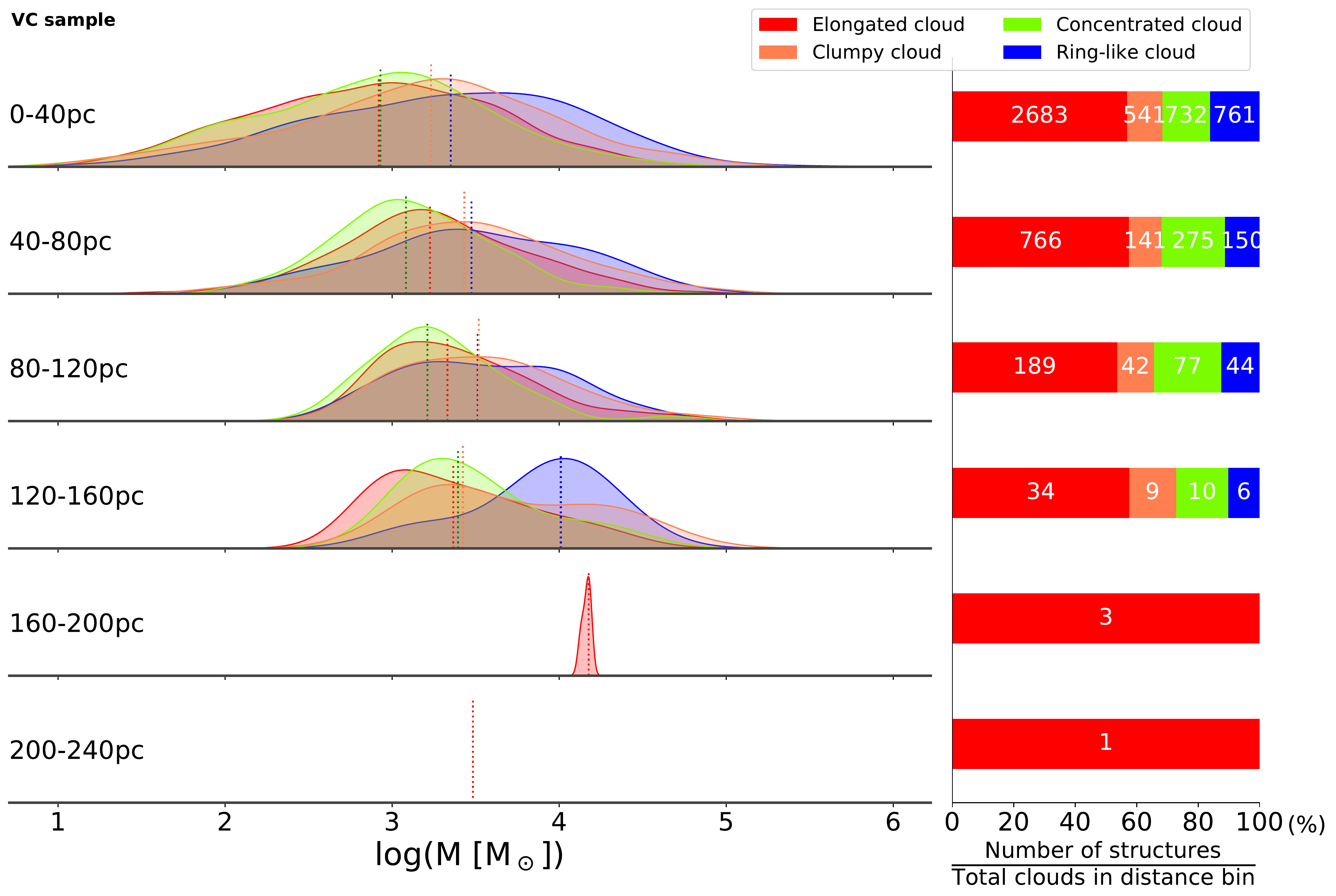}
    \caption{Mass ridge plot for the VC sample with $z_\mathrm{gal}$ bins.}
    \label{fig: SCIENCE mass ridge z bins}
\end{figure*}

\begin{figure*}
    \centering
    \includegraphics[width = 1\textwidth, keepaspectratio]{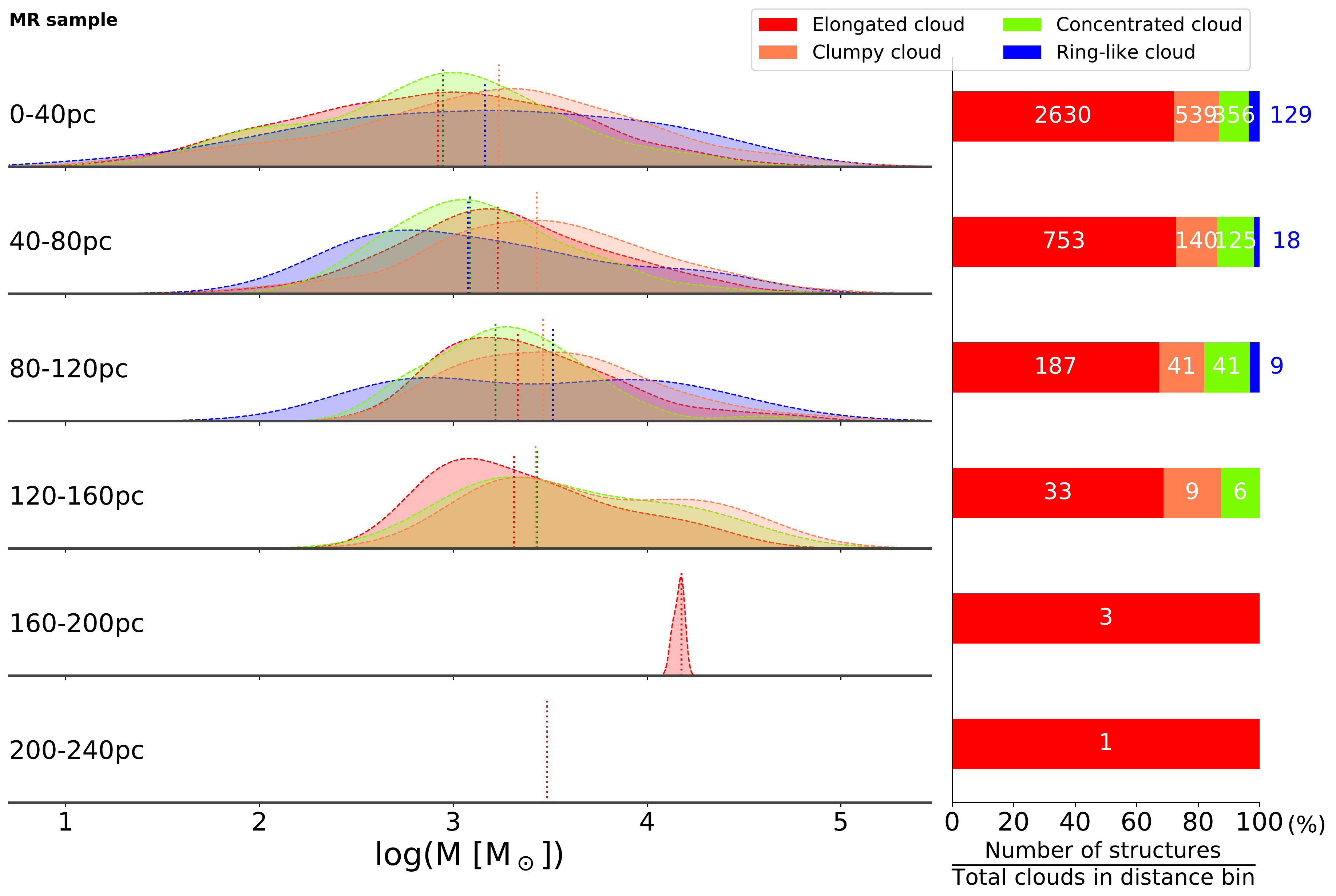}
    \caption{Mass ridge plot for the MR sample with $z_\mathrm{gal}$ bins.}
    \label{fig: morph mass ridge z bins}
\end{figure*}

\begin{figure*}
    \centering
    \begin{minipage}{\textwidth}
    \includegraphics[width = .5\textwidth, keepaspectratio]{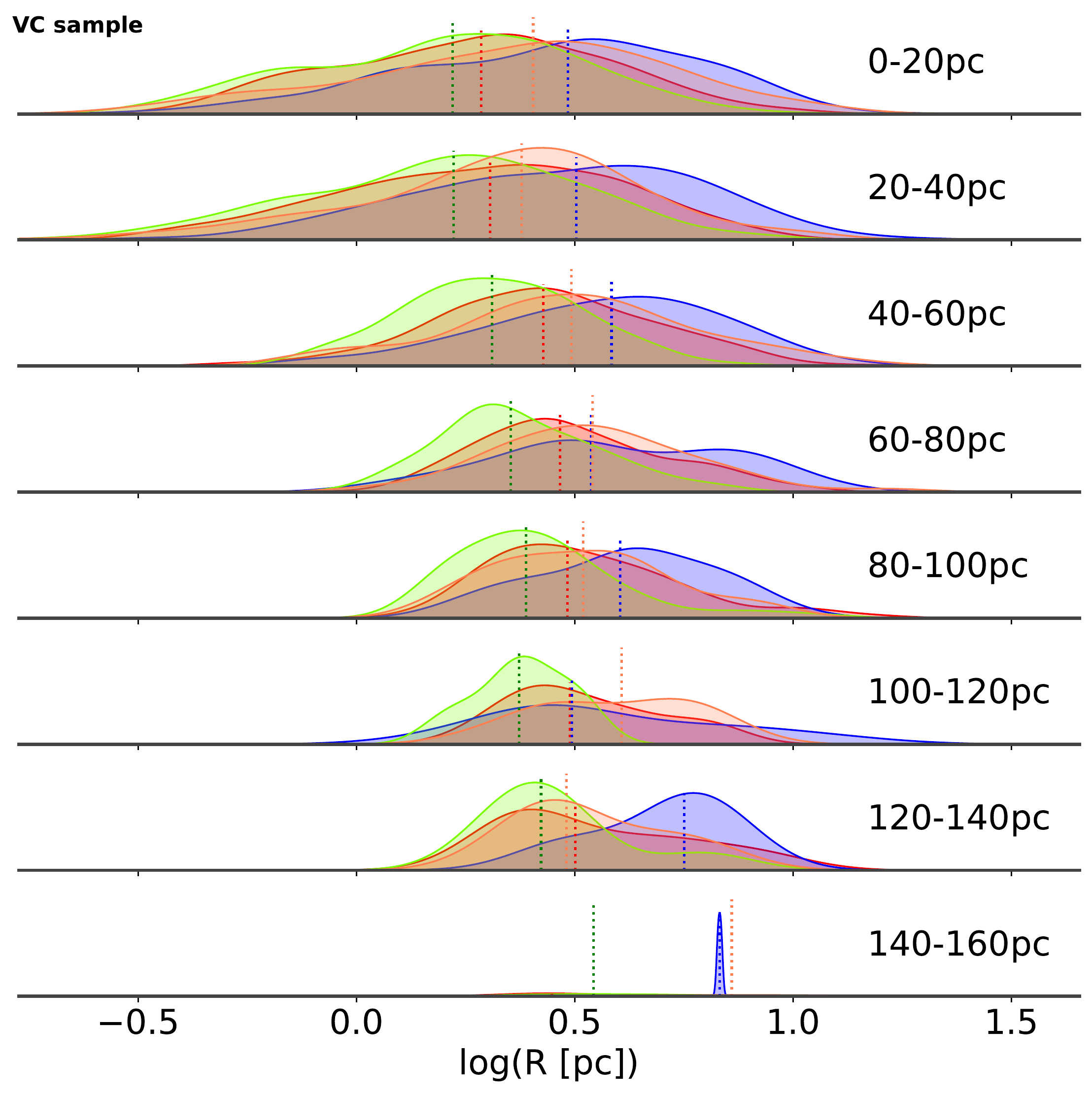}
    \includegraphics[width = .5\textwidth, keepaspectratio]{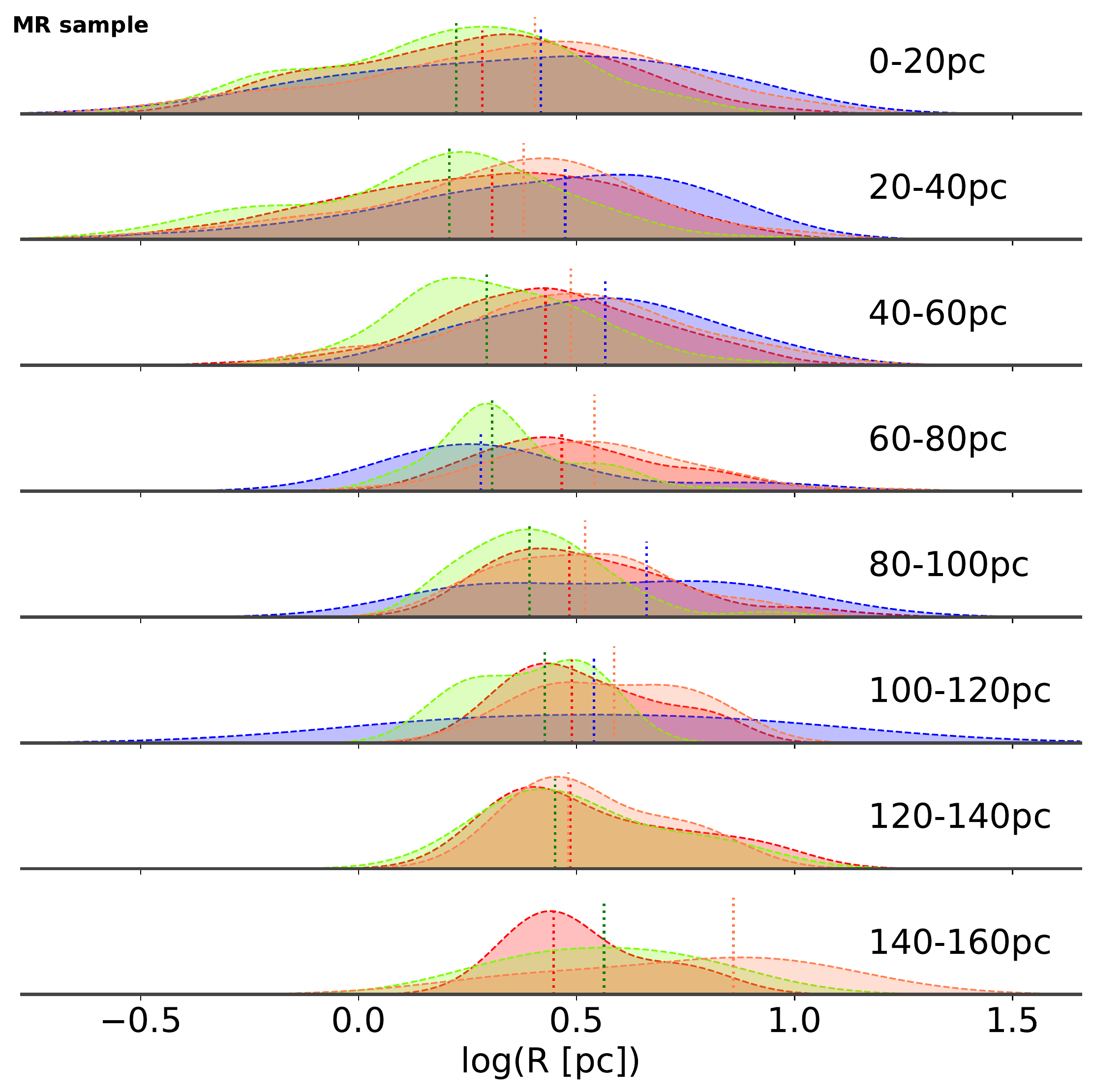}
    \caption{Radius ridge plots with $z_\mathrm{gal}$ bins. \textit{Left} (solid): VC sample, \textit{Right} (dashed): MR sample.}
    \label{fig: radius ridge plot z bins}
    \end{minipage}\hfill
\end{figure*}

\begin{figure*}
    \centering
    \begin{minipage}{\textwidth}
    \includegraphics[width = .5\textwidth, keepaspectratio]{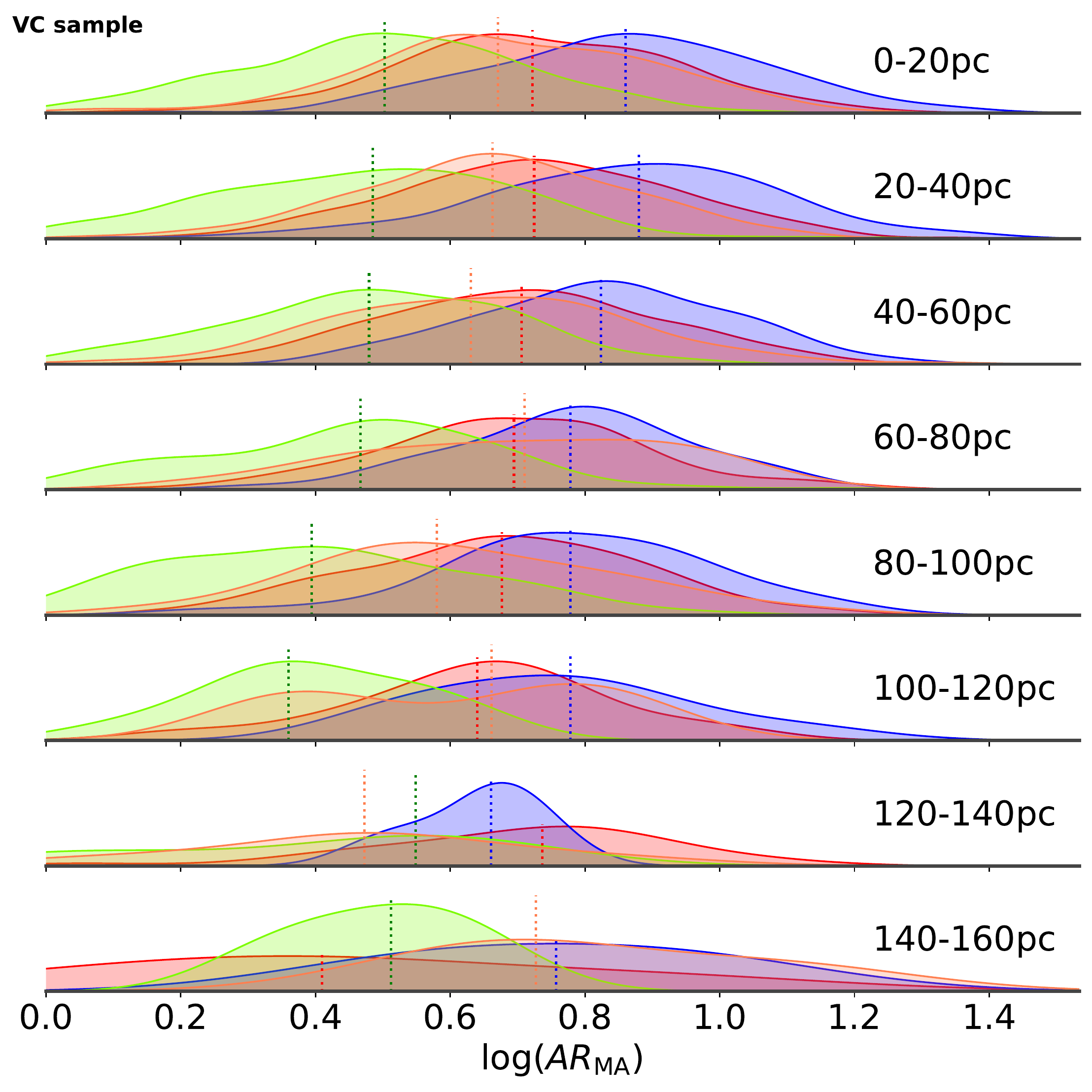}
    \includegraphics[width = .5\textwidth, keepaspectratio]{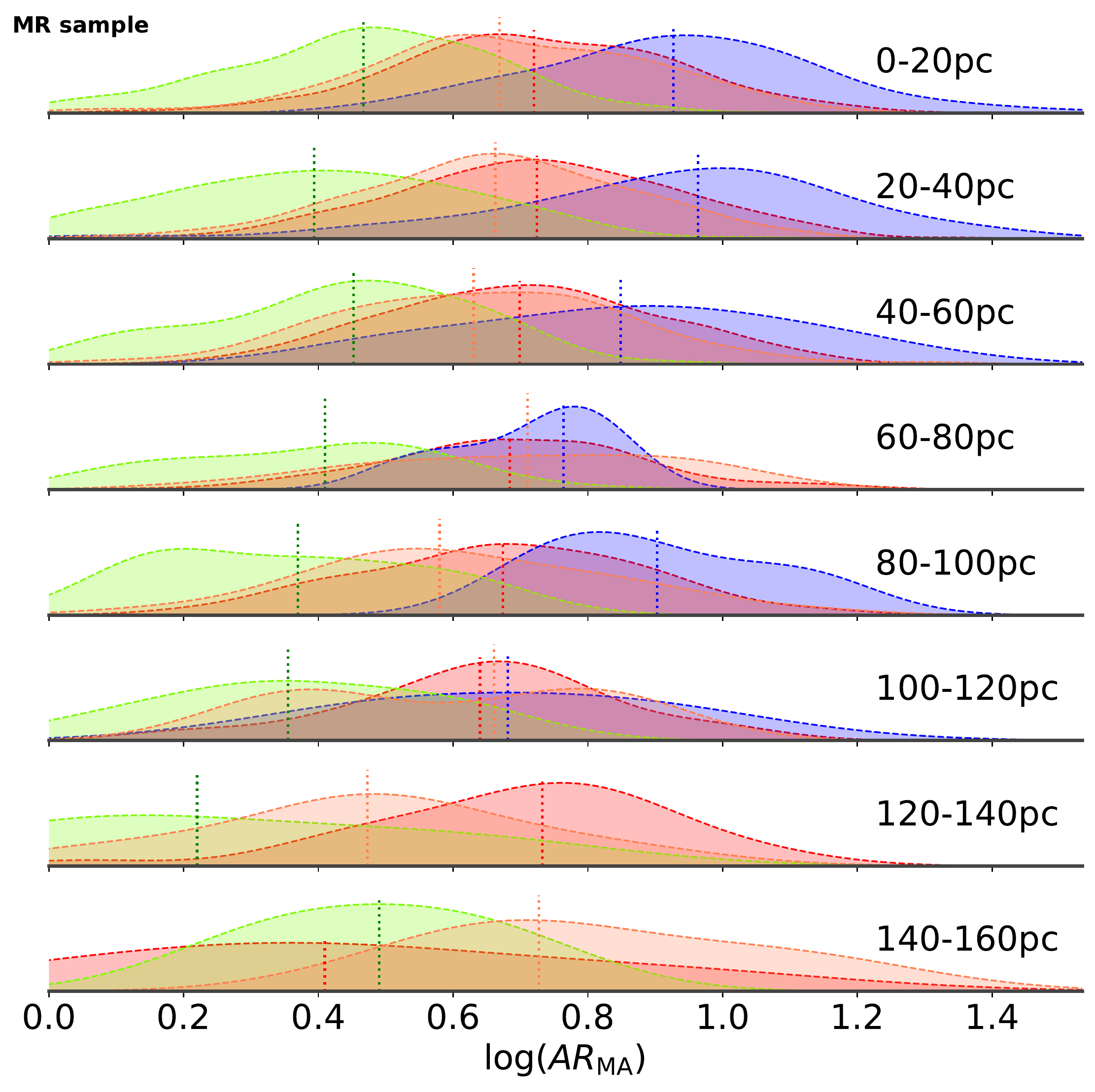}
    \caption{Aspect ratio ridge plots with $z_\mathrm{gal}$ bins. \textit{Left} (solid): VC sample, \textit{Right} (dashed): MR sample.}
    \label{fig: aspect ratio ridge plot z bins}
    \end{minipage}\hfill
\end{figure*}

\begin{figure*}
    \centering
    \begin{minipage}{\textwidth}
    \includegraphics[width = .5\textwidth, keepaspectratio]{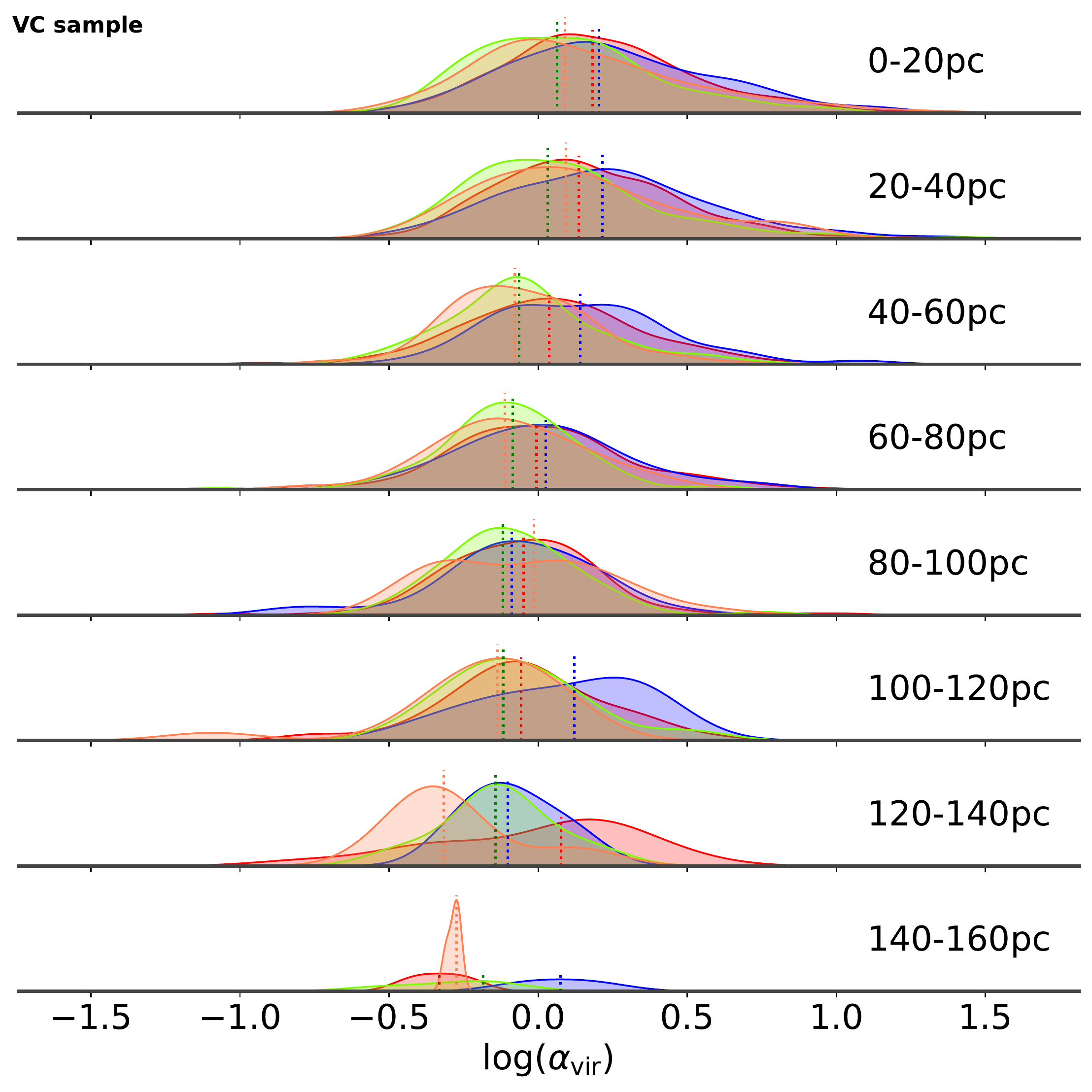}
    \includegraphics[width = .5\textwidth, keepaspectratio]{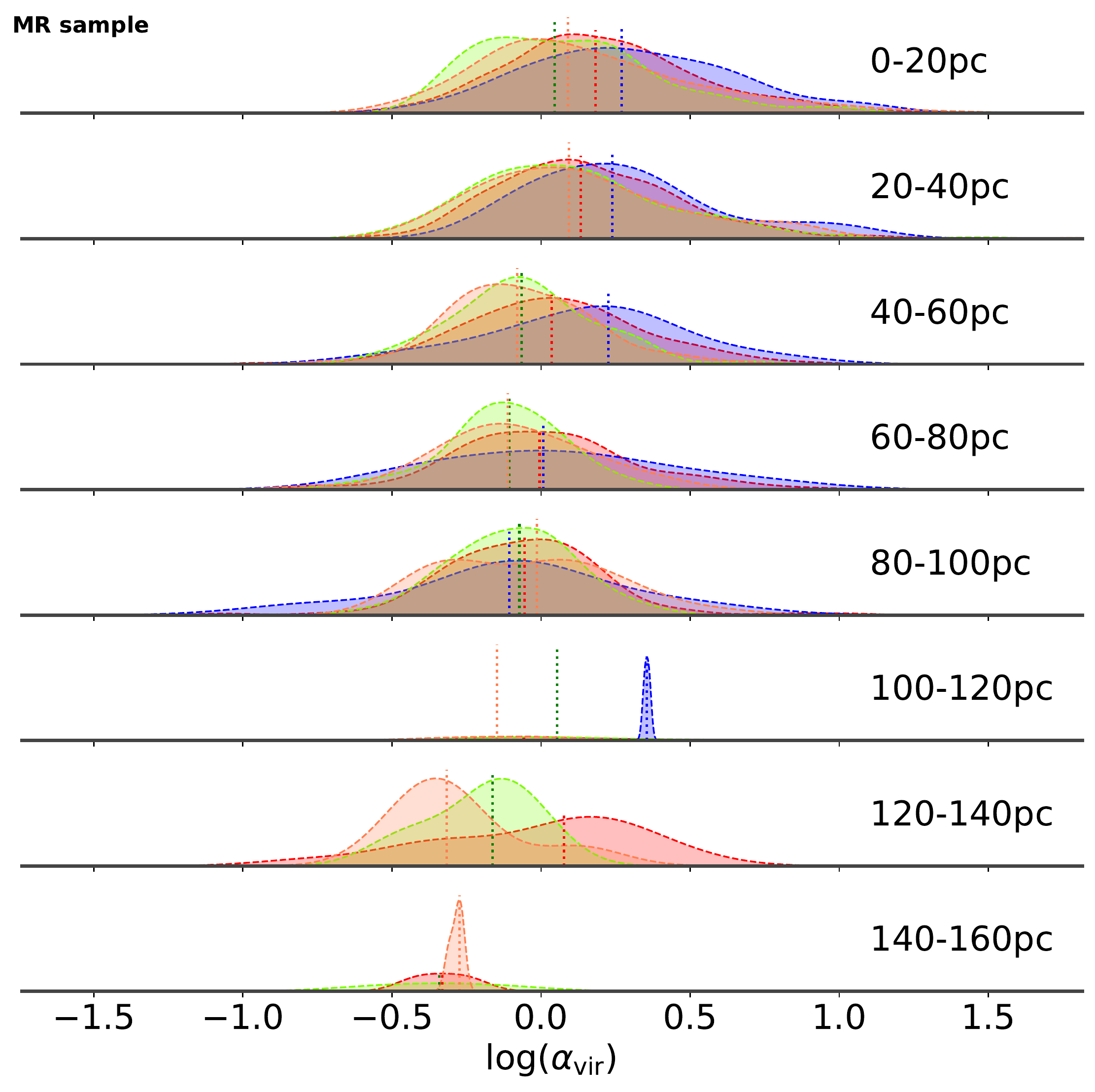}
    \caption{Virial parameter ridge plots with $z_\mathrm{gal}$ bins. \textit{Left} (solid): VC sample, \textit{Right} (dashed): MR sample.}
    \label{fig: virial parameter ridge plot z bins}
    \end{minipage}\hfill
\end{figure*}

\begin{figure*}
    \centering
    \begin{minipage}{\textwidth}
    \includegraphics[width = .5\textwidth, keepaspectratio]{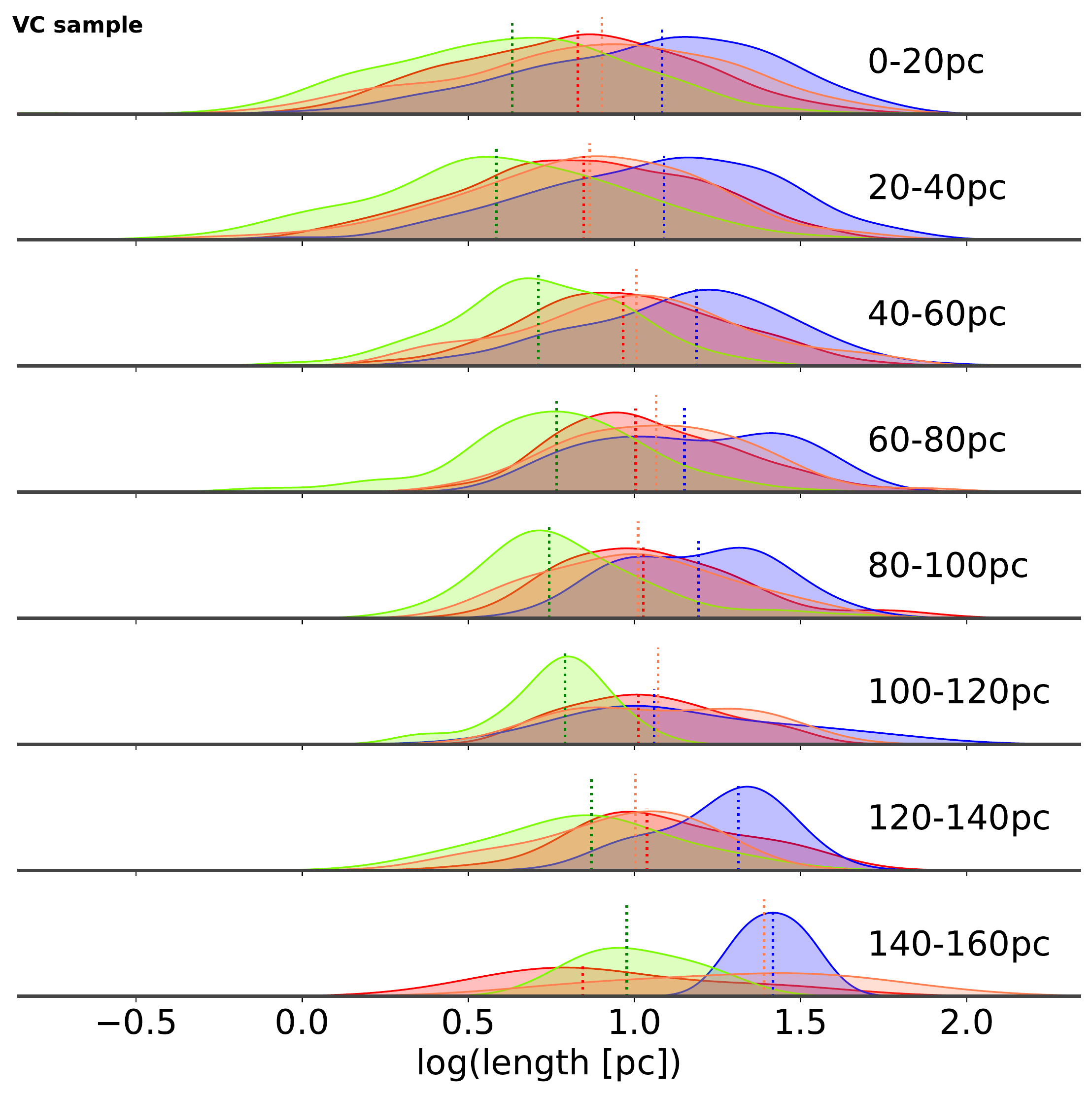}
    \includegraphics[width = .5\textwidth, keepaspectratio]{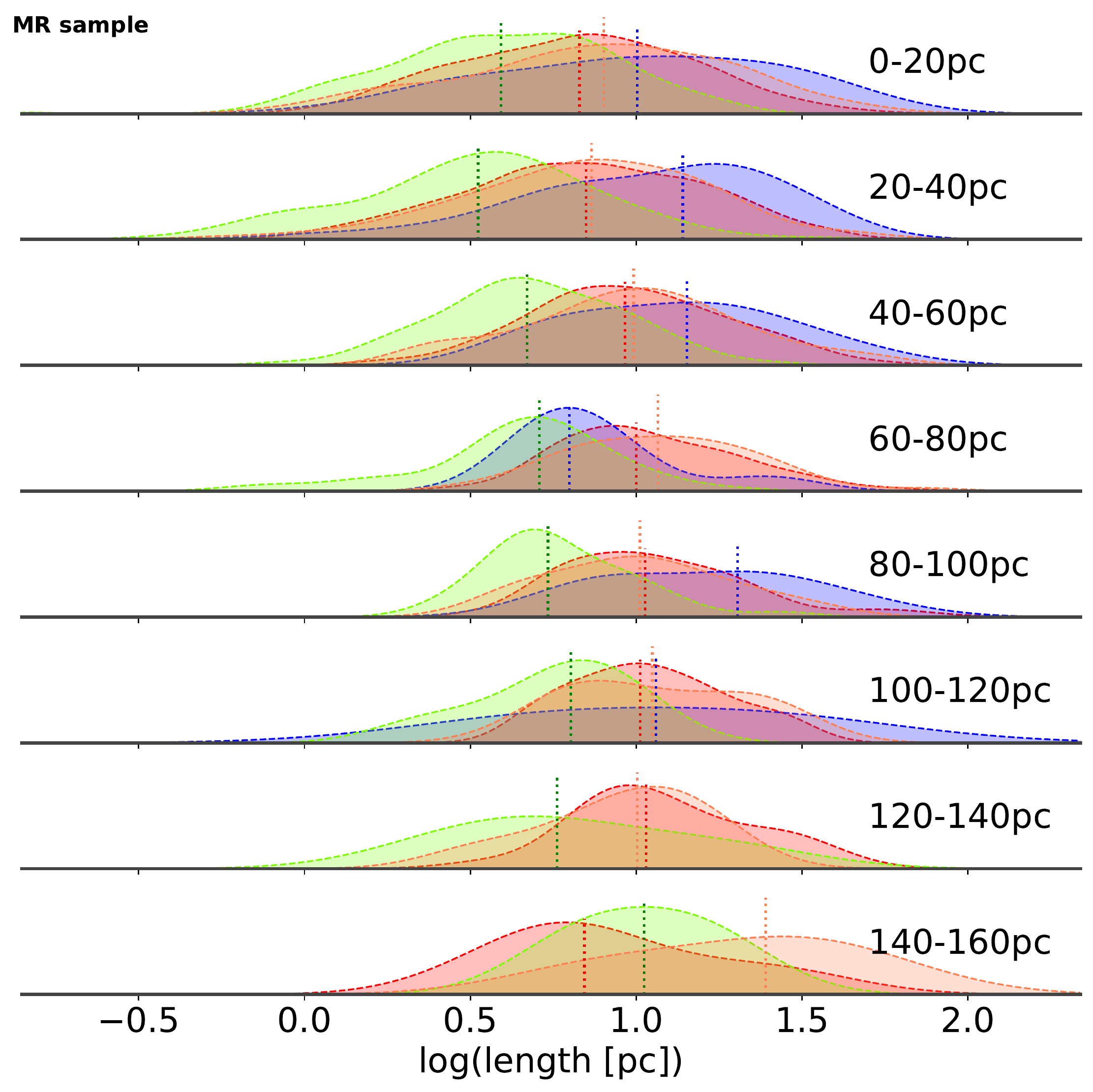}
    \caption{Length ridge plots with $z_\mathrm{gal}$ bins. \textit{Left} (solid): VC sample, \textit{Right} (dashed): MR sample.}
    \label{fig: length ridge plot z bins}
    \end{minipage}\hfill
\end{figure*}

\add{In this section, we study the clouds by distributing them into $z_\mathrm{gal}$ bins. The highest $z_\mathrm{gal}$ value for the clouds is $\approx$ 230 pc, however, < 1 $\%$ clouds lie above 160 pc. Thus, we bin the clouds into 8 bins of width 20 pc covering 0--160 pc. 
Clouds near the Galactic plane, have a larger span of sizes (both radius and length) as seen in Fig. \ref{fig: radius ridge plot z bins} and \ref{fig: length ridge plot z bins}. Smaller clouds at high Galactic heights might be obscured due to the foreground gas in the Galactic plane. The sensitivity and resolution of telescope could also prevent detection of smaller clouds at large distances.
Clouds at all Galactic height exhibit almost similar aspect ratio distributions (Fig. \ref{fig: aspect ratio ridge plot z bins}), showing that they show similar average shapes for a given morphology. 
We also see that less massive clouds are mostly located near the Galactic plane (Fig. \ref{fig: SCIENCE mass ridge z bins} and \ref{fig: morph mass ridge z bins}). The virial parameter distribution has a mild trend suggesting a lower average value with increasing Galactic height (Fig. \ref{fig: virial parameter ridge plot z bins}). In a nutshell, we detect a larger number of smaller and less massive cloud in the Galactic plane compared to large Galactic heights, but this could be due to the foreground molecular gas, telescope limitations or both.}

\begin{figure*}
    \centering
    \begin{minipage}{\textwidth}
    \includegraphics[width = .5\textwidth, keepaspectratio]{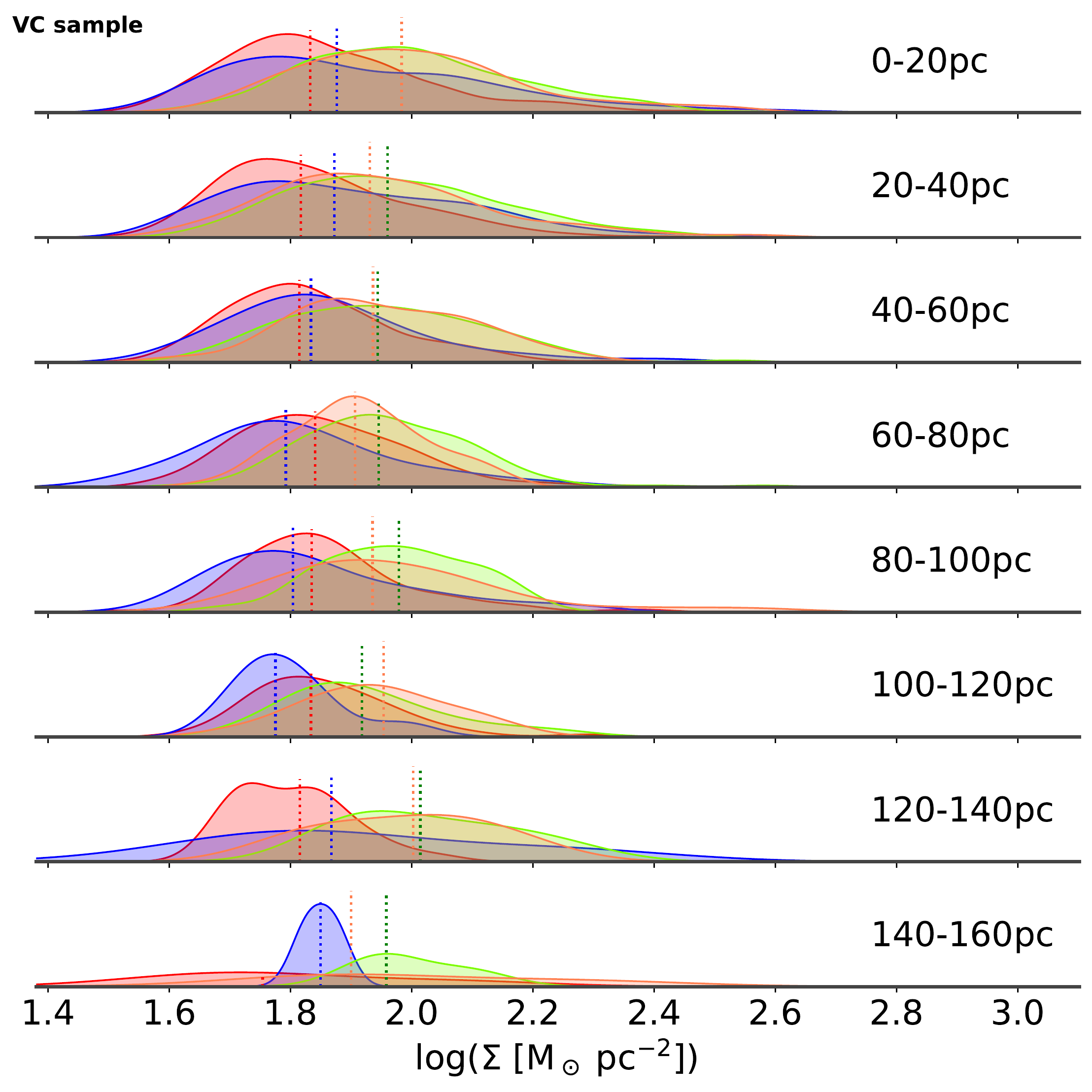}
    \includegraphics[width = .5\textwidth, keepaspectratio]{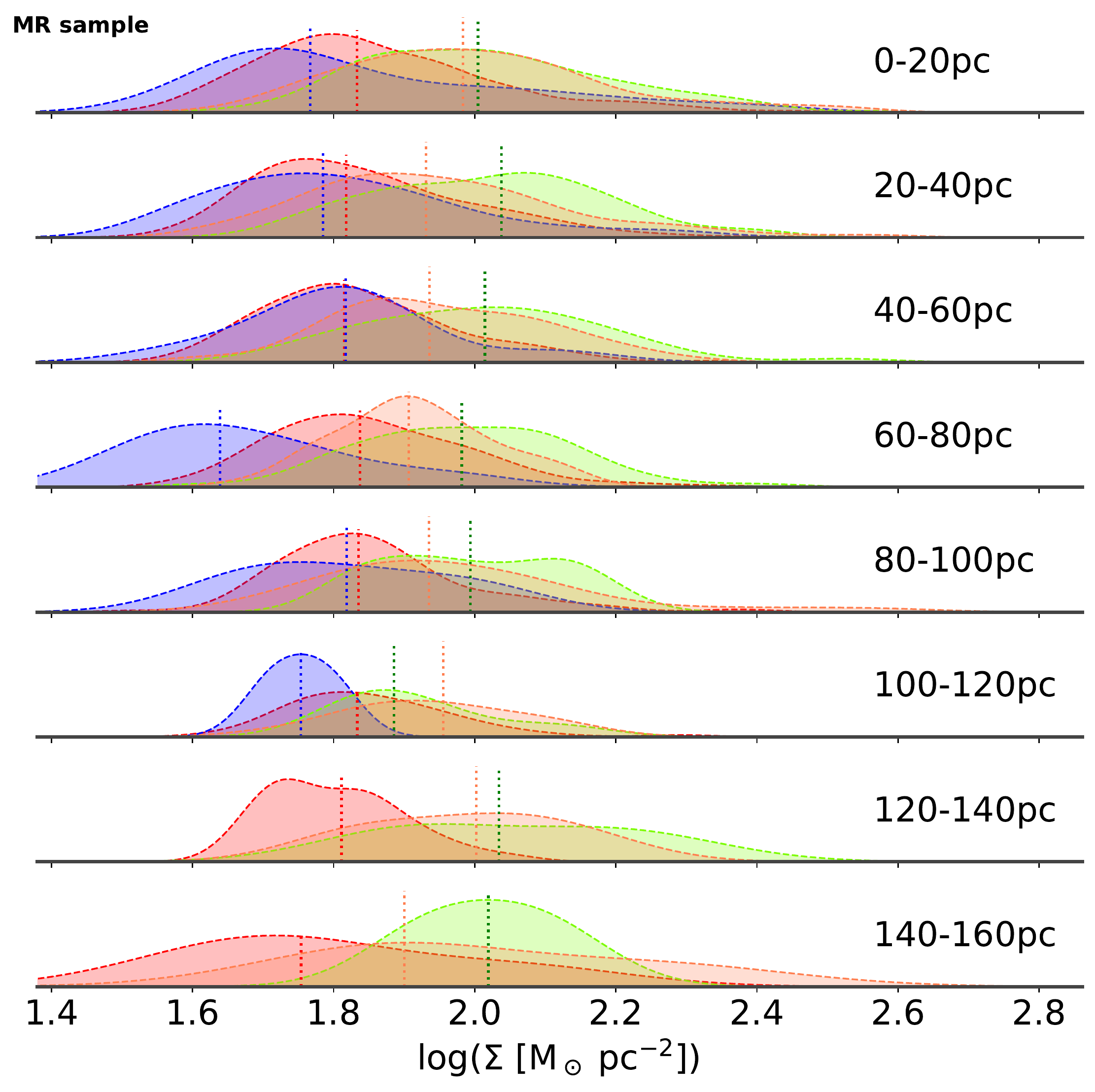}
    \caption{Surface density ridge plots with $z_\mathrm{gal}$ bins. \textit{Left} (solid): VC sample, \textit{Right} (dashed): MR sample.}
    \label{fig: Surface density ridge plot z bins}
    \end{minipage}\hfill
\end{figure*}

\begin{figure*}
    \centering
    \begin{minipage}{\textwidth}
    \includegraphics[width = .5\textwidth, keepaspectratio]{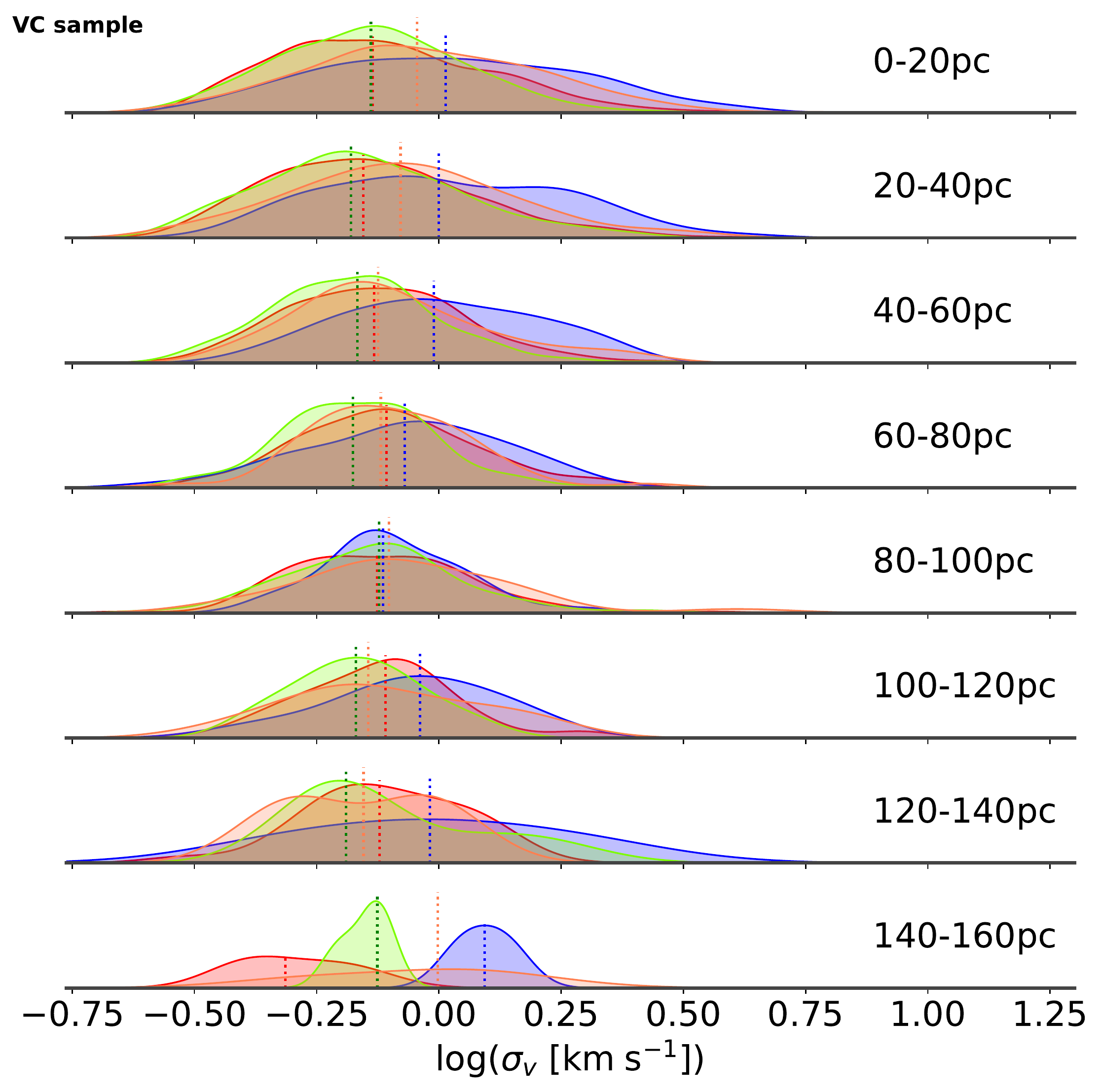}
    \includegraphics[width = .5\textwidth, keepaspectratio]{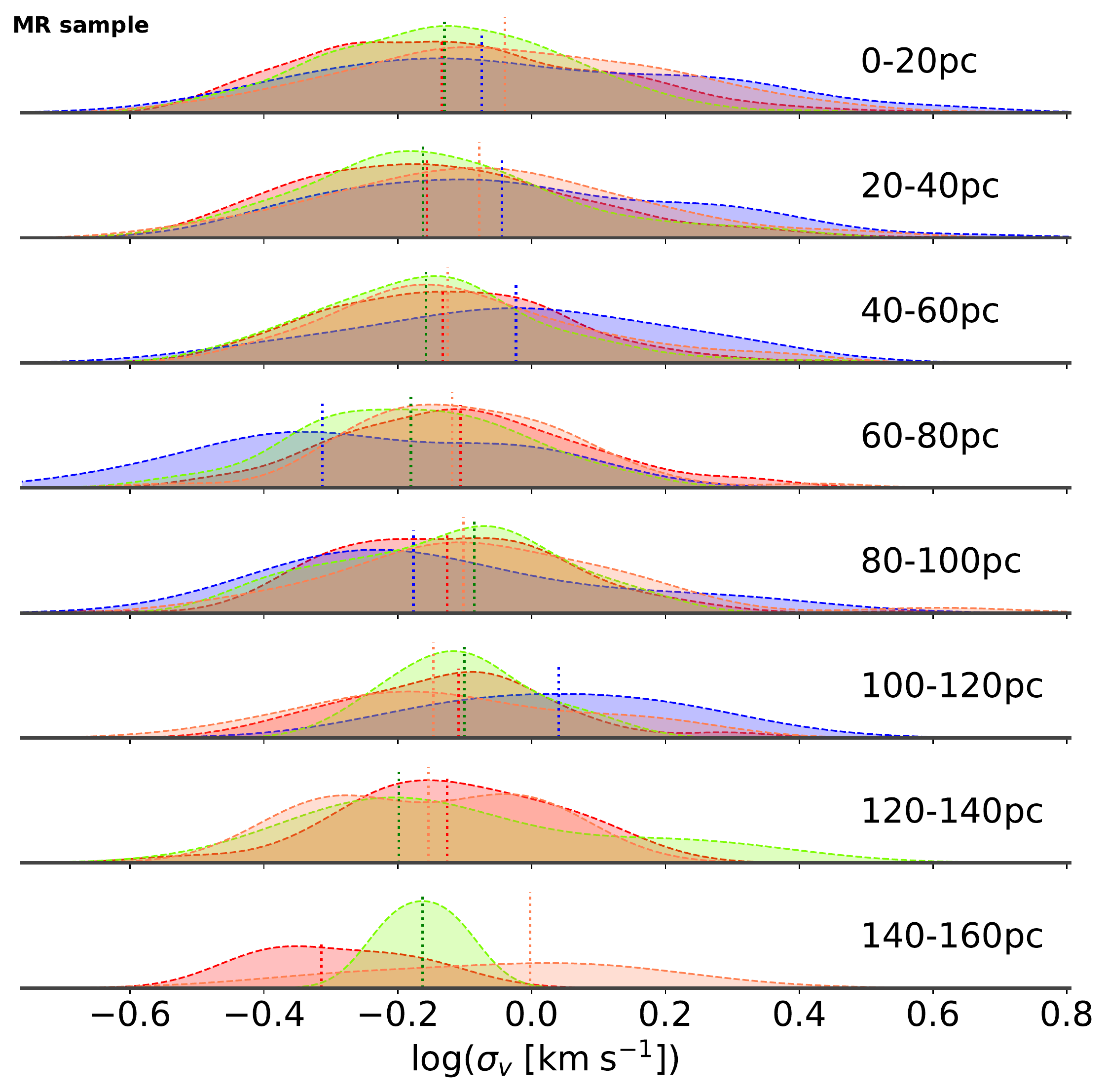}
    \caption{Velocity Dispersion ridge plots with $z_\mathrm{gal}$ bins. \textit{Left} (solid): VC sample, \textit{Right} (dashed): MR sample.}
    \label{fig: velocity dispersion ridge plot z bins}
    \end{minipage}\hfill
\end{figure*}

\end{appendix}

\end{document}